\documentclass[letterpaper,twocolumn,10pt]{article}
\PassOptionsToPackage{hyphens}{url}  
\usepackage{jsys}

\microtypecontext{spacing=nonfrench}

\usepackage{tikz}
\usepackage{amsmath}
\usepackage{amsthm}
\usepackage{amsfonts}
\usepackage{graphics}

\usepackage{algorithm}
\usepackage{algpseudocode}
\usepackage{algorithmicx}

\usepackage{bbding}
\AtBeginEnvironment{align}{\setcounter{equation}{0}} 
\usepackage{caption}
\usepackage{subcaption}

\definecolor{cerisepink}{rgb}{0.93, 0.23, 0.51}
\newcommand{\rb}[1]{{\textcolor{black}{ #1}}}
\newcommand{\arxiv}[1]{{\textcolor{black}{ #1}}}

\jsysfinaltrue

\usepackage{orcidlink}

\begin{document}

\date{}

\title{\arxiv{Rethinking Inference Placement for Deep Learning across \\ Edge and Cloud Platforms: A Multi-Objective Optimization Perspective and Future Directions}}

\author{
{\textsc ~Zongshun Zhang}\\
Department of Computer Science \\
Boston University
\and
{\textsc ~Ibrahim Matta}\\
Department of Computer Science \\
Boston University
} 

\maketitle

\thispagestyle{empty}

\begin{abstract}
Edge intelligent applications like VR/AR and language model based chatbots have become widespread with the rapid expansion of IoT and mobile devices.
However, constrained edge devices often cannot serve the increasingly large and complex deep learning (DL) models. 
To mitigate these challenges, researchers have proposed optimizing and offloading partitions of DL models among user devices, edge servers, and the cloud. 
In this setting, users can take advantage of different services to support their intelligent applications. 
For example, edge resources offer low response latency.
In contrast, cloud platforms provide low monetary cost computation resources for computation-intensive workloads. 
However, communication between DL model partitions can introduce transmission bottlenecks and pose risks of data leakage. 
Recent research aims to balance accuracy, computation delay, transmission delay, and privacy concerns. 
They address these issues with model compression, model distillation, transmission compression, and model architecture adaptations, including internal classifiers. 
\rb{This survey contextualizes the state-of-the-art model offloading methods and model adaptation techniques by studying their implication to a multi-objective optimization comprising inference latency, data privacy, and resource monetary cost.}
\end{abstract}

\section{Introduction}
\begin{figure*}
\captionsetup{justification=centering}
    \includegraphics[width=\textwidth]{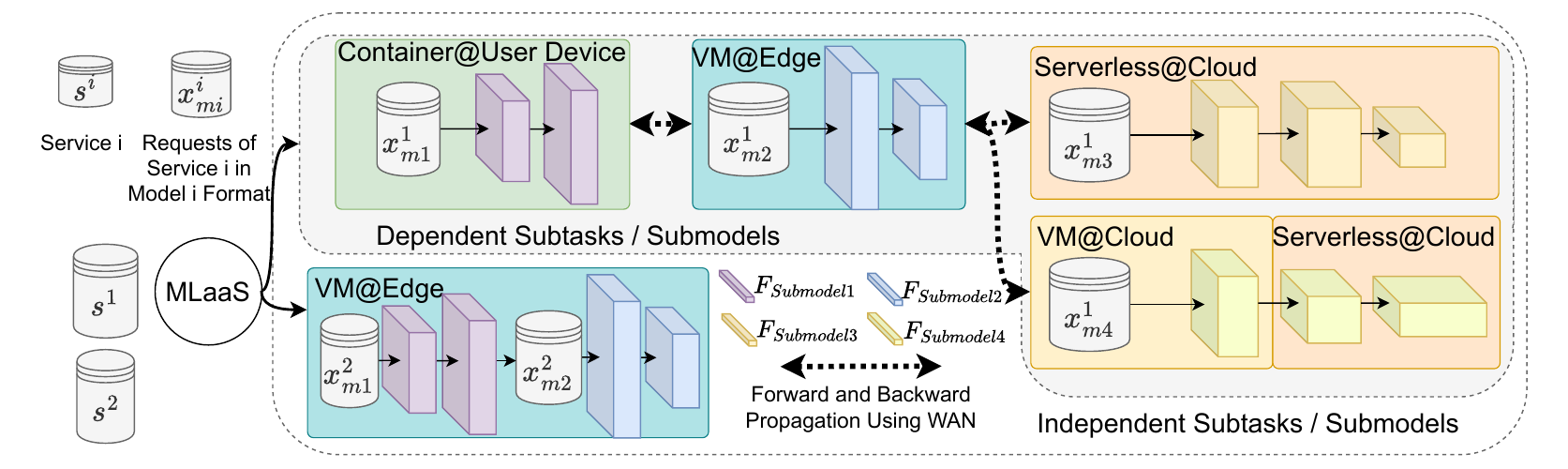}
    \caption{A MLaaS System Offloads Service/Applications $s^{i}$ across the Continuum of Device, Edge and Cloud Resources Using Different Resource Orchestration Platforms.
    }\label{fig:continuum}
\end{figure*}

In recent decades, large volumes of data have been generated on mobile and Internet-of-Things (IoT) devices.
Cloud computing resources remain more flexible in scaling and management than edge computing resources. 
However, edge computing can mitigate transmission bottlenecks over the wide-area network connecting users to the cloud~\cite{Redhat-White-Paper}.
Recent intelligent systems focus on offloading user applications across the continuum of personal device, edge and cloud resources~\cite{Edge-Systems-Survey, Edge-Systems-Survey-1, continuum}, leveraging resource orchestration services in the cloud and at the edge, while optimizing for latency, privacy, and monetary cost.
As exemplified in Fig.~\ref{fig:continuum}, a Machine Learning-as-a-Service (MLaaS) system can provision resources across the continuum of resources using various resource provisioning platforms -- container, Virtual Machine (VM), or Serverless function -- for a Machine Learning (ML) request $x^{1}$. 

ML applications have diverse performance requirements and face varying resource limits. 
For instance, mobile and IoT applications, including object recognition in housekeeping AIoT devices~\cite{AIoT-housekeeping} and localization in autonomous cars~\cite{CarMap}, are constrained by energy consumption~\cite{energy-edge-survey} or rely on emerging infrastructures such as 5G/6G base stations and smart city networks~\cite{AR-5G-edge-survey, AR-5G-edge-survey-1, AIoT-smartcity}. 

At the same time, the high monetary cost associated with computational resources for AI/ML training and inference creates a barrier, especially for research institutions and smaller companies. 
In contrast, large firms can afford to build extensive Deep Learning (DL) clusters. 
For example, pre-training LLaMA-3.1-8B (LLaMA-3.1-405B) requires $1.46$ million ($30.84$ million) in H100-80GB GPU hours~\cite{llama-3-1-model-card}, and ByteDance operates a cluster of over $10,000$ NVIDIA Ampere GPUs for their Large Language Model (LLM) workloads~\cite{jiang2024megascale}.

As AI/ML applications become popular, creating an infrastructure that is monetary cost-efficient, latency-optimized, and privacy-aware for architecture-optimized models has emerged as a critical area of research~\cite{survey-collaborate-EC, memory-wall, deepseek-r1}.
This survey studies the computation offloading problem across the client devices, edge and cloud resources, in particular for inference tasks.
Given a Neural Network (NN) model and a data source, we can split the model and provision its parts on cloud services, edge servers, or client devices using resource orchestration services (e.g., Virtual Machines, Containers, etc.)
As shown in Fig.~\ref{fig:continuum}, the model decomposition creates either an ensemble of independent submodels, or a tandem of dependent submodels (partitions of layers, given a multi-layered ML model).
Then, we consider an MLaaS system that places the submodels on resources across the continuum of user devices, edge and the cloud using different resource orchestration methods including containers, VMs, Serverless, etc.
This model decomposition research complements related work that further improves inference latency, source data privacy and resource monetary cost of ML systems through techniques such as quantization~\cite{NN-quant-survey}, weight pruning~\cite{Pruning-and-quantization-survey}, distillation~\cite{gou2021knowledge}, privacy preserving distributed learning~\cite{survey-trustworthy-dist-ai}, and monetary cost-based resource provisioning using edge and cloud resources~\cite{survey-ML-services-provisioning}.
We note that by treating each submodel as a general operator, we believe this research applies broadly to many applications.

\subsection{The Cost Model of Machine Learning-as-a-Service}
Machine Learning-as-a-Service (MLaaS) is a resource orchestration model that provides compute resources for ML training and inference on a pay-per-use basis. 
As exemplified in Fig.~\ref{fig:continuum}, an MLaaS system can provision resources across the continuum of user devices, edge, and cloud resources.
For example, for an ML service $s^{1}$ which prioritizes resource cost while maintaining low latency, an MLaaS framework can deploy the ML model in the cloud—leveraging the high parallelism of cloud resources—and offload the lightweight, heavily used portions to edge or user devices using containers or VMs.
On the other hand, for an ML service $s^{2}$ that prioritizes the privacy of source data, an MLaaS broker could provision the model using particular secure computation resources like client devices or VMs.

The MLaaS system abstracts away infrastructure management, so users only pay for what they consume.
Existing MLaaS systems provide managed services in the cloud and at the edge. 
Depending on how much configuration they allow, MLaaS systems expose resources at various levels of abstraction~\cite{MLaaS-complexity-performance}. 
For example, AWS provides a set of \textit{AI services}, including Amazon Rekognition~\cite{AWS-Rekognition}, which hide lower-level details like the ML models and computation resources from users. 
In contrast, their \textit{ML services}, such as Amazon Sagemaker~\cite{AWS-sagemaker}, allow users to define models, data sources, and resource orchestration across VM instances, serverless instances, S3 object storage, etc. 
Sagemaker Edge~\cite{AWS-sagemaker-edge} further extends this by deploying NN models on user-owned IoT devices and gathering data for inference and retraining.

MLaaS is vital for many ML workloads.
Although modern Large Language Model (LLM) training and inference demand enormous compute power and high inter-node bandwidth, typically available only in private GPU clusters~\cite{jiang2024megascale}, small teams can leverage parameter-efficient fine-tuning and inference on smaller models within public cloud or edge-based MLaaS environments.
For example, fine-tuning a $65B$ parameter model with a Low-Rank Adapter (LoRA) requires only a single $48GB$ GPU for less than $24$ hours~\cite{dettmers2023qlora}.
Likewise, running inference on a $Qwen2-7B-Instruct$ model using one $A100-80GB$ GPU achieves $41.20$ tokens per second~\cite{qwen2-speed-doc}.
Therefore, instead of building and maintaining their own compute clusters, smaller companies can act as MLaaS brokers, hosting decomposed model components -- e.g., LoRA adapters for LLMs or traditional Deep Neural Networks (DNNs) -- in virtualized cloud or edge environments to serve customers and support internal R\&D.

Recently, organizations such as Adobe~\cite{adobe-hybridcloud-infra-0, adobe-hybridcloud-infra-1} and Workday~\cite{workday-hybridcloud-infra}, have launched hybrid-cloud MLaaS platforms.
They combine public-cloud resources (VMs or containers on AWS, Azure, or GCP) with private clusters (at the edge close to the company users).
This approach involves tradeoffs and opportunities around service latency guarantee, user data privacy preservation, and savings in monetary cost of computation resources.

For example, provisioning $H100$ GPUs in the cloud offers fast processing, but incurs a high monetary resource cost and may increase data transmission time compared to private clusters or other edge resources located close to end users.
Conversely, allocating computation resources in the cloud instead of processing data securely at the edge, comprising the private cloud or user devices, can minimize the monetary cost of resource maintenance and processing delay, but increases the risk of data leakage.

Furthermore, some application-level objectives (accuracy, latency, privacy, and monetary cost) can be relaxed to improve other ones.
For example, federated learning on medical data has low tolerance to data leakage.
To meet this requirement, an MLaaS broker might run only a sensitive partition (e.g., shallow layers of a model) at the edge, so the raw data never leaves the device, while offloading the remaining partitions to the cloud.
Due to the limited edge capacity and transmission between the edge and cloud, the training time can be longer and overall monetary cost could be higher.
In contrast, clear data-sharing agreements can mitigate privacy concerns for less sensitive tasks and potentially allow the use of cheaper cloud resources~\cite{google-gen-privacy, openai2024privacypolicy}.

Investigating such tradeoffs in both traditional DNNs and LLMs yields valuable guidance for future MLaaS system design, helping smaller companies and institutions deploy large-scale ML systems with low latency, high privacy guarantee, and low monetary cost of resources.
Recent research has focused on computation-efficient and privacy-aware ML models or cost-efficient resource orchestration methods.
However, a study of the interactions between all three aspects of latency, privacy, and monetary cost remains an ongoing topic.
This section highlights recent work and open challenges in these dimensions via resource provisioning and model decomposition, leading to the contributions of this survey.

\paragraph{What are the limitations of existing MLaaS systems for ML Inference?}
Existing MLaaS services do not yet incorporate much of the existing research on model decomposition and resource provisioning to improve latency~\cite{DDNNs}, privacy~\cite{privacy-edge-survey}, and service cost~\cite{LLM-hybrid, MLaaS-federation}.
Sagemaker Edge compiles NN models to utilize the client's hardware architecture and memory access patterns for optimal ML inference speed~\cite{AWS-sagemaker-neo}, which represents only a small slice of possible model adaptations.
In contrast, integrating model decomposition, assigning each submodel to devices, edge, or cloud nodes, and provisioning compute, memory, and network resources for each, has not yet been adopted by mainstream MLaaS offerings.

Therefore, our research investigates model architecture optimization and resource provisioning strategies that could become part of future MLaaS services. 
In particular, we focus on model offloading in ML inference, including both independent and dependent model decomposition and resource orchestration with cloud and edge provisioning services, needed to meet strict latency, privacy, and monetary cost objectives.

\subsection{Opportunities and Challenges of DL Task offloading during Inference}
\paragraph{Why offloading Deep Learning (DL) tasks?}
ML applications that gather large volumes of data and use complex deep learning models often require low latency to meet quality-of-service (QoS) targets~\cite{wang2020survey, huang2022making, 10.1016/j.comnet.2021.108468, eie}.
Offloading selected layers to an edge or cloud tier can: (1) minimize inference latency by leveraging cloud compute capacity; (2) enhance source data privacy; and (3) reduce resource monetary cost via pay‐per‐use pricing models~\cite{serverless-in-the-wild}.
In this section, we examine one example: offloading user-interactive, latency-sensitive applications, such as Augmented Reality (AR) with partitions of a multi-layered ML model (dependent submodels), across edge and cloud environments to optimize inference latency and protect user data privacy.

For the threat model, we assume an \textbf{honest‐but‐curious} adversary who can:
\begin{itemize}
\item monitor activations $x_{interm}$ in transit between the edge and cloud;
\item train an offline reconstructor (auto‐encoder NN) $\mathcal{R}_{\text{MIA}}$ on public data;
\item output the reconstructed input data from observed intermediate data, i.e.,\ $\hat{x}_{input} = \mathcal{R}_{\text{MIA}}(x_{interm})$, evaluated by mean‐squared reconstruction error or misclassification rate on a target classifier.
\end{itemize}
This aligns with prior model inversion attack (MIA) and prompt inversion attack (PIA) formulations~\cite{model-inversion-attack-1,model-inversion-attack-4,prompt-inversion-attack-0,prompt-inversion-attack-1}.

\begin{table*}[!ht]
\small
\centering
\begin{tabular}{|| c | c | c | c | c | c ||} 
\hline
Model & Device & End-to-End & Pred & Task & Ref\\
 \hline
MNetV3~\cite{mobilenetv3}  & Rasp Pi $4B+$ & $595ms$ & \begin{tabular}{@{}c@{}}79.23\% \\ accuracy\end{tabular} & \begin{tabular}{@{}c@{}}$48*48$-pixel \\ RAF-DB~\cite{RAF-DB}\end{tabular} & \cite{huang2022making}\\
 \hline
MNetV2~\cite{mobilenetv2}  & Rasp Pi $4B+$ & $3571ms$ & \begin{tabular}{@{}c@{}}81.16\% \\ accuracy\end{tabular} & \begin{tabular}{@{}c@{}}$48*48$-pixel \\ RAF-DB~\cite{RAF-DB}\end{tabular} & \cite{huang2022making}\\
 \hline
\begin{tabular}{@{}c@{}}MNetV2 \\ +SSDLite~\cite{mobilenetv2}\end{tabular}  & Pixel 1 & $162ms$ & $22.1\%$ mAP & COCO~\cite{COCO} & \cite{mobilenetv3}\\
 \hline
\begin{tabular}{@{}c@{}}MNetV3 \\ +SSDLite~\cite{mobilenetv2}\end{tabular}  & Pixel 1 & $137ms$ & $22.0\%$ mAP & COCO~\cite{COCO} & \cite{mobilenetv3}\\
 \hline
YOLOv3~\cite{yolov3}  & Pixel 2 & $4500ms$ & $40\%$ IOU & \begin{tabular}{@{}c@{}}Imagenet \\ Video~\cite{ILSVRC12}\end{tabular} & \cite{MARLIN}\\
 \hline
Tiny-YOLO~\cite{tinyyolo}  & Pixel 2 & $1200ms$ & $40\%$ IOU & \begin{tabular}{@{}c@{}}Imagenet \\ Video~\cite{ILSVRC12}\end{tabular} & \cite{MARLIN}\\
 \hline
\end{tabular}
\caption{DNN performances at the edge in recent work}
\label{edge-model-performance}
\end{table*}
\begin{table*}[!ht]
\small
\centering
\begin{tabular}{|| c | c | c | c | c | c |} 
\hline
Model & Hardware & Processing & Pred & Task & Ref\\
 \hline
YOLOv4-608~\cite{yolov3}  & V100  & $16.1ms$ & $43.5\%$ COCOmAP & COCO~\cite{COCO} & \cite{yolov4}\\
 \hline
YOLOv3-608~\cite{yolov3}  & Titan X  & $57.9ms$ & $33\%$ COCOmAP & COCO~\cite{COCO} & \cite{yolov3}\\
 \hline
YOLOv2-544~\cite{tinyyolo}  & Titan X & NA & $21.6\%$ COCOmAP & COCO~\cite{COCO} & \cite{yolov3}\\
 \hline
\end{tabular}
\caption{DNN performances in the cloud in recent work}
\label{cloud-model-performance}
\end{table*}
\begin{table*}[!ht]
\small
\centering
\begin{tabular}{|| c | c | c | c | c | c | c | c ||} 
\hline
Model & Edge & Cloud & End-to-End & Pred & Bandwidth & Task & Ref\\
 \hline
\begin{tabular}{@{}c@{}c@{}}Faster \\ R-CNN \\ \cite{faster-r-cnn}\end{tabular}& \begin{tabular}{@{}c@{}}Jetson \\ TX2\end{tabular}  & \begin{tabular}{@{}c@{}}Titan \\ XP\end{tabular}  & $34.56ms$ & $70\%$ IoU & 82.8Mbps & \begin{tabular}{@{}c@{}}Xiph \\ \cite{Xiph}\end{tabular} & \begin{tabular}{@{}c@{}}Baseline \\ \cite{DRE+PSI+MvOT}\end{tabular}\\
\hline
\begin{tabular}{@{}c@{}c@{}}Faster \\ R-CNN \\ \cite{faster-r-cnn}\end{tabular}& \begin{tabular}{@{}c@{}}Jetson \\ TX2\end{tabular} & \begin{tabular}{@{}c@{}}Titan \\ XP\end{tabular}  & $22.96ms$ & $75.8\%$ IoU & 276Mbps & \begin{tabular}{@{}c@{}}Xiph \\ \cite{Xiph}\end{tabular} & \begin{tabular}{@{}c@{}}Baseline \\ \cite{DRE+PSI+MvOT}\end{tabular}\\
\hline
\begin{tabular}{@{}c@{}c@{}}Faster \\ R-CNN \\ \cite{faster-r-cnn}\end{tabular}& \begin{tabular}{@{}c@{}}Jetson \\ TX2\end{tabular} & \begin{tabular}{@{}c@{}}Titan \\ XP\end{tabular}  & $17.23ms$ & $86.4\%$ IoU & 82.8Mbps & \begin{tabular}{@{}c@{}}Xiph \\ \cite{Xiph}\end{tabular} & \begin{tabular}{@{}c@{}c@{}c@{}}DRE \\ +PSI \\ +MvOT \\ \cite{DRE+PSI+MvOT}\end{tabular}\\
\hline
\begin{tabular}{@{}c@{}c@{}}Faster \\ R-CNN \\ \cite{faster-r-cnn}\end{tabular}& \begin{tabular}{@{}c@{}}Jetson \\ TX2\end{tabular} & \begin{tabular}{@{}c@{}}Titan \\ XP\end{tabular} & $15.52ms$ & $91.1\%$ IoU & 276Mbps & \begin{tabular}{@{}c@{}}Xiph \\ \cite{Xiph}\end{tabular} & \begin{tabular}{@{}c@{}c@{}c@{}}DRE \\ +PSI \\ +MvOT \\ \cite{DRE+PSI+MvOT}\end{tabular}\\
\hline
\end{tabular}
\caption{End-to-end DNN performances combining edge and cloud resources}
\label{model-performance-e2e}
\end{table*}
Tables~\ref{edge-model-performance},~\ref{cloud-model-performance}, and~\ref{model-performance-e2e} summarize representative edge, cloud, and hybrid edge-cloud results that motivate our offloading study.
As shown in Table~\ref{edge-model-performance}, recent lightweight DNN models used in AR applications on edge devices, such as Raspberry Pi or smartphones, often struggle to meet the $30$ frames-per-second (FPS) video requirements or $100$ms human-sensible end-to-end (frame refresh) latency target~\cite{MARVEL, HCI-responsetime, CLONE, DVFS-0, HCI-responsetime-1}.
Edge platforms are constrained by limited compute power~\cite{BottleNet, AI-Accelerator-survey}, bandwidth~\cite{FL}, battery capacity~\cite{edge-battery}, and memory~\cite{CE-AFL}, making it difficult to sustain high performance.

Cloud resources can provide the extra processing capacity needed for demanding ML workloads, as shown in Table~\ref{cloud-model-performance}. 
However, relying on the public cloud raises privacy concerns~\cite{FSL-journal} and introduces transmission bottlenecks over the Internet~\cite{BBNet, BottleNet}, which prohibit transmitting source data to the cloud for various ML tasks.

To combine the advantages of both edge and cloud resources, recent research has focused on partitioning ML tasks and provisioning resources across the cloud and the edge, with customer data residing on edge servers or user devices to minimize data leakage.
In this approach, a portion of the ML task is performed on the client devices.
Only essential hidden/latent variables required for high-accuracy inference are transmitted to the cloud or edge server.
This paradigm keeps the source data on the client device, enhancing the efficiency and privacy of transmission.
Furthermore, to improve data privacy during inference and defend against MIA attacks, previous work introduces privacy-preserving training steps, including adding privacy-aware loss terms and differential privacy approaches to gauge the hidden variables that adversaries can leverage.

Then, to compensate for the potential added latency, this paradigm often ensures that data transmission occurs only when necessary, for example, by using early exits~\cite{Neurosurgeon} or dynamic region of interest encoding~\cite{DRE+PSI+MvOT}.
Specifically, a multi-layered/partitioned ML model can be augmented with ``internal classifiers" that can produce early output (prediction) and may not need to process data through all layers/partitions.
Similarly, the input data can be compressed/cropped based on the region of interest (RoI) to minimize data transmission. 
As shown in Table~\ref{model-performance-e2e}, some existing approaches based on this design achieve low latency and high model accuracy.

While the examples above focus on dependent submodels, there are also opportunities and challenges for independent submodel-based applications in achieving low latency, high privacy, and low monetary cost.
For example, offloading different independent submodels from the ensemble model to multiple VMs or serverless function instances enables efficient incremental training and minimizes inference time via parallel execution.
In this paper, we explore similar tradeoffs among latency, privacy, and monetary cost by leveraging diverse model decomposition techniques and resource-orchestration services both at the edge and in the cloud.

\paragraph{Challenges and Explorations in ML task offloading.}
Finding an optimal offloading plan for ML applications is not trivial.  
A na{\"\i}ve offloading plan can result in long transmission and processing delays, privacy breaches (Model Inversion Attack), or resource under- or over-provisioning.  
To capture these trade-offs, we formulate an optimization problem that balances latency, privacy, and monetary cost based on existing methods.
Previous surveys have addressed aspects of optimizing monetary cost, latency or privacy for AI applications (Table~\ref{related-survey}).
However, they do not formulate the optimization problem nor discuss monetary cost (\$) based approaches.
Detailed cost analysis using real-world cloud resources for low-cost (\$) ML serving remains limited.
Furthermore, while some existing surveys~\cite{survey-EE-SC, survey-collaborate-EC} provide valuable insights, they often lack comprehensive discussions on source data privacy in distributed inference systems.

\rb{In this work, grounded in a multi-objective optimization framework, we categorize prior work and study multi-objective optimization accounting for interactions among latency, privacy, and monetary cost. 
To narrow the scope of our literature review, the survey emphasizes model-inversion attacks~\cite{model-inversion-attack-1, model-inversion-attack-2, model-inversion-attack-3} and cloud pricing models~\cite{serverless-in-the-wild} for deep-neural-network inference pipelines composed of multiple machine-learning models.}

\rb{Our main contributions are:
\begin{itemize}
    \item An interaction-aware optimization framework evaluated via case studies
    \item Interaction analyses (e.g., IaaS vs.\ FaaS latency–cost trade-offs, privacy–latency tensions, and methods for jointly balancing latency, cost, and privacy) that turn the framework into actionable insights
    \item Identification of previously unstudied combinations of objectives under alternative threat models (e.g., prompt-inversion attacks in large language models), pricing models (e.g., serverless fine-grained billing vs.\ VM coarse-grained billing), and application domains
\end{itemize}
}



We organize the paper based on optimization objectives.
In Sec.~\ref{sec:prob-def}, we introduce the optimization problem by studying the challenges of ML task offloading given different optimization objectives, including Latency in Sec.~\ref{sec:problem-latency}, Privacy in Sec.~\ref{sec:problem-privacy}, and Monetary Cost (\$) in Sec.~\ref{sec:problem-cost}.
Then, in Sec.~\ref{sec:prob-form}, we formulate the optimization problem for Latency (Sec.~\ref{sec:prob-form-latency}), Privacy (Sec.~\ref{sec:prob-form-privacy}) and Monetary Cost (Sec.~\ref{sec:prob-form-cost}).
In Sec.~\ref{sec:sol}, we detail adaptive learning methods to deploy a DNN model across the spectrum of cloud, edge, and client resources by optimizing Latency (Sec.~\ref{sec:lat-sol}), Privacy (Sec.~\ref{sec:pri-sol}), and Monetary Cost (Sec.~\ref{sec:cot-sol}).
Next, Sec.~\ref{sec:coupling} discusses the interactions between optimization objectives in the formulation given certain use cases introduced by related works.
Sec.~\ref{sec:issues} discuss the open issues.
And Sec.~\ref{sec:conc} concludes the paper.

\begin{table*}[!ht]
\centering
\begin{tabular}{|| c | c | c | c | c | c | c | c | c | c ||} 
\hline
\begin{tabular}{@{}c@{}} Optimization \\ Formulation\end{tabular} & Monetary Cost (\$) & Latency & Privacy & DL & Placement & Scope & Inference & Training & Reference\\
\hline
\Checkmark & \Checkmark & \Checkmark & \Checkmark & \Checkmark & \Checkmark & \begin{tabular}{@{}c@{}} Edge Devices \& \\ Edge \& Cloud\end{tabular} & \Checkmark & $\circ$ &(Our Work)\\
\hline
\XSolid & \Checkmark & \Checkmark & \XSolid & \Checkmark & \Checkmark & \begin{tabular}{@{}c@{}} Graph in \\ Mobile \& Cloud\end{tabular} & \Checkmark & \Checkmark & 2020~\cite{survey-offload-EC}\\
\hline
\XSolid & \XSolid & \Checkmark & \Checkmark & \Checkmark & \Checkmark & \begin{tabular}{@{}c@{}} AIoT \& \\ Edge \& Cloud\end{tabular} & \Checkmark & \Checkmark & 2021~\cite{survey-AIOT}\\
\hline
\XSolid & \XSolid & \Checkmark & $\circ$ & \Checkmark & \Checkmark & \begin{tabular}{@{}c@{}} Early Exit in \\ Mobile \& Cloud\end{tabular} & \Checkmark & \Checkmark & 2022~\cite{survey-EE-SC}\\
\hline
\XSolid & \XSolid & \Checkmark & \Checkmark & \Checkmark & \Checkmark & \begin{tabular}{@{}c@{}} End Device \& \\ Edge \& Cloud\end{tabular} & \Checkmark & \Checkmark & 2023~\cite{survey-AI-EECS}\\
\hline
\XSolid & \XSolid & \Checkmark & $\circ$ & \Checkmark & \Checkmark & \begin{tabular}{@{}c@{}} End Device \& \\ Edge \& Cloud\end{tabular} & \Checkmark & \Checkmark & 2024~\cite{survey-collaborate-EC}\\
\hline
\XSolid & \Checkmark & \Checkmark & $\circ$ & \Checkmark & \XSolid & \begin{tabular}{@{}c@{}} LLM Prompt \\ Leakage\end{tabular} & \Checkmark & \Checkmark & 2024~\cite{survey-prompt-leakage}\\
\hline
\end{tabular}
\caption{Related survey comparison: \Checkmark indicates the corresponding survey covers up-to-date or more comprehensive discussion. $\circ$ indicates our work is more complementary or has different discussion than the corresponding work. \XSolid means the corresponding work does not discuss this aspect.}
\label{related-survey}
\end{table*}

\section{Problem Definition}\label{sec:prob-def}
In this section, we narrow down the specific challenges in the machine learning (ML) model offloading problem that this paper addresses by examining the deep neural network (DNN) offloading scenario (dependent submodels).
Recent studies have explored offloading a Deep Neural Network (DNN) model across the core cloud, edge, and client devices to satisfy resource constraints and privacy guarantees.
Given dynamic environments, each inference request can adaptively route through model partitions on the most capable resources to minimize latency and maintain privacy constraints in a cost-efficient manner.

However, partitioning the NN model, introduces new challenges.
Between model partitions, hidden variables and gradients transmitted during forward and backward propagation add additional transmission overhead~\cite{BBNet, BottleNet, Deep-Compressive-Offloading} and cause client data leakage~\cite{NoPeek, FSL-journal}.
Meanwhile, byproducts of running on the edge, for example, extra processing delays~\cite{CLIO, matsubara2021neural} and energy consumption~\cite{Neurosurgeon}, should be minimized.

\subsection{DNN Offloading Challenges}\label{sec:parting-study}
Existing MLaaS systems manage cloud resources~\cite{AWS-Rekognition} or user devices to run DL jobs~\cite{AWS-sagemaker-edge}.
Meanwhile, cloud-managed edge computing resources, including AWS Local Zones~\cite{AWS-local-zone} and Wavelength~\cite{AWS-wavelength}, and edge ML model optimizers have become important building blocks for ML services used by companies such as Holo-Light~\cite{aws-edge-use-case-1}, Netflix~\cite{aws-edge-use-case-2}, and SKT~\cite{aws-edge-use-case-3}.
With AWS Sagemaker Edge~\cite{AWS-sagemaker-edge} and AWS Greengrass~\cite{AWS-Greengrass}, a user can optimize their edge application by a compilation that targets their specific hardware (CPU architecture) and operating system.
In the future, we envision MLaaS service providers adopting more model and resource adaptations in their optimizers, improving {\em latency} of processing and transmission, \textit{privacy} of the source data, and the monetary cost of resources.
To enable such optimizers, we review the challenges of achieving high DNN performance when a DL model is partitioned between cloud, edge, and user devices.

\subsubsection{Latency}\label{sec:problem-latency}
In recent ML applications, including real-time object detection in AR/VR applications and LLM based ChatBots, latency has become an important concern considering the interactive manner of such applications, i.e, $\leq 100ms$ or $\geq 30fps$~\cite{HCI-responsetime, CLONE, DVFS-0, HCI-responsetime-1}.
However, it is non-trivial to achieve low latency given the resource demand of the highly-parametrized models.
In this section, we examine the challenges of minimizing ML inference latency.

The time spent in a distributed ML inference system can be decomposed into transmission and processing delays. 
When large volumes of data are sent between model partitions, transmission overhead can dominate inference latency~\cite{CE-AFL, FSL-journal, mudvari2024splitllmcollaborativeinferencellms}.
Meanwhile, offloading too many model parameters to constrained edge resources can also overwhelm user devices, resulting in long processing delays.
Ideally, a practical NN partitioning paradigm should optimize for both delays to ensure optimal latency performance.

\noindent {\bf Transmission.}
For a partitioned NN, hidden variables (or activations)\footnote{We use the terms ``hidden variables" and ``activations" interchangeably.} must be sent between partitions to complete forward propagation, often over the internet limited by bandwidth in IoT or mobile device-based systems.
In traditional DNN models (e.g., convolutional layers, fully connected layers, etc.), partitioning is often by layers.
The size of intermediate data is fixed and model inference is stateless, which allows straightforward estimation of transmission size and offloading strategies.
\begin{figure}
\captionsetup[subfigure]{justification=centering}
    \begin{subfigure}[t]{0.5\textwidth}
    \includegraphics[width=\textwidth]{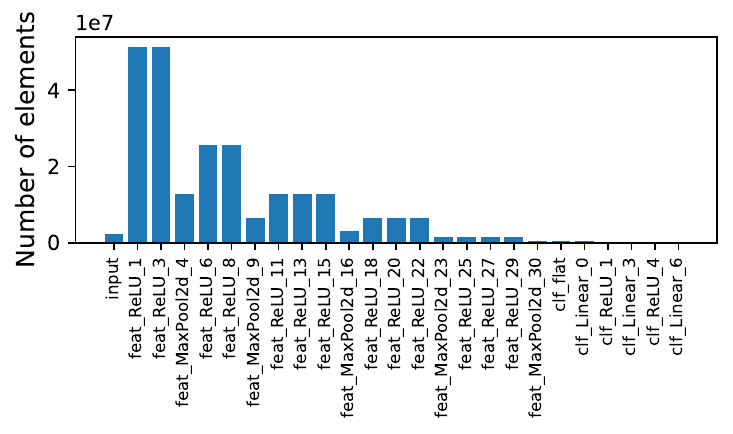}
    \end{subfigure}
    \caption{Hidden variable sizes of VGG16 with CiFAR-10 and batch size of $16$.
    }\label{example_output_size}
\end{figure}
As shown in Fig.~\ref{example_output_size}, we profiled the hidden variable sizes using a VGG-16 model~\cite{vgg} and CiFAR-10~\cite{cifar10} with a batch size of $16$.
The x-axis indicates the NN layer where the model is split, where the head part of the DNN (from the input layer up to and including the splitting layer) runs on a client device, and the tail part runs on an edge or cloud server.
The y-axis shows the output size of different splitting layers.
Different splitting layers yield different output sizes.
Therefore, the model splitting can be optimized for short delays~\cite{Neurosurgeon}.

For transformer-based models, the auto-regressive token generation is stateful.
When offloading decoder blocks, we also migrate the KV cache of previously generated tokens.
And the size of the KV cache can grow.
Overall, it complicates the transmission estimation as we also move the state of the model.
Model splitting and offloading methods can be applied with state migration~\cite{mudvari2024splitllmcollaborativeinferencellms,kafetzis2025largelanguagemodelpartitioning}.

On the other hand, previous work has explored the compression of intermediate data and model to reduce communication overhead.
By incorporating a bottleneck network, such as Auto-Encoder, at the splitting point, previous works select the key features for transmission~\cite{BBNet, BottleNet, LP-transmission-Edge-Cloud, Task-Oriented-bottleneck, Deep-Compressive-Offloading, matsubara2021neural, XAI-intermediate-data-compression}.
On the other hand, model trimming techniques, such as model distillation~\cite{hinton2015distilling} and quantization~\cite{CLIO, SmartSwitch-AI-Accelerator, bi-quan, jacob2018quantization}, can also minimize the size of hidden variables (activations) and gradients to send.

The sparse model activates a subset of the model parameters during inference.
For example, the Internal Classifier (IC) allows forward propagation to end in one partition, and no intermediate data transmission~\cite{DDNNs, SPINN}.
When a classifier gains confidence in the prediction, it emits the output, and no subsequent feature extraction is needed.

\noindent{\bf Processing.}
Another challenge for offloaded deep learning systems is the limited processing capacity of edge and client devices. 
As shown in Table~\ref{edge-model-performance}, edge devices such as Raspberry Pi~\cite{huang2022making} and mobile phones~\cite{MARLIN, mobilenetv3} often struggle to meet the latency or accuracy requirements demanded by machine learning applications.

To overcome these limitations, related work has explored model adaptations, including quantization~\cite{Pruning-and-quantization-survey}, pruning~\cite{darts, NN_Structure_Search, Dynamic-Neural-Networks, DDI}, and knowledge distillation~\cite{hinton2015distilling}.
Such methods reduce the complexity of the model, allowing applications deployed on edge devices or the cloud to meet QoS requirements.

In addition, other works have explored the use of cloud computing capacity to assist edge intelligence applications~\cite{DDNN, SPINN, LLM-hybrid, MMSL-LLM}.
However, this approach introduces challenges in privacy and transmission~\cite{XAI-intermediate-data-compression, FedMEC-conv-cli-dense-edge-dp}.

\subsubsection{Monetary Cost}\label{sec:problem-cost}

As cloud infrastructure has evolved, providers have introduced services beyond VMs that can be more cost-efficient. 
For example, different cloud offerings across providers exhibit individualized cost models~\cite{service-chain-mec-cost} and varying performance characteristics~\cite{MLaaS-federation}. 
These include Function-as-a-Service (FaaS) and Container-as-a-Service (CaaS) clusters in the cloud~\cite{AWS-lambda, AWS-ecs}, as well as AWS Lambda@Edge and Local Zones at the edge~\cite{AWS-lambda-edge, AWS-local-zone}, which provide fine-grained billing~\cite{serverless-in-the-wild}. 
An MLaaS system should therefore adaptively configure both its runtime environment and model architecture to optimize cost and performance across the device–edge–cloud continuum.



For highly dynamic inference workloads, slow scaling in the core cloud might result in under or overprovisioning of resources and consequently missing the QoS target~\cite{raza2021sok, LIBRA}.
Related works have explored dynamically directing workload to a deep NN in the cloud and a shallow NN at the edge for cost savings~\cite{LLM-hybrid}.
Other works deploy NN partitions using Function-as-a-Service (FaaS)~\cite{DNN-partition-SLS}. 
This approach leverages the pay-per-use nature of FaaS, where the user only pays for the actual computation time used, to avoid the costs of keeping VMs constantly running and provisioned, including node cold start and model loading time.

Furthermore, specific NN adaptations can enable resource provisioning for individual NN layers, achieving cost-effective QoS tracking.
By incorporating internal classifiers~\cite{SkipNet, ShallowDeepNet} or neuron skipping methods~\cite{SplitNet}, only a subset of the network's neurons is used for prediction.
Thus, users can minimize the monetary resource cost~(\$) based on different cloud resource pricing models~\cite{LIBRA}.
Specifically, low-workload layers can be provisioned on demand with FaaS platforms~\cite{AWS-lambda, AWS-lambda-edge, serverless-in-the-wild} without relying on reserved VMs, so there is less idle time for computation resources.
Such adaptations can be applied across different ML tasks.
For example, in an image classification task, the shallow layers might capture the contour of a banana, while the deep layers that focus on the details of the banana are less critical to some classifications~\cite{feat-selection-in-offloading, ShallowDeepNet}.
Consequently, these less frequently used deep layers are well-suited for FaaS.

\subsubsection{Privacy}\label{sec:problem-privacy}
\begin{table*}[!bt]
\small
\centering
\begin{tabular}{|| c | c | c | c | c | c | c |} 
\hline
Model & Hardware & Processing & Task & Privacy & Ref\\
 \hline
VGG-16~\cite{falcon}  & P100 & $57ms$ & \begin{tabular}{@{}c@{}}Tiny \\ ImageNet~\cite{le2015tiny}\end{tabular} & Plaintext & \cite{falcon}\\
 \hline
VGG-16~\cite{falcon}  & CPU & $1,300ms$ & \begin{tabular}{@{}c@{}}Tiny \\ ImageNet~\cite{le2015tiny}\end{tabular} & Plaintext & \cite{falcon}\\
 \hline
VGG-16~\cite{falcon}  & CPU(LAN) & $40,000ms$ & \begin{tabular}{@{}c@{}}Tiny \\ ImageNet~\cite{le2015tiny}\end{tabular} & SMPC(FALCON) & \cite{falcon}\\
 \hline
VGG-16~\cite{falcon}  & CPU(LAN) & $59,000ms$ & \begin{tabular}{@{}c@{}}Tiny \\ ImageNet~\cite{le2015tiny}\end{tabular} & SMPC(FALCON) & \cite{falcon}\\
 \hline
VGG-16~\cite{feat-selection-in-offloading}  & Titan Xp & $14.5ms$ & ImageNet~\cite{ILSVRC15} & Plaintext & \cite{feat-selection-in-offloading}\\
 \hline
VGG-16~\cite{feat-selection-in-offloading}  & CPU & $12,960ms$ & ImageNet~\cite{ILSVRC15} & SMPC(FALCON) & \cite{feat-selection-in-offloading}\\
 \hline
VGG-16~\cite{feat-selection-in-offloading}  & Titan Xp & $14.5ms$ & ImageNet~\cite{ILSVRC15} & Plaintext(Cloak) & \cite{feat-selection-in-offloading}\\
 \hline
\end{tabular}
\caption[Privacy-preserving DNN inference performance in the core cloud]{Privacy-preserving DNN inference performances in the core cloud in recent works}
\label{secure-cloud-model-performance}
\end{table*}
The privacy of source data has become a critical concern for DL systems.
Partitioning and offloading an NN to edge devices helps keep raw source data private, as user data are not sent over the network.
However, data breaches can still occur, as adversaries can exploit the information in the intermediate data.

Recent works~\cite{client-dp, NoPeek, FSL-journal, FedMEC-conv-cli-dense-edge-dp} discuss the use of Auto-Encoders~\cite{autoencoder, google-autoencoder} to reconstruct the source data from the intermediate data sent from the edge to the cloud during forward propagation for a deep neural network (DNN).
Such vulnerability can be exploited by the model inversion attack (MIA)~\cite{model-inversion-attack-1, model-inversion-attack-2, model-inversion-attack-3}.
The Auto-Encoder consists of an encoder neural network (NN) and a decoder NN, and uses a loss function, e.g., Mean Squared Error (MSE), to gauge the error between the source data and reconstructed data.
The encoder mirrors the architecture of the NN on client devices, while the decoder reflects the encoder structure to approximate matrix multiplication inversions.

As an example, for an ML application which infers the number in MNIST images of hand-written digits~\cite{mnist}, when the DNN is partitioned and offloaded to different edge or cloud resources, the hidden variables transmitted over the wire can be captured by an honest but curious adversary, as a man-in-the-middle attack, leading to data leakage.
To reveal the source data using the intermediate data (hidden variables), the adversary can train an Auto-Encoder model that can faithfully reconstruct a Kuzushiji-MNIST
~\cite{kmnist} dataset that has similar patterns in hand-writing and use the decoder NN to invert hidden variables to the source data.
Recent work discovered a similar attack that reveals the user prompt to an LLM leveraging the hidden variables transmitted over the network, namely Prompt Inversion Attack (PIA)~\cite{prompt-inversion-attack-0, prompt-inversion-attack-1}, which further emphasizes the pervasive nature of such attacks in real-world applications.
The formal definition of our attack model is presented in Sec.~\ref{sec:MIA-threat}.

This attack is not merely theoretical but poses a practical threat in real-world settings for two key reasons. First, the attacker's decoder architecture can be flexible and does not need to precisely mirror the client-side model to be effective~\cite{ResSFL, FSL-journal}.
Second, the proliferation of powerful open-source LLMs~\cite{llama-3-1-model-card, deepseek-r1} creates a significant vulnerability, particularly in collaborative edge-cloud inference systems~\cite{LLM-hybrid}. In such a setup, an honest-but-curious intermediate node can leverage the publicly known weights of a base model. Since many deployed models are fine-tuned versions of these public foundation models, the adversary can exploit this shared architectural knowledge to train a highly effective reconstructor (e.g., an Auto-Encoder~\cite{ResSFL}) and recover sensitive source data from the intermediate activations.

Privacy-preserving methods for model and user data are critical for an MLaaS system.
One approach involves encryption~\cite{smpc-1, he-1}, which can cause significant slowdowns~\cite{feat-selection-in-offloading, model-pruning-for-mia-privacy,prompt-inversion-attack-0,prompt-inversion-attack-1}.
As shown in Table~\ref{secure-cloud-model-performance}, without encryption, running VGG-16 on ImageNet takes $14.5\,$ms per inference on an NVIDIA Titan Xp GPU.
However, when using FALCON homomorphic encryption on CPUs, the same task takes $12{,}960\,$ms~\cite{feat-selection-in-offloading}.

The more popular privacy-preserving approach, however, 
mitigates MIA attacks by focusing on the training phase to build privacy-preserving models in the first place, rather than
adapting the model during inference time.
Such methods introduce a secondary loss function, e.g., distance correlation~\cite{NoPeek, PSL, FSL-journal}, to constrain the similarity between intermediate data and source data during model training.
Similar works also incorporate an Auto-Encoder to model training, using reconstruction error as privacy metric~\cite{ResSFL}.
Furthermore, previous works utilize DNN pruning with \textit{masks} to remove mutual information between the source and intermediate data~\cite{feat-selection-in-offloading, model-pruning-for-mia-privacy}.

Similarly, other privacy-preserving methods, instead of adjusting the loss function, apply perturbations to intermediate data during training~\cite{client-dp, FedMEC-conv-cli-dense-edge-dp}.
In these approaches, the intermediate activations retain minimal sensitive information, while the cloud-side neural network learns to extract the key features required for inference.


\section{Problem Formulation}\label{sec:prob-form}
We identify three key challenges in Deep Neural Network offloading -- covering both model decomposition and resource provisioning -- across the device-edge-cloud continuum, each tied to a different performance objective:
\begin{itemize}
    \item Latency:
    Achieving low inference time requires partitioning the neural network so that computation on devices and data transmission over the network are balanced.
    \item Monetary Cost:
    For sparsely activated models, we can map each submodel to the most cost-effective cloud service using a fine-grained resource-provisioning strategy while still meeting Service Level Objectives (SLOs).
    \item Privacy:
    Hidden variables transmitted over the network can be exploited by adversaries (for example, using auto-encoder networks) to reconstruct sensitive input data, creating a risk of data leakage.
\end{itemize}
In the rest of this section, we formulate a unified DL offloading formulation that integrates three subproblems -- one for each objective -- to determine how to place model partitions across edge and cloud resources.
Then, in Sec.~\ref{sec:sol}, we summarize the related works based on the optimization objectives.
And, in Sec.~\ref{sec:coupling}, we study the interactions between performance objectives by solving our optimization problems in certain use cases introduced in related works.

As a constrained multi‐objective optimization problem, we seek an offloading configuration that minimizes a weighted sum of latency, monetary cost, and privacy objectives:
\begin{align}
    min &(w^L{\cal L}^L + w^C{\cal L}^C + w^P{\cal L}^P)\\
    s.t.\ & \text{constraints on inference metrics}.
\end{align}
Here,
\begin{itemize}
  \item ${\cal L}^L$,${\cal L}^C$, and ${\cal L}^P$ are the latency, cost, and privacy loss functions, respectively,
  \item \(w^{L},w^{C},w^{P}\) are nonnegative weights reflecting their relative importance.
\end{itemize}
\rb{There are other constraints, including the battery capacity of edge devices~\cite{edge-battery}, developer tooling (for example, virtual machines or serverless platforms)~\cite{serverless-metric-benchmark} and wireless connection conditions~\cite{survey-AIOT}, that can influence the optimization. 
There are also application-specific metrics, e.g., Time-to-First-Token for LLMs~\cite{jiang2024minference}. 
We focus on a subset of the optimization targets and constraints to narrow this survey.}
Detailed constraints for each objective are discussed in Sec.~\ref{sec:prob-form-latency}, \ref{sec:prob-form-cost} and \ref{sec:prob-form-privacy}.

\subsection{Latency}\label{sec:prob-form-latency}
Balancing and minimizing transmission and processing delays are essential to DL inference tasks.
Arbitrary model partitioning can cause excessive data transmission.
In contrast, deploying too many layers on computation-limited edge devices yields a long processing time.
We first briefly introduce the latency optimization approaches.  
Then we present the latency formulation in Sec.~\ref{sec:latency-formulation}.

\begin{figure}
\captionsetup{justification=centering}
    \includegraphics[width=0.48\textwidth]{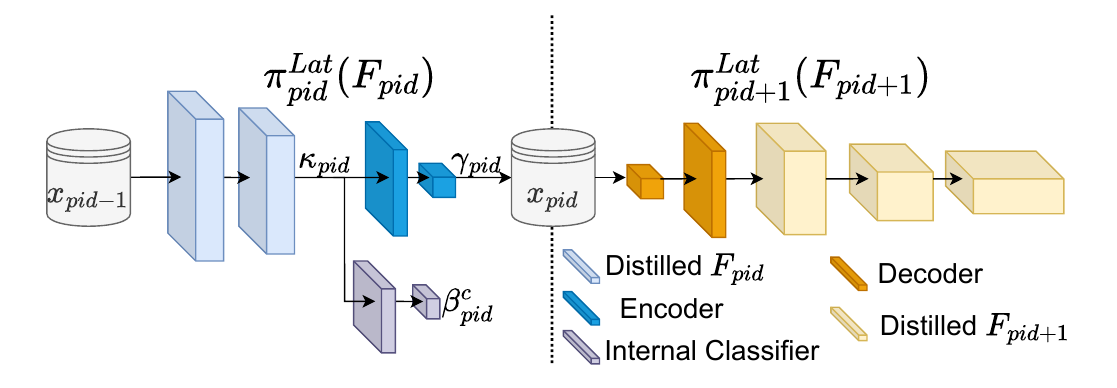}
    \caption{Illustration of latency optimization problem.
    }\label{lat_optimization}
\end{figure}

We focus on a DL model composed of $M$ partitions ($F_{pid}, pid \in {1,2,...,M}$ in Fig.~\ref{lat_optimization}) with the notations defined in Table~\ref{latency-notations}.
An individual model partition $pid$ can be offloaded to the edge or cloud based on the estimation of its inference time (denoted by $T_{pid}$, which is the sum of transmission delay $T_{pid}^{T}$ and computation delay $T_{pid}^{C}$), the hidden variable size ($Size(.)$) and the profiles of floating point operations performed by layers in the partition ($FLOPs_i^j(.)$).
In practice, because each partition’s output feeds the next, once the offloading decision for partition $k$ changes, the same decision is typically applied to all deeper partitions to avoid extra transmission overhead~\cite{Neurosurgeon, DDNNs, LP-transmission-Edge-Cloud}.

We now overview the orthogonal approaches, such as early‐exit, transmission compression, and model quantization, that an MLaaS broker can apply independently of privacy or cost‐driven adaptations. 
The related works are detailed in Sec.~\ref{sec:sol} and the interactions among latency, privacy, and cost optimizations are discussed in Sec.~\ref{sec:coupling}.

\paragraph*{Early-exit adaptation.}
A model partition can be adapted to reduce inference time ($\pi_{pid}$ and $\pi_{pid+1}$ on both sides of the dashed line in Fig.~\ref{lat_optimization}).
Each partition $pid$ can be accelerated by attaching $Q_{pid}$ internal classifiers, where each classifier $c$ has the confidence threshold $\alpha_{pid}^{c}$, request exit rate $\beta_{pid}^{c}$, and the observed test metrics $A_{\pi_{pid}}^{c}$, including precision and recall~\cite{CLIO, SPINN, li2019edge}.
We denote the sum of the exit percentages for partition $pid$ as $\beta_{pid}$.
With $\beta_{pid}$ of requests finishing at a shallow partition, internal classifiers reduce the running time of requests by proportionally cutting the communication cost of deeper partitions: only the remaining $(1-\beta_{pid})$ fraction of queries must transmit their activations, so the expected transmission delay for partition $pid$ is scaled by the same $(1-\beta_{pid})$ factor in the term $T_{pid}^{T}$ of Eq.~(\ref{trans_time}).

\paragraph*{Transmission and model compression.}
Partitions can also be adapted with transmission and model compression methods to mitigate both transmission and computation overhead.
For each partition $pid$, we denote by $x_{pid}$ the hidden variable output ($x_0$ refers to the source data) and two model compression ratios: (1)~$\gamma_{pid}$ for latent space compression layers (e.g., in Fig.~\ref{lat_optimization}, an Auto-Encoder Neural Network consists of dark blue layers representing an encoder and dark orange layers representing the decoder), and (2)~$\kappa_{pid}$ for model compression, including model size reduction approaches like knowledge distillation, neuron pruning and quantization~\cite{Task-Oriented-bottleneck, Deep-Compressive-Offloading, matsubara2021neural, Auto-tuning-Quantization-NN, XAI-intermediate-data-compression, PieSlicer} as exemplified by light blue and light orange layers in Fig.~\ref{lat_optimization}.

\paragraph*{Transmission compression.}
Within partition $pid$, the encoder–decoder pair
$\langle\mathcal{E}_{pid},\mathcal{D}_{pid}\rangle$ replaces $x_{pid}$ with a latent vector
\[
\tilde{x}_{pid}= \mathcal{E}_{pid}\!\bigl(x_{pid}\bigr),\qquad
\lvert\tilde{x}_{pid}\rvert = (1-\gamma_{pid})\,\lvert x_{pid}\rvert,\; 0\le\gamma_{pid}\le 1,
\]
so that only a $(1-\gamma_{pid})$ fraction of the original bytes is sent over the network.
After arrival, the decoder reconstructs the activation map,
\[
\hat{x}_{pid}= \mathcal{D}_{pid}\!\bigl(\tilde{x}_{pid}\bigr),
\]
before the forward pass resumes.
Because only the remaining $(1-\beta_{pid})$ requests continue beyond partition~$pid$, the expected transmission delay for this segment is modulated by the product
$(1-\beta_{pid})(1-\gamma_{pid})$ in the term $T_{pid}^{T}$ of constraint~\eqref{trans_time} in the formulation.

\paragraph*{Model compression.}
Model size reduction approaches like knowledge distillation, neuron pruning and quantization shrink the computation and output footprint of $F_{pid}$:
\begin{itemize}
  \item \emph{Distillation} trains a student network $\widetilde{F}_{pid}$ that mimics $F_{pid}$ with fewer parameters and narrower channel widths, reducing FLOPs and activation size to $(1-\kappa_{pid})$ of the original.
  \item \emph{Pruning} filters redundant neurons, leading to a $(1-\kappa_{pid})$ reduction in both $T_{pid}^{C}$ and the tensor passed to the next hop.
  \item \emph{Quantization} stores weights and activations in low-bit-width integers. We fold its effect into the same factor~$\kappa_{pid}$ for brevity.  An 8-bit model, for instance, halves memory traffic and doubles the effective SIMD throughput on hardware that packs two 8-bit multiply-accumulate (MAC) operations into the slot of a single 16-bit MAC, thereby improving parallelization and further shortening $T_{pid}^{C}$.
\end{itemize}
Together, these techniques scale the compute term $T_{pid}^{C}$ in constraint~\eqref{comp_time_i1} and the downstream transmission term $T_{pid}^{T}$ by the multiplicative factor $(1-\kappa_{pid})$ as shown in Eq.~(\ref{trans_time}).

\subsubsection{Latency Formulation}\label{sec:latency-formulation}
We formulate a constrained multi-objective optimization that jointly minimizes processing and transmission delays subject to an accuracy constraint.
\begin{align}
    \begin{split}
        {\cal L}^L &= \min_{pid, \alpha_{pid}^{c}, \kappa_{pid}, \gamma_{pid}}       (\xi_{0}^{T}T_{0}^{T}+\sum_{pid=1}^{M}T_{pid}) \\
    \end{split}\label{optimization_goal}\\
    s.t.\ & \sum_{pid=1}^{M}\sum_{c=1}^{Q_{pid}} \beta_{pid}^{c}*A_{\pi_{pid}}^{c} \geq A_{tar} \label{accuracy_constrain}\\
    \beta_{pid} &= \sum_{c=1}^{Q_{pid}}Pr(\alpha_{pid}^{'c} > \alpha_{pid}^{c}) = \sum_{c=1}^{Q_{pid}}\beta_{pid}^{c} \label{alpha}\\ 
    x_{pid} &= \pi_{pid}(F_{pid})(x_{pid-1}) \label{lat_output_def}\\
    T_{pid}^{T} &= \frac{(1-\kappa_{pid})(1-\beta_{pid})(1-\gamma_{pid})Size(F_{pid}(x_{pid-1}))}{bandwidth} \label{trans_time}\\
    T_{0}^{T} &= \frac{(1-\gamma_{0})Size(x_{0})}{bandwidth} \label{trans_time_i0}\\
    T_{pid}^{C} &= \frac{FLOPs_{pid}^{pid}(x_{pid-1},\pi_{pid})}{\mu_{pid}} \label{comp_time_i1}\\
    T_{pid} &=  \xi_{pid}^{C}T_{pid}^{C}+\xi_{pid}^{T}T_{pid}^{T} \label{i_and_i1}\\
    &\alpha_{pid}^{c} \in [0,1], 
    \kappa_{pid} \in [0,1],\\
    &\gamma_{pid} \in [0,1],
    \xi \in \mathbb{R}^+
\end{align}

\begin{table}[t]
\centering
\begin{tabular}{|| c | c ||} 
\hline
Notation & Definition\\
 \hline
  $\pi_{pid}^{Lat}(.)$ & \begin{tabular}{@{}c@{}}Adapt partition $F_{pid}$ \\ to minimize inference latency\end{tabular}\\
 \hline
 $\alpha_{pid}^{c}$ & \begin{tabular}{@{}c@{}}Confidence thresholds \\ for classifier $c$ in partition $pid$\end{tabular}\\
 \hline
 $\kappa_{pid}$ & \begin{tabular}{@{}c@{}}Output compression rate \\ of model knowledge distillation\end{tabular}\\ 
 \hline
 $\gamma_{pid}$ & \begin{tabular}{@{}c@{}}Output compression rate \\ of compression layers(encoder\&decoder)\end{tabular}\\
 \hline
 $\gamma_{0}$ & \begin{tabular}{@{}c@{}}Source data compression rate \\ of compression layers(encoder\&decoder)\end{tabular}\\
 \hline
 $\beta_{pid}^{c}$ & \begin{tabular}{@{}c@{}}Percentage of request exit \\ at classifier $c$ in partition $pid$\end{tabular}\\
 \hline
 $\beta_{pid}$ & Percentage of request exit in partition $pid$\\
 \hline
 $x_{0}$ & Source data\\
 \hline
 $Q_{pid}$ & Quantity of classifier in partition $pid$\\
 \hline
 $A_{\pi_{pid}}^{c}$ & Observed model accuracy after adaptation\\
 \hline
 $A_{tar}$ & User-defined model target accuracy\\
 \hline
 $T_{pid}^{T}$ & Estimated transmission time for activations and states\\
 \hline
 $T_{pid}^{C}$ & Estimated computation time\\
 \hline
 $FLOPs_{i}^{j}(.)$ & FLOPs from layer $i$ to layer $j$ inclusive\\
 \hline
\end{tabular}
\caption{Latency Optimization Formulation Notations}
\label{latency-notations}
\end{table}

In line~\ref{accuracy_constrain}, $A_{tar}$ is a user-defined model accuracy constraint and $\beta_{pid}^c$ denotes the percentage of requests leaving the internal classifier $c$ in partition $pid$.
In line~\ref{alpha}, $\alpha_{pid}^{'c}$ is the profiled mean confidence during inference for the internal classifier $c$ in partition $pid$, $\alpha_{pid}^{c}$ is the confidence threshold for the internal classifier $c$ in partition $pid$, and $\beta_{pid}$ indicates the percentage of requests leaving partition $pid$ during inference.

In line~\ref{lat_output_def}, we define the output of partition $pid$ as $\pi_{pid}(F_{pid})(x_{pid-1})$, where the model partition $F_{pid}$ adapted with $\pi_{pid}(.)$ takes $x_{pid-1}$ as input.
Notice that the model adaptations include the introduction of internal classifier(s) and transmission and model compression.
To quantify the effect of those adaptations in inference latency, in line~\ref{trans_time}, we estimate the transmission delay from the adapted partition $pid$ to $pid+1$ based on input size $Size(x_{pid-1})$, the early exiting ratio $\beta_{pid}$ and the two compression ratios ($\gamma_{pid}$ and $\kappa_{pid}$).
Then, in line~\ref{trans_time_i0}, we estimate the transmission time for the source data to the location of the first NN partition.
$\gamma_{0}$ is the compression ratio of the source data.

In line~\ref{comp_time_i1}, $FLOPs_{pid}^{pid}(x_{pid-1},\pi_{pid})$ is the profiled count of FLOPs (FLoating-point OPerations) for partition $pid$, given input ($x_{pid-1}$) and adaptation ($\pi_{pid}$).
The subscript and superscript of $FLOPs_{pid}^{pid}(.)$ indicates the start and end partitions to count FLOPs.
When we focus on one partition, the subscript and superscript are the same.

\subsection{Monetary Cost}\label{sec:prob-form-cost}
\begin{figure}
\captionsetup{justification=centering}
    \includegraphics[width=0.48\textwidth]{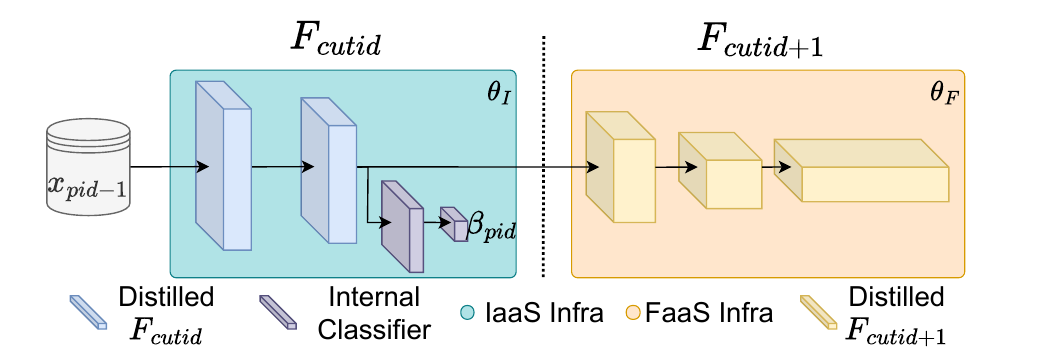}
    \caption{Illustration of cost(\$) optimization problem.
    }\label{cost_optimization}
\end{figure}
In this section, we categorize cost optimization approaches and introduce our monetary cost formulation (Sec.~\ref{sec:cost-formulation}).
Resource provisioning approaches based on resource cost (\$) for DL inference tasks remain underexplored.
Using detailed cost models of different cloud and edge services, an MLaaS broker can determine cost-efficient resource provisioning strategies for different workloads.
In particular, for decomposable sparsely activated ML models where each portion faces a varying workload, fine-grained resource provisioning and load balancing for submodels are essential to achieve cost-efficient ML inference.

\begin{table}[t]
\centering
\begin{tabular}{|| c | c ||} 
\hline
Notation & Definition\\
 \hline
 $Latency$ & Latency bound\\
 \hline
 $T_{I}$ & Observed Mean IaaS Time\\
 \hline
 $T_{F}$ & Observed Mean FaaS Time\\ 
 \hline
 $T_{cold}^{cutid}$& FaaS function cold start time\\
 \hline
 $T_{trans}^{cutid}$& Transmission delay from IaaS to FaaS\\
 \hline
 $C_{I}(.)$ & Unit cost of IaaS given VM capacity\\
 \hline
 $C_{F}(.)$ & Unit cost of FaaS given function capacity\\
 \hline
 $\theta_{I}$ & VM capacity\\
 \hline
 $\theta_{F}$ & Function capacity\\
 \hline
\end{tabular}
\caption{Cost(\$) Optimization Formulation Notations}
\label{cost-notations}
\end{table}

In our formulation, we minimize the combined costs of provisioning Infrastructure‐as‐a‐Service (IaaS) and Function‐as‐a‐Service (FaaS) platforms, regardless of whether they are deployed at the edge or in the cloud.  
FaaS charges users only for the actual execution time of the deployed model, making it particularly cost‐efficient for low‐rate or bursty workloads -- for example, holiday traffic spikes in DNN‐based vehicle localization~\cite{serverless-in-the-wild, DNN-partition-SLS, FaaS-dnn-serve, CarMap}.  
Conversely, IaaS platforms automatically scale virtual machines (VMs), which incur longer cold‐start delays (for hardware and OS provisioning) and are billed for the entire time the resources are reserved.  
Because VM scaling is coarse‐grained, users typically provision based on service‐level objectives (SLOs) rather than real‐time workload fluctuations, often leading to over‐provisioning~\cite{LIBRA}.  
However, when a workload maintains high utilization, characterized, for instance, by a steady ingestion rate around the mean $\pm$ one standard deviation under a Poisson model~\cite{faas-vs-iaas-ml-training, LIBRA}, IaaS can be more economical than serverless functions.

\subsubsection{Monetary Cost Formulation}\label{sec:cost-formulation}
Motivated by recent advances in sparsely activated models~\cite{efficient-moe} and the inability of edge devices to host large neural networks~\cite{Neurosurgeon}, we represent a sequential DNN with internal classifiers as a dependent acyclic graph of submodels (Fig.~\ref{cost_optimization}).
For inference, partition $F_{pid}$ produces an early exit for a fraction $\beta_{pid}$ of requests with its internal classifier.
The remaining $(1-\beta_{pid})$ portion continues to partition $F_{pid+1}$.
Shallow partitions thus face a steady, high-rate workload, whereas deeper partitions see a lower request rate.
Consequently, for a constant arrival rate of $N$ requests/s, we provision VMs to handle up to $r_{\max}$ of those requests at the shallow partitions, ensuring high VM utilization, while routing the remaining requests to FaaS to avoid under-utilizing any individual VM.

\begin{align}
    {\cal L}^C = \min_{cutid, \theta_{F}, \theta_{I}, \alpha_{pid}^{k}} \;&
    C_{I}(\theta_{I})T_{I}\sum_{pid = 1}^{cutid}\beta_{pid} \notag \\
    &+ C_{F}(\theta_{F})T_{F}\sum_{pid = cutid+1}^{M}\beta_{pid} \label{cost_optimization_goal} \\
    s.t.\quad Latency &\geq T_{I} + T_{F} \label{SLA} \\
    \sum_{pid=1}^{M}\sum_{c=1}^{Q_{pid}} & \beta_{pid}^{c} A_{\pi_{pid}}^{c} \geq A_{tar} \label{cost_accuracy_constrain} \\
    \beta_{pid} &= \sum_{c=1}^{Q_{pid}} Pr(\alpha_{pid}^{'c} > \alpha_{pid}^{c}) \notag \\
    &= \sum_{c=1}^{Q_{pid}} \beta_{pid}^{c} \label{Fload} \\
    T_{F} &= \frac{FLOPs_{cutid+1}^{M}(x_{cutid})}{\theta_{F}} + T_{cold}^{cutid} \label{meanFT} \\
    T_{I} &= \frac{FLOPs_{1}^{cutid}(x_{0})}{\theta_{I}} + T_{trans}^{cutid} \label{meanIT} \\
    cutid &\in [1,M],\; \alpha_{pid}^{k} \in [0,1], \\
    \theta_{F} &\in \{\text{FaaS Capacities}\}, \\
    \theta_{I} &\in \{\text{IaaS Capacities}\}
\end{align}

The above optimization targets a steady arrival rate of $r_{\max}$ DNN inference requests per second, enough to fully utilize the VMs.  
It chooses between IaaS, FaaS, or \emph{hybrid offloading} -- offloading some requests from IaaS to FaaS to complete their deep-layer processing.  
The decision variables are the FaaS configuration ($\theta_{F}$), the IaaS configuration ($\theta_{I}$), the internal classifier thresholds ($\alpha_{pid}^{c}$), and the partition index ($cutid$, assuming two partitions), all subject to the latency constraint in line~\ref{SLA}.

$C_{I}$ and $C_{F}$ map resource configurations to monetary cost.
Then, we estimate the running cost, based on the average durations $T_{I}$ (VM reservation time) and $T_{F}$ (FaaS execution time) profiled for each forward pass.
In line~\ref{Fload}, $\beta_{pid}$ is the fraction of requests exiting at partition $pid$, determined by confidence thresholds $\alpha_{pid}^{c}$.  
Lines~\ref{meanFT} and \ref{meanIT} define the profiled mean durations for FaaS and IaaS, respectively.  
For a given $cutid$, motivated by previous work~\cite{ShallowDeepNet}, the duration is computed by dividing the required FLOPs ($FLOPs_{cutid+1}^{M}$ from partition $cutid+1$ to $M$) by the provisioned capacity ($\theta_{F}$ or $\theta_{I}$), assuming once offloaded to FaaS, execution does not revert to IaaS.  
This may overestimate utilization since some requests exit early.  
To account for early exits, we weigh $T_{F}$ by $\sum_{pid = cutid+1}^{M}\beta_{pid}$ and $T_{I}$ by $\sum_{pid = 1}^{cutid}\beta_{pid}$ in Eq.~\eqref{cost_optimization_goal}.  
We also include the hidden‐variable transmission delay $T_{trans}^{cutid}$ in $T_{I}$ (line~\ref{meanIT}), since the serverless function has not yet been invoked and thus does not incur FaaS billing until execution.  
Conversely, the serverless cold‐start delay $T_{cold}^{cutid}$ is included in $T_{F}$ (line~\ref{meanFT}), as it involves model loading and hardware setup that are billable.\footnote{Major FaaS providers keep containers warm to minimize cold starts~\cite{serverless-in-the-wild, aws_lambda_coldstart, azure_functions_hosting, gcf_instance_lifespan}, so this term may overestimate FaaS cost.  VM cold starts are omitted since we assume long‐running VMs.}

Next, given the VM configuration for $r_{\max}$ requests/s, we optimize load‐balancing so that $r_{\max}$ requests are served by shallow partitions on fully utilized VMs, with the remainder directed to FaaS.
\rb{Note that the sparsity of early-exit models is highly dependent on the input-data distribution and must be profiled offline.}
\rb{When the probability of FaaS provisioning increases, resulting in higher costs, we must update our workload distribution profile and resource configurations.}



\subsection{Privacy}\label{sec:prob-form-privacy}
\begin{figure}
\captionsetup{justification=centering}
    \includegraphics[width=0.48\textwidth]{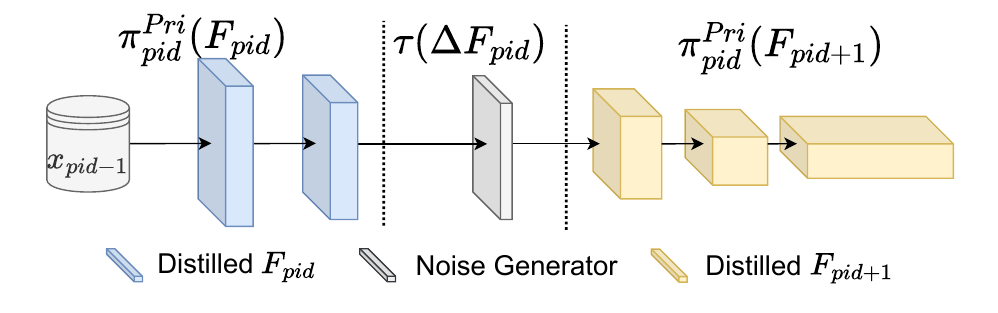}
    \caption{Illustration of privacy optimization problem.
    }\label{pri_optimization}
\end{figure}
We study the privacy of source data in distributed DNN inference applications.
This section overviews the concepts -- accessible by an MLaaS broker -- for defense against model inversion attacks (MIA)~\cite{model-inversion-attack-1, model-inversion-attack-2, model-inversion-attack-3}.
We formulate the privacy optimization problem in Sec.~\ref{sec:privacy-formulation}.
Many of the techniques discussed are applied during model training, rather than only at inference
Practical countermeasures for MIA are presented in detail in Section~\ref{sec:pri-sol}.

\subsubsection{\rb{Threat Model: Model Inversion Attack (MIA)}}\label{sec:MIA-threat}
\rb{We consider an \emph{honest-but-curious} adversary who observes the hidden variables~$x_{pid}$ transmitted between two partitions of a distributed DNN.  
The adversary stores these activations and trains an auto-encoder (or any other reconstructor) $\mathcal{R}_{\text{MIA}}$ \emph{offline} with public data, then outputs $\hat{x}_{0}=\mathcal{R}_{\text{MIA}}(x_{pid})$ at inference time.
We measure privacy leakage by the expected reconstruction error, e.g., mean square error (MSE),
\[
\text{Leak}_{pid}
   \;=\;
   \operatorname*{\mathbb{E}}_{x_{0}\sim\mathcal{D}}
   \bigl[
     \lVert x_{0}-\hat{x}_{0}\rVert_{2}^{2}
   \bigr],
\]
where $\mathcal{D}$ is the distribution of user inputs.
A \emph{smaller} MSE implies a \emph{stronger} attack.
Such attacks have been well explored in the research community~\cite{erdogan2022unsplit, ResSFL, gao2023pcat, xu2024stealthy} and in particular the Prompt Inversion Attack (PIA) for LLMs~\cite{prompt-inversion-attack-0, prompt-inversion-attack-1}.}

\subsubsection{Privacy-oriented Adaptation}\label{sec:privacy-only}
To mitigate MIA, an MLaaS broker can leverage two orthogonal defenses -- \textbf{regularization} and
\textbf{perturbation} -- applied \emph{independently} of any latency or monetary cost-driven optimizations:
\begin{enumerate}
  \item \emph{Regularization.}  We fine-tune each partition with a privacy-aware objective, denoted by $\pi_{pid}^{\text{pri}}$, that explicitly penalizes the attacker’s reconstruction loss~\cite{NoPeek,ResSFL,depth-dropout-privacy,FSL-journal}.
  \item \emph{Perturbation.}  We add a noise layer $\tau(\Delta F_{pid})$ based on the output sensitivity $\Delta F_{pid}$ and bounded by a scalar $\lambda$ \cite{depth-dropout-privacy,Shredder-noise-tensor}.
\end{enumerate}
Figure~\ref{pri_optimization} illustrates the three stages (separated by dashed lines):
partition~$pid$ (left), the noise layer $\tau(\cdot)$ (center), and
partition~$pid\!+\!1$ (right).
However, we note that the effectiveness of such remedies in handling PIA for LLMs remains under-explored due to the prohibitive model training overhead.
We next formulate the privacy optimization problem.

\subsubsection{Privacy Formulation}\label{sec:privacy-formulation}
With the notations in Table~\ref{privacy-notations}, we formulate the \emph{privacy} objective:
\begin{align}
    \begin{split}
        {\cal L}^P = \min_{\pi_{pid}^{pri}, \Delta, \lambda}(&w_{CE}CE(\hat{y},y)-\sum_{pid=1}^{M}w_pMSE(F_{pid}^{-1}(x_{pid}), x_{pid-1})) 
    \end{split} \label{pri_optimization_goal} \\
    s.t.\ CE(\hat{y}, y) &\geq Thr_{CE}  \label{CE_thr}\\
    x_{pid} &= \lambda\pi_{pid}^{pri}(F_{pid})(x_{pid-1})+\tau(\Delta F_{pid}) \label{pri_output_def}\\
    \forall pid &> 1
\end{align}
\medskip
\begin{table}[t]
\centering
\begin{tabular}{|| c | c ||} 
\hline
Notation & Definition\\
 \hline
 $\pi_{pid}^{pri}(.)$            & \begin{tabular}{@{}c@{}}Fine-tune partition $F_{pid}$ \\ for better privacy guarantee\end{tabular}\\
 \hline
 $\tau(.)$                       & Noise generator method\\ 
 \hline
 $\lambda$                       & Output bounding parameter\\
 \hline
 $\Delta F_{pid}$                & Sensitivity of $F_{pid}$’s output\\
 \hline
 $w_{CE},\,w_{p}$                & Weights for loss terms \\
 \hline
 $Thr_{CE}$                      & Maximum allowed cross-entropy loss \\
 \hline
 $F_{pid}^{-1}(\cdot)$           & Learned inverse (attacker’s surrogate) \\
 \hline
\end{tabular}
\caption{Privacy Optimization Formulation Notations}
\label{privacy-notations}
\end{table}

Equation~\eqref{pri_optimization_goal} \emph{maximizes} the attacker’s error ($-$sign) while keeping model accuracy above a threshold \eqref{CE_thr}.  
In this formulation, $CE(\hat{y},y)$ denotes the cross-entropy of the model predictions, and $MSE(F_{pid}^{-1}(x_{pid}), x_{pid-1})$ quantifies the fidelity of the attacker’s reconstruction.
The forward rule \eqref{pri_output_def} shows how bounding ($\lambda$) and random noise
$\tau(\cdot)$ are injected between partitions.
The interaction with latency and monetary cost is discussed in Sec.~\ref{sec:coupling}; here we isolate privacy.

\section{Problem Solutions}\label{sec:sol}
In the preceding section, we formulated an optimization problem for deploying a dependent acyclic graph of submodels, accounting for latency, source data privacy, and resource monetary cost (\$). 
In this section, we detail the solution techniques to the optimization sub-objectives, such as early exits, compression, and privacy-preserving inference approaches. 
Based on the existing works, we then discuss the interactions among sub-objectives in Sec.~\ref{sec:coupling}.
Furthermore, we identify open issues in Sec.~\ref{sec:issues}.
Overall, these solution techniques can serve as valuable control mechanisms for ML service providers, improving Quality of Service (QoS), and increasing revenue.


\subsection{Latency}\label{sec:lat-sol}
The end-to-end latency of a neural network model comprises both processing and transmission delays. 
Building on earlier discussions, existing work dynamically minimizes the transmission of excessive hidden variables and combines capable cloud services.
This section begins by exploring dynamic deep neural network offloading~\cite{Neurosurgeon, EOP}. 
Then, we discuss internal classifiers which allow early exit and save computation for deep layers~\cite{SPINN, Edgent, li2019edge, ee-part-opt-form, combine-DNN-with-EE}.
Next, we examine transmission data and model compression approaches~\cite{BBNet, BottleNet, Task-Oriented-bottleneck, CLIO, SPINN, PieSlicer, matsubara2021neural, Self-Distillation}. 

\subsubsection{Dynamic Partitioning}
When dynamically partitioning a neural network, the computation ($FLOPs_{i}^{j}(.)$) and activation size ($Size(.)$) can be estimated based on the model weights and input size~\cite{flops-calculation, ptflops}. 
Thus, delays, especially inference durations ($T_{pid}^{T}$ and $T_{pid}^{C}$) can be modeled using regression methods, by profiling the NN model, across the cloud and resource-limited edge environments~\cite{Neurosurgeon, EOP}.
Previous work estimates transmission and processing delays for various model configurations, factoring in computation resources and input sizes to devise a deployment plan for $M$ NN partitions that minimizes latency and energy consumption~\cite{Neurosurgeon}. 
However, feasible solutions may not always exist for a given model architecture or environment, for example, when resource availability is constrained.
Next, we explore orthogonal methods to reduce demands on transmission and processing resources.

\subsubsection{Early Exits}
\noindent {\bf Background.}
Deep layers of a DNN model tend to focus on fine details, which, however, can result in misclassifications.
While, shallow layers extract high-level features, which can be sufficient for accurate request classification.
Such observation is described as \textit{overthinking}~\cite{ShallowDeepNet}.
Related research~\cite{ShallowDeepNet, Zero-Time-Waste, BertLosePatience} addresses this concern, proposing the reuse of features extracted from various layers to improve the inference latency through \textit{internal classifiers}~\cite{BranchyNet},

These classifiers share a structure similar to traditional NN classifier layers, comprising feature reduction (pooling) layers, fully connected layers, and a softmax activation function.
Except, internal classifiers are attached to the hidden layers.
To trigger early exits, one can configure a threshold~\cite{ShallowDeepNet} for the Bayesian probabilities of class predictions at each classifier~\cite{NN-bayesian-prob, guo2017calibration}.

\begin{figure}
\captionsetup[subfigure]{justification=centering}
    \begin{subfigure}[t]{0.5\textwidth}
    \includegraphics[width=\textwidth]{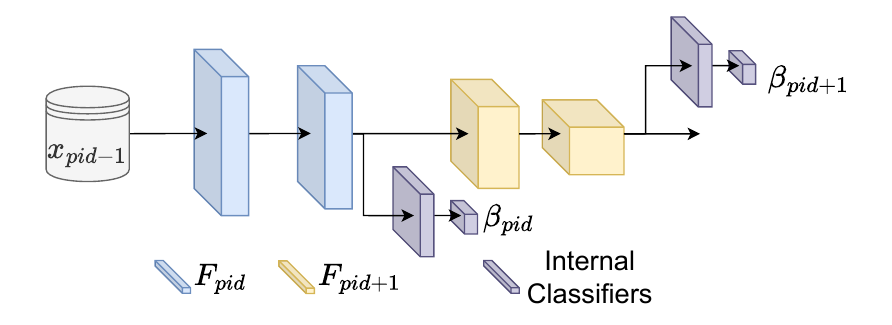}
    \end{subfigure}
    \caption{Internal Classifier Architecture: Each internal classifier allows requests to exit in the middle of an NN. For example, \(\beta_{pid}\) of requests exit at NN partition \(F_{pid}\).
    }\label{SDN_Architecture}
\end{figure}

In deep learning (DL) tasks, incorporating early exits and residual connections at various internal layers of a DL model allows for better utilization of insights during inference, leading to improved accuracy and latency.
Early exits prevent excessive forwarding of requests (hidden variables) to deep layers for classification. 
As exemplified in Fig.~\ref{SDN_Architecture}, an internal classifier allows $\beta_{pid}$ portion of requests to exit the model partition $F_{pid}$, highlighted in blue.

\noindent {\bf Methods.}
Recent research~\cite{DDNNs, SPINN, CLIO} models the relationship between the confidence threshold ($\alpha$) of internal classifiers and the proportion of early exits ($\beta$) when formulating inference latency. 
We incorporate such relationship in our latency optimization framework, which allows tuning $\beta$ through $\alpha$ to meet a specific mean latency target (lines~\ref{accuracy_constrain} and \ref{alpha} in Sec.~\ref{sec:latency-formulation}.)

Lowering the confidence threshold for each inference can also negatively impact the inference accuracy.
To maintain high accuracy, previous studies explore multi-objective optimization including confidence threshold and dynamic layers offloading between the edge and cloud based on network conditions, as characterized by lines~\ref{lat_output_def}, \ref{trans_time}, and \ref{comp_time_i1} in our latency formulation.
For example, SPINN~\cite{SPINN} empirically demonstrates that under high and stable WAN bandwidth, more layers can be offloaded to cloud nodes. 
The increased computational capacity compensates for additional communication delays, reducing overall latency, which allows high confidence threshold and high accuracy.
In contrast, when network bandwidth is limited, the approach shifts more layers to resource-limited edge nodes. 
Despite an increase in processing time at the edge, overall latency is optimized by minimizing reliance on WAN communication, without significantly compromising accuracy.

\subsubsection{Input and output compression}\label{data-compression-method}
\noindent {\bf Background.}
During inference, not all features are necessary for a classification task. 
Apart from traditional statistical or heuristic methods~\cite{jpeg}, Deep Neural Networks (DNNs), specifically Auto-Encoder NNs, can facilitate feature selection to preserve prediction performance~\cite{Task-Oriented-bottleneck}. 
An Auto-Encoder NN consists of two components: an encoder, which transforms inputs to a compressed representation, and a decoder, responsible for inverting the dimensionality reduction~\cite{autoencoder}.
On the other hand, model compression methods can also reduce feature size.
These methods will be explored further in Sec.~\ref{model-compression-methods}.

\noindent {\bf Methods.}
In our latency optimization (${\cal L}^{L}$), we denote the cropping and compression of input data with rate $\gamma_{0}$. 
The compression rate of intermediate data achieved through model compression is denoted as \(\kappa_{pid}\), while the rate achieved through feature engineering methods such as Auto-Encoder is represented as \(\gamma_{pid}\).
We encapsulate the computation overhead of Autoencoder NN in the model transformation $\pi_{pid}$.

Heuristic-based compression methods, such as JPEG for image inputs, can help reduce feature dimension. 
In particular, certain activation functions, for example relu~\cite{relu}, produce zero or near-zero outputs, allowing compression from a dense matrix into a sparse matrix that is storage and transmission efficient~\cite{CLIO}.
Moreover, related works~\cite{CLIO, SmartSwitch-AI-Accelerator, bi-quan, jacob2018quantization} explore the quantization of weights and intermediate data representations. 
Rather than using double precision floats, these methods consider 8-bit~\cite{jacob2018quantization} or in the extreme case single-bit~\cite{bi-quan} approximations.

Several prior works investigate content-aware transmission compression.
For example, in an \textit{AMBER Alert} system, if the model only requires identifying a car or person in the scene, the edge device only transmits cropped images focusing on Region of Interest (RoI) to the cloud for analysis, thereby reducing bandwidth and latency~\cite{Llama}. 
\rb{Likewise, RoI extraction in multimodal LLM pipelines enables processing of high-resolution imagery by isolating the most informative patches for downstream analysis~\cite{GPT4RoI, chen2025visualinstructiontuningchain}. 
For text-based LLMs, prompt-compression techniques prune tokens and sentences that are considered uninformative with respect to the query, improving responsiveness without necessarily sacrificing relevance~\cite{zhang2025spargeattn, zhang2025sageattention, zhang2024sageattention2, jiang-etal-2023-llmlingua, jiang-etal-2024-longllmlingua, pan-etal-2024-llmlingua, yuan-etal-2025-native, context-aware-sentence-pruning, jiang2024minference, li2025scbench, li2025mminference}.}
Although these approaches generally reduce transmission delay, their impact on accuracy depends on the quality of the cropping or pruning mechanism
Related study suggests that concentrating on relevant data can both cut transmission costs and improve predictive performance~\cite{PieSlicer}.

Previous work has also applied ML-based dimensionality reduction to reduce transmission data and maintain accuracy.
One idea is to insert a \textit{bottleneck} between two neural network partitions using an Auto-Encoder NN~\cite{BBNet, BottleNet, LP-transmission-Edge-Cloud, Task-Oriented-bottleneck, Deep-Compressive-Offloading, matsubara2021neural}.
The Auto-Encoder is trained by minimizing the Mean Squared Error (MSE) loss between the input and output data (the reconstruction).
In this setup, the encoder projects the intermediate data into a more space-efficient latent space, effectively reducing the channels, width, and height. 
The decoder, which serves as an approximation of an inverted encoder function, reconstructs the input of the previous partition using the compressed intermediate data. 
This approach leads to a compact representation of intermediate data, enhancing efficiency minimal accuracy degradation.

\rb{For attention-based architectures, low-rank approximation (LoRA) techniques have been proposed to compress sparse attention matrices~\cite{lora-sparse}. Because attention matrices scale quadratically with sequence length, approximating them via low-rank factors significantly lowers both computational and memory costs.}

Other approaches jointly optimize the backbone and an intermediate autoencoder (AE) to better preserve end-to-end accuracy~\cite{Task-Oriented-bottleneck, LP-transmission-Edge-Cloud, Deep-Compressive-Offloading}.
By training the AE together with the task model, the compressed latent space tends to retain features that are most relevant for the downstream prediction task.
However, finding an optimal compressed feature space for both high accuracy and high $\gamma_{pid}$ remains a challenging task that requires extensive hyperparameter tuning. 
To address this, recent research~\cite{XAI-intermediate-data-compression} uses explainable AI techniques, including Integrated Gradients~\cite{Integrated-Gradients}, to construct an intermediate data space that emphasizes features with the greatest impact on predictions.

\noindent \textbf{Summary of data compression methods:} 
In our latency optimization framework, we adjust the data compression rates $\gamma_{pid}$ and $\kappa_{pid}$ to minimize transmission overhead.
The introduction of data compression adds FLOPs in each partition ($\pi_{pid}$ in line~\ref{lat_output_def}), creating a trade-off among reduced transmission overhead, increased computation overhead, and potentially compromised model accuracy.
Various data compression methods are detailed in Table~\ref{Comparing-Data-Compression-Methods}, highlighting the practicality of both Auto-Encoder and quantization techniques as they are broadly model and data agnostic.
However, an Auto-Encoder offers greater flexibility compared to quantization, which enables fine-tuning the compression model (encoder), inversion model (decoder), and feature ranking techniques (such as explainable AI tools) to optimize latency and accuracy based on the user's specific use case. 
While an Auto-Encoder introduces additional computational overhead, a quantized model and activations also require specialized training tools due to the discrete space. 
For example, stochastic gradient descent (SGD) must be adapted to accommodate the discrete space~\cite{NN-quant-survey, jacob2018quantization}.

\begin{table*}[t]
\centering
\begin{tabular}{|| c | c | c | c | c | c ||} 
\hline
 \begin{tabular}{@{}c@{}}\textbf{Activation} \\ \textbf{Compression Method}\end{tabular}& 
\begin{tabular}{@{}c@{}}\textbf{Pre-Processing} \\ \cite{PieSlicer}\end{tabular} & 
\begin{tabular}{@{}c@{}}\textbf{Heuristic} \\ \cite{CLIO}\end{tabular} &  
\begin{tabular}{@{}c@{}}\textbf{Quantization} \\ \cite{CLIO, SmartSwitch-AI-Accelerator, bi-quan, jacob2018quantization} \end{tabular} & 
\begin{tabular}{@{}c@{}} \textbf{AE} \\ \cite{BBNet, BottleNet, LP-transmission-Edge-Cloud, Task-Oriented-bottleneck, Deep-Compressive-Offloading, matsubara2021neural}\end{tabular} & 
\begin{tabular}{@{}c@{}} \textbf{AE(XAI)} \\ \cite{LP-transmission-Edge-Cloud, Deep-Compressive-Offloading, matsubara2021neural, Task-Oriented-bottleneck, XAI-intermediate-data-compression} \end{tabular} \\
 \hline
 Computation & low & low & low  & medium & medium\\
 \hline
 Compression & medium & medium & medium  & low & low\\ 
 \hline
 Accuracy & high & high & medium & medium & high\\
 \hline
 Practicality & low & low & high & high & medium\\
 \hline
\end{tabular}
\caption{
Comparison of Input and Intermediate Data Compression Methods:
The accuracy of the pre-processing method~\cite{PieSlicer} depends on the ability of the algorithm to accurately identify and crop the features of interest before sending data to the model (low practicality, high accuracy given good cropping algorithm, low extra computation for cropping input, and overall medium compression rate for cropping).
Heuristic-based compression algorithms, like clustering for zeros, rely on user expertise (low practicality, high accuracy, low extra computation, and overall medium compression rate depending on the inputs and heuristics applied).
Intermediate data quantization shortens data representation but may impact accuracy (high practicality, medium accuracy, and medium compression rate compared to other task-oriented methods) and demands an adapted optimization method for discrete space (low extra computation).
The Auto-Encoder (AE) can be applied to various data representations (high practicality) and can be adapted to different ML tasks by using shallower layers to minimize computation overhead (medium computation overhead) or designing smaller latent spaces to create a narrow bottleneck that tradeoffs accuracy (medium accuracy and low compression rate).
Na\"ive input or intermediate data compression can significantly compromise model accuracy if the features selected for transmission are suboptimal. 
In contrast, AE approaches leveraging explainable AI (XAI) tools selectively transmit crucial features for classifications, reducing transmission delay while maintaining high accuracy (overall medium practicality based on feature selection methods, overall medium computation demand with AE, high model accuracy, and low compression rate).}
\label{Comparing-Data-Compression-Methods}
\end{table*}

\subsubsection{Model Compression and Knowledge Distillation}\label{model-compression-methods}
\noindent {\bf Background.}
Deep Learning (DL) models can be customized to meet specific Machine Learning (ML) tasks and computational constraints. 
For latency-sensitive applications, simplifying the model can enable faster response times. 
Such simplification can be achieved through quantization~\cite{Pruning-and-quantization-survey}, layer skipping~\cite{darts}, adding, removing, or editing the layer blocks of a neural network~\cite{NN_Structure_Search}, and knowledge distillation~\cite{hinton2015distilling}. 

A common tradeoff of reducing model parameters is potential accuracy degradation.
However, such a drawback might be tolerable given the use case or in certain settings the model would not suffer a significant accuracy drop, as model simplification can be considered a form of regularization that mitigates overfitting and discourages shortcut learning~\cite{geirhos2020shortcut, Trustee}. 
Thus, model simplification is considered a versatile approach applicable across many different ML systems, achieving short processing times without significant accuracy loss.

\rb{For example, in large ML systems, such as large language models (LLM), the Mixture of Experts model architecture~\cite{shazeer2017, gross2017hard} decomposes a high-dimensional model into smaller \textit{experts} with a router NN selecting a subset of submodels for each request, reducing computational demands.
In certain LLM-based chatbot applications, chats generated by simplified LLMs on user devices can be enhanced by constraining the output space or leveraging cached outputs from full-sized LLMs deployed in the cloud.
This allows the final chatbot responses to match full-sized LLM quality while maintaining high throughput on the device~\cite{on-device-llm-reuse, Guidance}.}

\rb{On the other hand, distilled or quantized LLMs can assist original LLMs in speeding up chat generation.
Traditionally, chat generation proceeds sequentially token-by-token, resulting in low throughput.
Instead, a smaller LLM can speculatively generate the next $t$ tokens, which the original LLM then verifies and either accepts or rejects them in parallel based on the speculatively generated context.
This \textit{speculative decoding} approach significantly boosts chatbot throughput~\cite{speculative-decoding-0, speculative-decoding-1, speculative-llm}.}


We first discuss simplifications for standalone ML systems and then generalize them to the distributed setting.

Quantization or compression of neural network (NN) weights and activations is a popular technique to reduce computation complexity during inference~\cite{Pruning-and-quantization-survey}. 
Previous work introduces Post-Training Quantization (PTQ) and Quantization-Aware Training (QAT).
For small models or large models with aggressive quantization, recent studies have shown that using lower precision during training, for example $8$ bits~\cite{jacob2018quantization} and $1.58$ bits~\cite{ma2025bitnetb1_58_2b4t}, for NN weights can achieve accuracy comparable to higher precision representations.
In contrast, for very large models including LLMs, quantization training introduces significant overhead.
In AWQ~\cite{lin2023awq}, SmoothQuant~\cite{xiao2023smoothquant} and OPTQ~\cite{frantar2023optq}, researchers adjust the \textit{Scale} and \textit{Zero Point} (origin) for weights in pretrained models with static analysis of activation and weights.
Quantization improves inference latency and reduces storage requirements for various deep learning (DL) inference tasks. 
For model compression, in EIE~\cite{eie}, the authors introduce a novel representation and matrix multiplication algorithm that omits most common values in activations, optimizing computation and storage efficiency.

We can also dynamically skip layers or make predictions before the neural network (NN) model completes its full pass without modifying the base model. 
This approach, known as \textit{dynamic inference}~\cite{Dynamic-Neural-Networks, gating-DRL}, involves training additional gating networks to determine which layers or channels within a layer to skip, indicated by $0$ (skip) or $1$ (use). 
In DDI~\cite{DDI} and SkipNet~\cite{SkipNet}, the authors use a Long Short-Term Memory (LSTM) NN to make skipping decisions.

Methods focusing on weight and layer changes can be suboptimal in reducing latency, as they introduce minimal structural changes to the model.
Knowledge Distillation~\cite{hinton2015distilling, gou2021knowledge} trains a lightweight high-performance model (Student Model) using the inputs and outputs (including hidden variables or logits) of a more complex model (Teacher Model). 
As the Teacher Model is often over-parameterized, the Student Model can attain similar accuracy with reduced computational complexity. 
Furthermore, the Teacher Model can also be further refined with weighted outputs from the Student Models (\textit{soft labels}) and ground truth labels in a \textit{student-student} knowledge distillation setting~\cite{Self-Distillation, label-smoothing-regularization-KD, label-smoothing-bias-variance-KD}.
In this process, the output of the student model serves as a form of regularization to prevent overfitting.


\noindent {\bf Methods.}
In our latency optimization formulation in a distributed setting (Sec.~\ref{sec:prob-form-latency}), we use \(\pi_{pid}\) to represent all adaptations, including model compression. 
The function \(FLOPs_{i}^{j}(.)\) estimates the number of floating point operations given different partition configurations \(\pi_{pid}\) and partition input size $Size(x_{pid-1})$. 
In a dense NN, the FLOP count is linearly proportional to the number of model weights and the size of input data, so more layers or larger input sizes lead to longer processing times. 
To minimize computation latency, we can employ model compression or knowledge distillation methods for each NN partition. 
As a beneficial byproduct, model compression can also reduce the size of hidden variables which improves \(\kappa_{pid}\) discussed in Sec.~\ref{data-compression-method}.


Quantization~\cite{LP-transmission-Edge-Cloud, Auto-tuning-Quantization-NN, NN-quant-survey} and neuron skipping~\cite{BBNet, CLIO, Bottle-Reconstruct-Coop} can be applied to each NN partition.
Intuitively, weights close to zero contribute little to classification and can be pruned to save processing time without significantly affecting accuracy. 
In CLIO \cite{CLIO}, a certain percentage of weights, sorted by distance to zero, are ignored. 
However, with a higher compression rate, such approaches suffer from low prediction performance.

To maintain accuracy in cases of significant compression for resource-constrained edge partition NNs, Lee et al.~\cite{Bottle-Reconstruct-Coop} suggest the use of a deep decoder NN at the cloud node. 
The deep decoder with high inversion approximation capabilities compensates for the aggressively compressed NN partition, helping to preserve the model's accuracy.



Instead of compressing a base model, training a lightweight model replacement can effectively scale model size down to fit the capacity constraints.
With limited edge capacity, applying knowledge distillation to partitions could save processing time with minimal loss in accuracy~\cite{matsubara2021neural}.

\noindent \textbf{Summary of model compression methods:} 
In our latency optimization framework (Sec.~\ref{sec:prob-form-latency}), \(\pi_{pid}\) represents model adaptations including both model compression and distillation, which reduce model complexity yielding a placement with minimal processing time. 
This section reviewed the application of quantization, knowledge distillation, and weight pruning, each with strengths and weaknesses. 
While these methods are orthogonal and should be evaluated together to optimize model complexity and placement, they vary in terms of practicality and computational overhead. 
Table~\ref{Comparing-Model-Compression-Methods} presents a summary and comparison of these model compression techniques.

Among these techniques, knowledge distillation is the most configurable, offering various student model designs and distillation approaches to achieve high model accuracy.
Thus, it is considered the most practical, but with the highest computation overhead. 
In contrast, quantization is less configurable, so we position it at medium practicality but with the lowest computation overhead. 
The effectiveness of weight-pruning depends largely on the underlying data distribution. 
For example, pruning weights close to zero is a well-explored method to maintain accuracy while reducing computational complexity. 
However, mask-based pruning methods may require training specific to each source data distribution~\cite{model-pruning-for-mia-privacy}, leading to medium practicality but low computation overhead for each mask.

\begin{table*}[t]
\centering
\begin{tabular}{|| c | c | c | c | c ||} 
\hline
 \textbf{Model Compression Method} & \textbf{Quantization}~\cite{Auto-tuning-Quantization-NN, LP-transmission-Edge-Cloud} & \textbf{KD}~\cite{matsubara2021neural} & \textbf{Weight-Pruning}~\cite{BBNet, CLIO, Bottle-Reconstruct-Coop}\\
 \hline
 Computation & low & high & low\\
 \hline
 Accuracy & medium & high & high\\
 \hline
 Practicality & medium & high & medium\\
 \hline
\end{tabular}
\caption{Comparing Model Compression and Knowledge Distillation (KD) Methods: 
KD preserves essential weights to ensure high model accuracy, which involves model training (high computation). However, it can be applied to models of any size (high practicality).
Quantization reduces the precision of all weights.
While it is generally task-agnostic, model accuracy can degrade (medium accuracy and practicality).
The process quantizes the representation and adapts the optimization method, which is lightweight (low computation).
The effectiveness of weight pruning depends on the distribution of weights and the specific task (medium practicality, high accuracy, and medium weight size).
Heuristic-based pruning method also has low computation complexity.
}
\label{Comparing-Model-Compression-Methods}
\end{table*}

\subsection{Cost(\$)}\label{sec:cot-sol}
Deep learning tasks require substantial resources due to their high computation and memory demands. 
For example, the recent deep transformer models encounter memory bottlenecks when loading and saving attention layers~\cite{memory-wall}.
Previous work discusses splitting model states~\cite{zero}, kernel fusion techniques~\cite{flashattention}, and sparse attention mechanisms~\cite{ye2025flashinfer} to reduce GPU memory demands and transmission between CPU and GPU memory.

Previous works emphasize low-level DL task scheduling.
Instead, this survey focuses on the higher level aspects of resource provisioning.
Organizations developing intelligent applications using an MLaaS system often face budget constraints when provisioning resources from cloud providers.
For instance, the daily running cost for ChatGPT can reach up to $\$700,000$~\cite{ChatGPT-Cost}. 
Furthermore, cloud services have different cost models that factor in billing granularity, scaling speed, availabilities, etc.
A cost-efficient MLaaS system should strategically choose, configure, and load balance the cloud resources to optimize expenses while meeting performance demands.

\subsubsection{Methods}
To bridge this gap and promote ML systems with low monetary cost of cloud resource usage, various organizations provide services to construct intelligent applications on diverse infrastructures, with an emphasis on minimizing costs.
For example, Redhat OpenShift AI~\cite{Openshift-AI-press} provides a container-based Machine Learning as a Service for on-demand model serving.

Previous research has examined the dynamic scheduling of neural network and model partitions across generic edge and cloud resources~\cite{mono-service-mec-cost, service-chain-mec-cost, model-freshness-deploy, CLIO, SPINN}.
Later work also considers specific cost models according to the cloud services, such as Infrastructure-as-a-Service (IaaS) and Function-as-a-Service (FaaS)~\cite{LIBRA, DNN-partition-SLS}, to balance processing and transmission demands while minimizing expenses.
However, detailed cost analyses using real-world cloud resources for low-cost (\$) ML serving remain limited. 
Many studies model resource expenses on the edge and in the cloud using generic unit costs~\cite{mono-service-mec-cost, service-chain-mec-cost, model-freshness-deploy}. 
However, the specific provisioning factors for each edge and cloud service, including container cold starts and billing time granularity, are critical to minimizing real-world cloud usage costs.

\noindent \textbf{\rb{Summary of cost-saving methods:}}
\rb{Based on our cost formulation (Sec.~\ref{sec:cost-formulation}), the accumulated running time of heterogeneous resources introduces heterogeneous resource-usage costs. 
From the resource-orchestration perspective, the various pricing components of provisioning services, including cold-start time ($T_{cold}^{cutid}$), transmission time ($T_{trans}^{cutid}$), and processing time ($\frac{FLOPs_{cutid+1}^{M}(x_{cutid})}{\theta_{F}}$ and $\frac{FLOPs_{1}^{cutid}(x_{0})}{\theta_{I}}$), significantly affect provisioning decisions. 
We summarize related work in Table~\ref{tab:comparing-cost-analysis}. 
Overall, coarse-grained cost analyses that abstract away application- or model-specific details are more generalizable. 
While fine-grained methods can better approach cost-optimal solutions, they often rely on specialized model features, such as internal classifiers.}

\begin{table*}[t]
\centering
\begin{tabular}{|| l | c | c ||} 
 \hline
 \textbf{Analysis Granularity} & \textbf{Coarse}~\cite{mono-service-mec-cost, service-chain-mec-cost, model-freshness-deploy, CLIO, SPINN} & \textbf{Fine-Grained}~\cite{DNN-partition-SLS, LIBRA}\\
 \hline
 Cost Saving Optimality & Low & High \\
 \hline
 Model Structure Flexibility & High & Low \\
 \hline
\end{tabular}
\caption{
\rb{A comparison of cost analysis methodologies for ML systems.
\textit{Coarse-Grained Analysis} treats the ML model as a monolith. This approach offers high \textit{Model Structure Flexibility} but overlooks potential savings by not optimizing resources for individual components, leading to lower \textit{Cost Saving Optimality}. Conversely, \textit{Fine-Grained Analysis} models the cost of individual layers or partitions. This allows for highly optimized resource allocation (e.g., VMs vs. FaaS for different parts of a model), but is less flexible and requires a more detailed system model.}
}
\label{tab:comparing-cost-analysis}
\end{table*}

\subsection{Privacy}\label{sec:pri-sol}
Distributed DL systems processing sensitive personal data raise data leakage concerns.
Private data should be inaccessible outside the customer's infrastructure or protected from reconstruction during transmission over wide area networks (WAN).
As described in Section~\ref{sec:problem-privacy}, an adversary could reconstruct the intermediate data transmitted between DNN partitions~\cite{NoPeek} using an Auto-Encoder Neural Network. 
To protect against this vulnerability for model inference, previous research adds \textit{Perturbation} to intermediate data~\cite{depth-dropout-privacy, Shredder-noise-tensor} or incorporates a \textit{Regularization} step during training~\cite{NoPeek, ResSFL, depth-dropout-privacy, FSL-journal}. 
Such methods preserve only essential features for ML tasks and remove sensitive information.

\subsubsection{Perturbation}
\noindent {\bf Background.}
\rb{Differential privacy (DP) has been used to improve privacy in statistical databases by adding noise to query outputs proportional to the \textit{sensitivity} of the query~\cite{dwork2014dp, FedMEC-conv-cli-dense-edge-dp}. 
Consider a query \(f: D \rightarrow R\) on a dataset \(D\) with samples \(x, x^{'} \in D\). 
The global sensitivity \(\Delta f\) of this query is defined as:
\begin{align}
    \Delta f = \max_{x,x^{'}} \|f(x)-f(x^{'})\|
\end{align}
Users can set the \textit{Privacy Budget} \(\epsilon\).
Then, noise can be drawn from a Laplace distribution, \(X \sim Laplace(\frac{\Delta f}{\epsilon})\), to achieve the desired level of privacy based on various privacy definitions. 
More specifically, the probability density function (PDF) is $p(x) = \frac{1}{2b}e^{\frac{-\|x\|}{b}}$, where $b=\frac{\Delta f}{\epsilon}$.
The value of \(\epsilon\) can be determined by a grid search against the attack model.
With a smaller \(\epsilon\), we spread the PDF and introduce more diverse noise to the output, so less information is preserved.}

\rb{In practice, finding the global sensitivity $\Delta f$ is challenging as it requires testing all inputs.
Instead, previous work bounds the sensitivity in model partition output~\cite{dp-sgd, FedMEC-conv-cli-dense-edge-dp}.
\begin{align}
    x_{pid}' = \frac{x_{pid}}{max(1,\frac{\|x_{pid}\|}{C})} 
\end{align}
where $C$ is the clipping threshold.
In this way, $\|x_{pid}'\| < C$.
Notice that clipping modifies the hidden variables which leads to accuracy degradation.
To optimize $C$, the common practice is to set the median of $x_{pid}$ based on the training dataset~\cite{dp-sgd}.}

\rb{Previous studies apply this practical DP implementation in DP-SGD~\cite{dp-sgd, opacus} during training to mitigate the risk of reconstructing training datasets from the served models.
They add Gaussian noise to gradients, reducing the model's sensitivity to individual training samples. 
As a result, the distribution of prediction confidences for training dataset samples is similar to other samples, preventing over-concentration on the true label.
This approach complicates membership inference attacks, in which adversaries deduce whether a sample was part of the training data based on prediction logits~\cite{model-inversion-attack-0}.}

\rb{In edge inference settings, recent work suggests injecting noise to hidden variables that obscure sensitive information, for example, race, age, or gender, transmitted over the Internet~\cite{depth-dropout-privacy, Shredder-noise-tensor}. }

\noindent {\bf Methods.}
\rb{In our privacy optimization formulation (Sec.~\ref{sec:prob-form-privacy}), in line~\ref{pri_output_def}, an MLaaS system can inject noise to intermediate data (\(\tau(\Delta f)\)) based on its sensitivity \(\Delta f\). 
With differential privacy, recent studies~\cite{FedMEC-conv-cli-dense-edge-dp, client-dp, Shredder-noise-tensor, depth-dropout-privacy} have developed fitted noise layers that either sample noise from a distribution or nullify specific entries. 
This approach is highly flexible, allowing users to choose different noise layers to append to the final layer on the edge device when the source data distribution changes. 
The noise injected during training and inference complicates the inversion approximation used by the attacker at model serving time. 
Meanwhile, the model retains its capacity to extract relevant information for accurate predictions.}

\subsubsection{Regularization}
\noindent {\bf Background.}
We can also solve the privacy of the source data as an optimization problem. 
One approach is to incorporate source data privacy as a secondary objective by adding a regularization term to the loss function. 
Thus, we encourage the model to preserve only the features that contribute to prediction.
On the other hand, deep edge neural networks (NNs) with non-invertible hidden variables, such as rectangular matrices, are harder to approximate with an inversion matrix.
Therefore, we can optimize the placement of NN partitions to maximize the privacy level of the source data.

\noindent {\bf Methods.}
In our privacy optimization formulation (Sec.~\ref{sec:prob-form-privacy}), we incorporate the privacy loss, exemplified as \(MSE(F_{pid}^{-1}(x_{pid}), x_{pid-1})\) into the loss function in line~\ref{pri_optimization_goal}.
The mean square error gauges differences between the reconstructed and original source data.
Then, we can tune the privacy level of model inference by specifying hyperparameters \(w_{CE}\) and \(w_{p}\) for training~\cite{NoPeek, FSL-journal, ResSFL} and model partition placements~\cite{depth-dropout-privacy, FSL-journal}.

There are ways to incorporate privacy objectives into model training.
For example, we can include a distance correlation loss function, comparing intermediate data and source data in addition to the Cross-Entropy loss~\cite{NoPeek}.
Alternatively, additional training epochs can be dedicated to optimizing the privacy objective~\cite{FSL-journal}.
For more task-specific solutions, ResSFL~\cite{ResSFL} introduces a privacy loss function that compares the source data and the reconstructed data derived from intermediate data using a \textit{decoder} following the threat model in model inversion attacks.
Thus, by designing decoders with different capacity, the model can defend against different model inversion attacks.

\noindent \textbf{\rb{Summary of privacy-preserving methods:}}
\rb{As our privacy formulation (Sec.~\ref{sec:privacy-formulation}) shows, to mitigate model-inversion attacks (MIA, or prompt-inversion attacks, PIA), prior work aims to limit the information encoded in shallow layers so that an adversary cannot reconstruct the source data. 
Related methods are typically applied during training, and also introduce adaptations at inference time. 
We summarize these works in Table~\ref{tab:comparing-privacy-methods}. 
Overall, regularization approaches typically require end-to-end model training.
They tend to better preserve accuracy but incur greater training overhead. 
By contrast, perturbation-based methods generally introduce less overhead because they involve fewer trainable parameters or can be applied post hoc. 
However, they may offer weaker guarantees in accuracy or privacy.
Importantly, the efficacy and practicality of these methods for large language models remain insufficiently studied, and adapting them to the scale and tokenized representations of modern LLMs is an open challenge.}

\begin{table*}[t]
\centering
\begin{tabular}{|| l | c | c ||} 
 \hline
 \textbf{Defense Method} & \textbf{Perturbation}~\cite{FedMEC-conv-cli-dense-edge-dp, client-dp, Shredder-noise-tensor, depth-dropout-privacy} & \textbf{Regularization}~\cite{NoPeek, FSL-journal, ResSFL, depth-dropout-privacy}\\
 \hline
 Training Overhead & Medium & High \\
 \hline
 Required Model Retraining & Partial or Complete & Complete \\
 \hline
 Influence on Accuracy & High & Medium \\
 \hline
\end{tabular}
\caption{
\rb{A comparison of two common defenses against model-inversion attacks. 
\emph{Perturbation} methods add noise to intermediate data, while \emph{Regularization} methods add a privacy-penalty term to the training loss. 
\emph{Perturbation} often has lower \emph{training overhead} because noise-generating components can sometimes be trained separately~\cite{Shredder-noise-tensor}. 
However, this training paradigm can lead to a significant accuracy drop (e.g., differential privacy vs. CPA-DC in Federated Split Learning~\cite{FSL-journal}). 
In contrast, \emph{Regularization} requires end-to-end retraining, which incurs higher overhead but typically affects accuracy less. 
Although these techniques have been studied for traditional MIA, their substantial training cost makes them largely impractical for defending against \emph{prompt-inversion attacks} (PIA) in large-scale LLMs.}
}
\label{tab:comparing-privacy-methods}
\end{table*}

\section{\rb{Multi-Objective Optimization Case Studies}}
\label{sec:coupling}
\rb{Optimizing for latency ($\mathcal{L}^{L}$), monetary cost ($\mathcal{L}^{C}$), and data privacy ($\mathcal{L}^{P}$) simultaneously presents a significant challenge, as these objectives often conflict.
An improvement in one area can adversely affect another.
For instance, prior work on CIFAR-10 has shown that reducing the edge model's depth from $7$ to $4$ layers -- a change that could decrease latency -- degrades privacy, causing the attacker's top-1 misclassification rate on reconstructed images to fall from approximately $80\%$ to below $40\%$~\cite{FSL-journal}.
Similarly, techniques designed to accelerate inference, such as internal classifiers ($\beta_{pid}$) or latent compression ($\gamma_{pid}$), can lead to unpredictable resource usage.
This unpredictability may result in the under-utilization of provisioned virtual machines, thereby increasing the effective monetary cost per request.}

\rb{However, these same mechanisms can also be used to co-optimize multiple objectives when applied in the right context.
For example, early exits reduce traffic to deeper model layers, allowing a small set of always-on VMs to handle the ``hot path'' (frequent, low-latency requests) while offloading less frequent ``cold'' traffic to serverless functions (FaaS).
This approach can simultaneously reduce both latency and cost.
In this section, we therefore explore the tradeoffs inherent in our formulation through a series of use cases, analyzing scenarios where a subset of objectives is relaxed to maximize performance on others.}

\subsection{\rb{Latency \& Cost}}\label{sec:coupling-lat-cost}
\rb{In this section, we analyze scenarios where data privacy is a relaxed constraint, allowing us to focus on the trade-off between inference latency and monetary cost.
This situation is common for large models like LLM chatbots (e.g., ChatGPT~\cite{OpenAIChatGPT} and Gemini~\cite{geminiteam2023gemini}), which are too computationally intensive to be deployed on user devices and must run on remote servers.
Accordingly, we now examine how our formulation can be used to balance these two objectives.}

\rb{Our optimization of latency (${\cal L}^L$) is guided by the objective in line~\ref{optimization_goal}, which minimizes total inference time:
\begin{align*}
    \begin{split}
        {\cal L}^L &= \min_{pid, \alpha_{pid}^{c}, \kappa_{pid}, \gamma_{pid}}       (\xi_{0}^{T}T_{0}^{T}+\sum_{pid=1}^{M}T_{pid}) \\
    \end{split}
\end{align*}
This total time is composed of transmission delay ($T_{pid}^{T}$), defined in line~\ref{trans_time}, and processing delay ($T_{pid}^{C}$), defined in line~\ref{comp_time_i1}:
\begin{align*}
    T_{pid}^{T} &= \frac{(1-\kappa_{pid})(1-\beta_{pid})(1-\gamma_{pid})\text{Size}(F_{pid}(x_{pid-1}))}{bandwidth} \\
    T_{pid}^{C} &= \frac{FLOPs_{pid}^{pid}(x_{pid-1},\pi_{pid})}{\mu_{pid}}
\end{align*}
Similarly, our optimization of monetary cost (${\cal L}^C$) is guided by the objective in line~\ref{cost_optimization_goal}. 
This function models the combined cost of IaaS resources up to a partition point, \texttt{cutid}, and FaaS resources thereafter:
\begin{align*}
    {\cal L}^C = \min_{cutid, \theta_{F}, \theta_{I}, \alpha_{pid}^{k}} \;&
    C_{I}(\theta_{I})T_{I}\sum_{pid = 1}^{cutid}\beta_{pid} \notag \\
    &+ C_{F}(\theta_{F})T_{F}\sum_{pid = cutid+1}^{M}\beta_{pid}
\end{align*}
This cost is weighted by the probability that a request will reach a given layer, $\beta_{pid}$, which is derived from the workload distribution as shown in line~\ref{Fload}:
\begin{align*}
    \beta_{pid} &= \sum_{c=1}^{Q_{pid}} \text{Pr}(\alpha_{pid}^{'c} > \alpha_{pid}^{c}) = \sum_{c=1}^{Q_{pid}} \beta_{pid}^{c}
\end{align*}}

\rb{Although latency and cost are coupled, reducing one does not automatically reduce the other.
Techniques that lower latency, such as data compression ($\kappa_{pid}$, $\gamma_{pid}$) and early exits ($\beta_{pid}$), directly influence the total execution time ($T_{I}$, $T_{F}$), which in turn determines monetary cost.
However, this relationship is complex.
For example, while IaaS platforms may offer more cost-efficient hardware than FaaS, dedicating a full VM to a model is not always cheaper.
In a DNN with internal classifiers, many requests may exit early, leading to short, unpredictable inference times~\cite{ShallowDeepNet, SPINN}.
\rb{And in a hybrid LLM setup where easy queries are handled by a local LLM while complex one needs more sophisticated cloud LLM~\cite{LLM-hybrid}.
In particular, in Hybrid LLM, with $1\%$ drop in BART score of responses (i.e., use smaller edge LLM model), they reduced $22\%$ traffic to the cloud LLM (i.e., api calls to GPT-3.5-turbo)~\cite{LLM-hybrid}.}
These short jobs can under-utilize a provisioned IaaS VM, making it less cost-efficient.
In contrast, a FaaS platform, which bills based on precise execution time, is often more cost-effective for these variable, short-running requests.}

\rb{To jointly optimize inference latency and monetary cost, prior research has explored architectural-aware resource provisioning and model offloading.
One prominent approach involves \textbf{hybrid edge-cloud systems}, which route computationally simple queries to local edge devices while offloading complex requests to more powerful, highly-parameterized models (e.g., Large Language Models) in the cloud~\cite{LLM-hybrid, UniLCD-RL}.
Other works leverage \textbf{internal classifiers} and ensemble model architectures, allowing for on demand fine-grained resource provisioning for layers deployed on the edge or in the cloud~\cite{DDNN, SPINN, praxipaas, efficient-moe, efficient-transformer-infer, SplitNet}.}

\rb{When considering resource costs rather than per-call API fees for services or models, the unique pricing models of cloud providers also affect ML resource-provisioning decisions. 
In LIBRA~\cite{LIBRA}, the authors identify a \textit{Cost Indifference Point} showing high steady-rate traffic is best served by reserved VMs and low-rate traffic can be handled by FaaS for cost-efficiency.
They achieved $20\%$ cost reduction by load balancing.
Serverless (FaaS) platforms provide fine-grained resource provisioning with response times measured in milliseconds, making them ideal for dynamic or transient workloads and for minimizing idle resource costs~\cite{Spock}.
In contrast, reserved VMs, although slower to respond to QoS targets, offer a lower cost per request for sustained workloads that fully utilize the VMs~\cite{serverless-in-the-wild}. }


\rb{For Large Language Models (LLMs), where output lengths are highly unpredictable, recent work has focused on the unique characteristics of the auto-regressive decoding phase.
Specifically, because the Key-Value (KV) cache of previous tokens remains static during the generation of a new token, it is possible to migrate the KV cache to a different node with sufficient GPU memory~\cite{Llumnix}.
This technique enables \textbf{low-downtime migration} of ongoing LLM inference jobs, which is critical for maintaining low 99th-percentile latency.
Furthermore, this low-overhead migration method facilitates dynamic resource allocation, allowing for minimal over-provisioning and thereby reducing monetary costs.}

\subsection{\rb{Latency \& Privacy}}\label{sec:coupling-lat-priv}
\rb{In our second scenario, we examine privacy-sensitive applications where minimizing inference latency is critical, while monetary cost is a less stringent constraint.
Such applications are common in environments like smart hospitals~\cite{privacy-iot-survey} or in enterprise use cases for LLM chatbots, where prompts may contain proprietary information such as company names or contact details~\cite{prompt-inversion-attack-0, prompt-inversion-attack-1}. }

\rb{In these contexts, strict data privacy requirements often mandate that source data cannot leave the user's local facility.
Despite this on-premise processing constraint, the machine learning system is still required to deliver high-quality outputs with low inference latency, creating a significant technical challenge.}

\rb{As defined in Sec.~\ref{sec:MIA-threat}, our privacy formulation centers on a \textbf{Model Inversion Attack (MIA)}.
We consider an environment where a DNN is partitioned across a directed acyclic graph, and the threat goal is to infer the source data from the intermediate activations exchanged between partitions.}

\rb{In this threat model, an honest-but-curious adversary trains a \textbf{reconstructor model} ($\mathcal{R}_{\text{MIA}}$) that takes an intermediate activation, $x_{pid}$, as input and attempts to generate a reconstruction of the original source data, $\hat{x}_{0}=\mathcal{R}_{\text{MIA}}(x_{pid})$.
The resulting privacy leakage from any partition \texttt{pid} can be quantified by the Mean Squared Error (MSE) between the original data and its reconstruction:
\[
\text{Leak}_{pid} \;=\; \operatorname*{\mathbb{E}}_{x_{0}\sim\mathcal{D}} \bigl[ \lVert x_{0}-\hat{x}_{0}\rVert_{2}^{2} \bigr]
\]
Accordingly, our privacy-optimization objective, introduced in line~\ref{pri_optimization_goal} of privacy formulation (Sec.~\ref{sec:privacy-formulation}), is formulated as:
\begin{align*}
    \begin{split}
        {\cal L}^P = \min_{\pi_{pid}^{pri}, \Delta, \lambda}(&w_{CE}CE(\hat{y},y)-\sum_{pid=1}^{M}w_pMSE(F_{pid}^{-1}(x_{pid}), x_{pid-1})) 
    \end{split}
\end{align*}
This objective function seeks to strike a balance between two competing goals: maintaining model accuracy, measured by Cross-Entropy (CE), and preserving source data privacy by minimizing the invertibility of intermediate representations, measured by MSE.}

\rb{The privacy-enhancing techniques we consider alter the intermediate data, $x_{pid}$, exchanged between partitions, as shown in line~\ref{pri_output_def}:
\begin{align*}
    x_{pid} = \lambda\pi_{pid}^{pri}(F_{pid})(x_{pid-1})+\tau(\Delta F_{pid}) 
\end{align*}
These alterations, introduced by the privacy-aware training approaches ($\pi_{pid}^{pri}(\cdot)$ and $\tau(\cdot)$), directly impact the transmission and processing delays ($T_{pid}^{T}$ and $T_{pid}^{C}$) in our latency formulation.
The function $\pi_{pid}^{pri}(\cdot)$ is analogous to the $\pi_{pid}(\cdot)$ function used for latency optimization (lines~\ref{lat_output_def} and~\ref{comp_time_i1}), as both represent paradigms including model compression, data compression, and partition offloading.}

\rb{For traditional DNNs used in tasks like image classification (e.g., VGG), the overhead from these privacy measures can be minimal. 
Techniques such as regularization~\cite{ResSFL,depth-dropout-privacy, FSL-journal} or data perturbation~\cite{depth-dropout-privacy,Shredder-noise-tensor} often introduce few architectural changes during inference.
\rb{For example, ResSFL~\cite{ResSFL} introduce an inversion network of roughly $76.2,$M FLOPs together with a client model of $21.5,$M FLOPs during model training, while maintaining an MSE of $0.02$ between source and reconstructed data during inference.}
Furthermore, any resulting increase in processing or transmission overhead can frequently be compensated for by applying model compression techniques~\cite{Deep-Compressive-Offloading,DRE+PSI+MvOT}.}

\rb{In the context of modern Large Language Models (LLMs), the threat of model inversion has evolved into the \textbf{Prompt Inversion Attack (PIA)}.
The motivation for such attacks is strong: user prompts can contain sensitive personal information, while system prompts often represent valuable intellectual property.
To mitigate these risks, some works propose offloading schemes that partition the LLM between the edge and the cloud, preventing the raw prompt from being transmitted over the network while also managing GPU memory constraints.
However, this does not solve the core problem, as an honest-but-curious adversary can still attempt to reconstruct the prompt by capturing the intermediate activations between model partitions~\cite{prompt-inversion-attack-0,prompt-inversion-attack-1}.}

\rb{At first glance, inverting LLMs appears more difficult than inverting traditional CNNs like VGG.
LLMs are significantly deeper and more non-linear, and their inputs -- discrete token embeddings -- are theoretically harder to reconstruct than continuous image data.
Despite these challenges, recent research has demonstrated successful prompt reconstruction by leveraging open-source base model weights and LoRA adapters~\cite{prompt-inversion-attack-lora}.
Furthermore, traditional privacy-preserving techniques like regularization and data perturbation are generally impractical for models of this scale, as they introduce prohibitive training overhead and can cause significant performance degradation.
Therefore, developing effective and efficient methods to guarantee the privacy of prompt inputs remains a critical and open research challenge.}

\subsection{\rb{Latency, Cost \& Privacy}}\label{sec:coupling-lat-priv-cost}
\rb{In our third and final scenario, we address the most comprehensive challenge: the simultaneous optimization of ML inference latency, source data privacy, and monetary cost.
This setting is crucial for deploying practical, privacy-sensitive applications that must also operate under tight latency and budget constraints.
While our previous sections analyzed pairwise trade-offs, the economic realities of modern large-scale models necessitate a holistic approach.
The exponential growth in inference expenses has made per-request cost a critical metric~\cite{deepseek-r1, CachedAttention}, meaning a truly viable solution must co-optimize all three objectives.}

\rb{Achieving this is exceptionally challenging, as improvements in one area often create complex, competing effects on the others.
Consider the impact of a single architectural decision designed to enhance privacy against a Model Inversion Attack (MIA).
A straightforward strategy is to increase the number of layers processed locally on a user's device (the head partition) before offloading to a remote server.
This strengthens privacy by applying more non-linear transformations but introduces a multifaceted trade-off:
\begin{itemize}
    \item \textbf{Transmission Latency ($T^{T}_{pid}$)} may decrease, as the resulting intermediate data sent to the server is often smaller.
    \item \textbf{Processing Latency ($T^{C}_{pid}$)} on the user's device will increase due to the heavier computational load.
    \item \textbf{Monetary Cost} may be reduced, as less computation is required from paid server resources.
\end{itemize}
This single example illustrates the difficulty finding a global optimum across latency, privacy, and cost.}

\rb{Furthermore, the choice of cloud service for each model partition is a critical factor due to differing pricing models (e.g., AWS EC2~\cite{amazon_ec2} vs. AWS Lambda~\cite{AWS-lambda}). 
As established in our cost formulation (Sec.~\ref{sec:cost-formulation}), we can model a hybrid system that uses a cost-efficient EC2 instance for the initial, privacy-sensitive partitions and a more expensive, serverless Lambda function for the deeper layers.}

\rb{In this model, architectural decisions directly translate to monetary cost. 
For instance, one can shift more layers of computation onto the EC2 instance (as long as the VM is fully utilized). 
While this increases the runtime on the cheaper VM, it proportionally reduces the execution time billed for the more expensive Lambda function. 
Provided that end-to-end latency and data privacy requirements are met, this strategic shift of computation from a high-cost to a low-cost resource can significantly reduce the total cost per inference. 
This demonstrates that resource selection and model partitioning must be carefully co-optimized to achieve a true balance among latency, privacy, and cost.}

\rb{Despite progress in optimizing individual objectives, developing a unified framework to systematically balance inference latency, monetary cost, and privacy against Model Inversion Attacks (MIA) remains a significant challenge. 
The current landscape reveals critical gaps in research.
For instance, much of the existing work adopts a narrow definition of privacy, focusing on ensuring the physical locality of source data (\rb{e.g.,Multi-tier Multi-node Scheduling of LLM, etc}~\cite{model-freshness-deploy, MMSL-LLM}) while overlooking the more subtle threat of information leakage from intermediate activations during inference (i.e., MIA and Prompt Inversion Attacks).}

\rb{Furthermore, while the danger of Prompt Inversion Attacks (PIA) in Large Language Models is increasingly recognized, the practicality and efficiency of potential countermeasures are still largely unexplored.
The remedies suggested by our privacy formulation, especially for LLMs, demand rigorous investigation before they can be considered viable.
Drawing from these gaps in the literature and the challenges highlighted by our framework, the following section outlines several key open research issues.}

\section{\rb{Open Issues}}\label{sec:issues}
Drawing from our multi-objective optimization framework, this section summarizes the inherent challenges in coupling latency, cost, and privacy, motivating future research directions.

\subsection{Monetary Cost and Latency Optimization via Fine-Grained Resource Orchestration}\label{sec:issue:practical-cost-reduction}
Optimizing the monetary cost of hybrid edge-cloud ML systems introduces significant challenges in resource provisioning.
Previous work, as discussed in Sec.~\ref{sec:coupling-lat-cost}, has primarily focused on load balancing at the \textbf{granularity of entire requests}.
For instance, LIBRA~\cite{LIBRA} routes traffic to reserved VMs or serverless platforms based on traffic patterns by identifying a Cost Indifference Point, while Hybrid LLM~\cite{LLM-hybrid} routes queries to different-sized models based on prompt difficulty.

We propose, however, that a more fine-grained approach, orchestrating resources at the level of \textbf{individual neural network partitions}, allows substantial cost savings, particularly for models featuring internal classifiers.
When a DNN is partitioned, it forms a Directed Acyclic Graph (DAG) where each partition processes features for subsequent layers.
The presence of early exits means that many requests may terminate at shallow partitions, leading to a significantly reduced and more variable workload for deeper ones.

Consequently, provisioning a single, powerful Virtual Machine (VM) for the entire inference path is inefficient.
The resource remains idle for short-running requests but is billed as if fully utilized.
A more cost-effective strategy would be to employ high-granularity, pay-per-use resources like FaaS for the deeper partitions.
This allows the system to scale resources dynamically to match the fluctuating workload, ensuring that the monetary cost accurately reflects the actual computation performed.

\subsection{Defending against Prompt Inversion Attacks under Latency and Cost Constraints}\label{sec:issue:practical-privacy-preservation}
\rb{While our privacy formulation focuses on the general principles of Model Inversion Attacks (MIA), its application to Large Language Models (LLMs) is known as the \textbf{Prompt Inversion Attack (PIA)}.
As LLM-based chatbots become popular, their prompts, often containing sensitive user data or proprietary business logic, have become valuable targets~\cite{prompt-inversion-attack-0, prompt-inversion-attack-1}.
However, traditional MIA remedies like regularization and perturbation are generally impractical for models of this scale, creating a substantial open attack surface for distributed LLM applications.}

\rb{A difficulty in defending against PIA lies in the \textbf{unpredictable nature of privacy-preserving hyperparameter tuning}.
As shown in our privacy formulation (Sec.~\ref{sec:prob-form-privacy}), the final balance between model accuracy ($CE(\hat{y},y)$) and privacy (invertibility of $x_{pid-1}$) for any given set of hyperparameters $\{\pi_{pid}^{pri}, \Delta, \lambda\}$ can only be evaluated after a full model training cycle converges.}

\rb{This leads to two critical problems. First, the tuning process is \textbf{prohibitively time-consuming}, requiring numerous, resource-intensive training runs.
Second, the tuning phase itself creates a \textbf{window of vulnerability}.
In a distributed setting, the iterative exchange of gradients and intermediate activations during tuning exposes data \textit{before} an effective privacy configuration has been found.
While heuristics may offer some guidance (e.g., higher compression often degrades inversion accuracy), systematically and safely identifying optimal privacy hyperparameters for LLMs remains an unsolved research problem.}

\rb{Some related work has attempted to mitigate these risks in traditional DNNs by using transfer learning~\cite{ResSFL}.
The strategy is to pre-train a privacy-aware model on a public dataset and then fine-tune it on the private data, hoping to establish a secure baseline.
However, this approach often fails when there is a significant domain mismatch between the public and private tasks (e.g., pre-training on CIFAR-10 for a CIFAR-100 task).
In such cases, extensive fine-tuning on the private data is required to achieve acceptable accuracy, which reintroduces the original risks of high computational cost and data leakage during the prolonged tuning phase.
Consequently, an efficient and secure methodology for tuning privacy-aware LLMs is critical.}

\section{Conclusion}
\label{sec:conc}

This survey has contextualized state-of-the-art model offloading and adaptation methods as critical "control knobs" within a multi-objective optimization framework that balances latency, cost, and privacy.
We traced the evolution from traditional full-model offloading to the sophisticated, partitioned neural network architectures required by demanding applications like LLM chatbots.
The increasing prevalence of these applications highlights the significant potential for future edge-cloud collaborative Machine-Learning-as-a-Service (MLaaS) systems, particularly for organizations leveraging managed cloud infrastructure~\cite{workday-hybridcloud-infra, adobe-hybridcloud-infra-0, adobe-hybridcloud-infra-1}.

By analyzing common challenges such as transmission delays, processing overhead, and privacy vulnerabilities, this work provides a structured perspective on developing next-generation MLaaS platforms.
The open issues identified, especially in minimizing monetary cost and strengthening guarantees against attacks like MIA and PIA, represent ground for future research.
We believe this survey serves as a starting point to guide the future advancements in the field of collaborative edge-cloud intelligence.

\bibliographystyle{plainurl}
\bibliography{references}

\begin{thebibliography}{100}

\bibitem{dp-sgd}
Martin Abadi, Andy Chu, Ian Goodfellow, H.~Brendan McMahan, Ilya Mironov, Kunal Talwar, and Li~Zhang.
\newblock {Deep Learning with Differential Privacy}.
\newblock In {\em Proceedings of the 2016 ACM SIGSAC Conference on Computer and Communications Security}, CCS '16, page 308–318, New York, NY, USA, 2016. Association for Computing Machinery.
\newblock \href {https://doi.org/10.1145/2976749.2978318} {\path{https://doi.org/10.1145/2976749.2978318}}.

\bibitem{adobe-hybridcloud-infra-0}
Adobe.
\newblock {Commerce Cloud Infrastrcture Overview}, 2023.
\newblock URL: \url{https://experienceleague.adobe.com/en/docs/commerce-operations/implementation-playbook/infrastructure/cloud/overview}.

\bibitem{relu}
Abien~Fred Agarap.
\newblock {Deep Learning using Rectified Linear Units (ReLU)}, 2019.
\newblock \href {http://arxiv.org/abs/1803.08375} {\path{arXiv:1803.08375}}.

\bibitem{survey-prompt-leakage}
Divyansh Agarwal, Alexander Fabbri, Ben Risher, Philippe Laban, Shafiq Joty, and Chien-Sheng Wu.
\newblock Prompt leakage effect and mitigation strategies for multi-turn {LLM} applications.
\newblock In Franck Dernoncourt, Daniel Preo{\c{t}}iuc-Pietro, and Anastasia Shimorina, editors, {\em Proceedings of the 2024 Conference on Empirical Methods in Natural Language Processing: Industry Track}, pages 1255--1275, Miami, Florida, US, November 2024. Association for Computational Linguistics.
\newblock URL: \url{https://aclanthology.org/2024.emnlp-industry.94/}, \href {https://doi.org/10.18653/v1/2024.emnlp-industry.94} {\path{https://doi.org/10.18653/v1/2024.emnlp-industry.94}}.

\bibitem{CarMap}
Fawad Ahmad, Hang Qiu, Ray Eells, Fan Bai, and Ramesh Govindan.
\newblock {CarMap: Fast 3D Feature Map Updates for Automobiles}.
\newblock In {\em Proceedings of the 17th Usenix Conference on Networked Systems Design and Implementation}, NSDI'20, page 1063–1082, USA, 2020. USENIX Association.

\bibitem{adobe-hybridcloud-infra-1}
Amazon.
\newblock {AWS and Adobe}, 2024.
\newblock URL: \url{https://aws.amazon.com/partners/adobe/}.

\bibitem{amazon_ec2}
{Amazon Web Services}.
\newblock {Amazon EC2}.
\newblock \url{https://aws.amazon.com/ec2/}, 2024.
\newblock Accessed: 2025-09-09.

\bibitem{AWS-ecs}
{Amazon Web Services}.
\newblock {Amazon ECS}.
\newblock \url{https://aws.amazon.com/ecs/}, 2024.
\newblock Accessed: 2025-09-09.

\bibitem{AWS-lambda}
{Amazon Web Services}.
\newblock {Amazon Lambda}.
\newblock \url{https://aws.amazon.com/lambda/}, 2024.
\newblock Accessed: 2025-09-09.

\bibitem{AWS-lambda-edge}
{Amazon Web Services}.
\newblock {Amazon Lambda Edge}.
\newblock \url{https://aws.amazon.com/lambda/edge/}, 2024.
\newblock Accessed: 2025-09-09.

\bibitem{AWS-Rekognition}
{Amazon Web Services}.
\newblock {Amazon Rekognition}.
\newblock \url{https://docs.aws.amazon.com/rekognition/index.html}, 2024.
\newblock Accessed: 2025-09-09.

\bibitem{AWS-sagemaker}
{Amazon Web Services}.
\newblock {Amazon SageMaker}.
\newblock \url{https://aws.amazon.com/sagemaker/}, 2024.
\newblock Accessed: 2025-09-09.

\bibitem{AWS-sagemaker-edge}
{Amazon Web Services}.
\newblock {Amazon SageMaker Edge}.
\newblock \url{https://aws.amazon.com/sagemaker/edge/}, 2024.
\newblock Accessed: 2025-09-09.

\bibitem{AWS-sagemaker-neo}
{Amazon Web Services}.
\newblock {Amazon SageMaker NEO}.
\newblock \url{https://aws.amazon.com/sagemaker/neo/}, 2024.
\newblock Accessed: 2025-09-09.

\bibitem{AWS-Greengrass}
{Amazon Web Services}.
\newblock {AWS IoT Greengrass}.
\newblock \url{https://aws.amazon.com/greengrass/}, 2024.
\newblock Accessed: 2025-09-09.

\bibitem{AWS-local-zone}
{Amazon Web Services}.
\newblock {AWS Local Zones}.
\newblock \url{https://aws.amazon.com/about-aws/global-infrastructure/localzones/}, 2024.
\newblock Accessed: 2025-09-09.

\bibitem{AWS-wavelength}
{Amazon Web Services}.
\newblock {AWS Wavelength}.
\newblock \url{https://aws.amazon.com/wavelength/}, 2024.
\newblock Accessed: 2025-09-09.

\bibitem{aws_lambda_coldstart}
{Amazon Web Services}.
\newblock {Lambda Runtime Environment: Cold Start Latency}.
\newblock \url{https://docs.aws.amazon.com/lambda/latest/dg/lambda-runtime-environment.html#cold-start-latency}, 2025.
\newblock Accessed: 2025-05-01.

\bibitem{privacy-iot-survey}
Ons Aouedi, Thai-Hoc Vu, Alessio Sacco, Dinh~C. Nguyen, Kandaraj Piamrat, Guido Marchetto, and Quoc-Viet Pham.
\newblock A survey on intelligent internet of things: Applications, security, privacy, and future directions.
\newblock {\em IEEE Communications Surveys \& Tutorials}, 27(2):1238--1292, 2025.
\newblock \href {https://doi.org/10.1109/COMST.2024.3430368} {\path{https://doi.org/10.1109/COMST.2024.3430368}}.

\bibitem{MARLIN}
Kittipat Apicharttrisorn, Xukan Ran, Jiasi Chen, Srikanth~V. Krishnamurthy, and Amit~K. Roy-Chowdhury.
\newblock {Frugal Following: Power Thrifty Object Detection and Tracking for Mobile Augmented Reality}.
\newblock In {\em Proceedings of the 17th Conference on Embedded Networked Sensor Systems}, SenSys '19, page 96–109, New York, NY, USA, 2019. Association for Computing Machinery.
\newblock \href {https://doi.org/10.1145/3356250.3360044} {\path{https://doi.org/10.1145/3356250.3360044}}.

\bibitem{yolov4}
Alexey Bochkovskiy, Chien-Yao Wang, and Hong-Yuan~Mark Liao.
\newblock {YOLOv4: Optimal Speed and Accuracy of Object Detection}, 2020.
\newblock \href {http://arxiv.org/abs/2004.10934} {\path{arXiv:2004.10934}}.

\bibitem{survey-AIOT}
Zhuoqing Chang, Shubo Liu, Xingxing Xiong, Zhaohui Cai, and Guoqing Tu.
\newblock {A Survey of Recent Advances in Edge-Computing-Powered Artificial Intelligence of Things}.
\newblock {\em IEEE Internet of Things Journal}, 8(18):13849--13875, 2021.
\newblock \href {https://doi.org/10.1109/JIOT.2021.3088875} {\path{https://doi.org/10.1109/JIOT.2021.3088875}}.

\bibitem{speculative-decoding-0}
Charlie Chen, Sebastian Borgeaud, Geoffrey Irving, Jean-Baptiste Lespiau, Laurent Sifre, and John Jumper.
\newblock {Accelerating Large Language Model Decoding with Speculative Sampling}, 2023.
\newblock URL: \url{https://arxiv.org/abs/2302.01318}, \href {http://arxiv.org/abs/2302.01318} {\path{arXiv:2302.01318}}.

\bibitem{MARVEL}
Kaifei Chen, Tong Li, Hyung-Sin Kim, David~E. Culler, and Randy~H. Katz.
\newblock {MARVEL: Enabling Mobile Augmented Reality with Low Energy and Low Latency}.
\newblock In {\em Proceedings of the 16th ACM Conference on Embedded Networked Sensor Systems}, SenSys '18, page 292–304, New York, NY, USA, 2018. Association for Computing Machinery.
\newblock \href {https://doi.org/10.1145/3274783.3274834} {\path{https://doi.org/10.1145/3274783.3274834}}.

\bibitem{chen2025visualinstructiontuningchain}
Yixin Chen, Shuai Zhang, Boran Han, and Bernie Wang.
\newblock {Visual Instruction Tuning with Chain of Region-of-Interest}, 2025.
\newblock URL: \url{https://arxiv.org/abs/2505.06840}, \href {http://arxiv.org/abs/2505.06840} {\path{arXiv:2505.06840}}.

\bibitem{kmnist}
Tarin Clanuwat, Mikel Bober-Irizar, Asanobu Kitamoto, Alex Lamb, Kazuaki Yamamoto, and David Ha.
\newblock Deep learning for classical japanese literature, 2018.
\newblock URL: \url{https://arxiv.org/abs/1812.01718}, \href {http://arxiv.org/abs/cs.CV/1812.01718} {\path{arXiv:cs.CV/1812.01718}}.

\bibitem{flashattention}
Tri Dao, Daniel~Y Fu, Stefano Ermon, Atri Rudra, and Christopher Re.
\newblock {FlashAttention: Fast and Memory-Efficient Exact Attention with IO-Awareness}.
\newblock In Alice~H. Oh, Alekh Agarwal, Danielle Belgrave, and Kyunghyun Cho, editors, {\em Advances in Neural Information Processing Systems}, 2022.
\newblock URL: \url{https://openreview.net/forum?id=H4DqfPSibmx}.

\bibitem{survey-ML-services-provisioning}
Shuiguang Deng, Hailiang Zhao, Binbin Huang, Cheng Zhang, Feiyi Chen, Yinuo Deng, Jianwei Yin, Schahram Dustdar, and Albert~Y Zomaya.
\newblock {Cloud-Native Computing: A Survey From the Perspective of Services}.
\newblock {\em Proceedings of the IEEE}, 2024.

\bibitem{dettmers2023qlora}
Tim Dettmers, Artidoro Pagnoni, Ari Holtzman, and Luke Zettlemoyer.
\newblock {QLoRA: Efficient Finetuning of Quantized LLMs}.
\newblock In {\em Thirty-seventh Conference on Neural Information Processing Systems}, 2023.
\newblock URL: \url{https://openreview.net/forum?id=OUIFPHEgJU}.

\bibitem{LLM-hybrid}
Dujian Ding, Ankur Mallick, Chi Wang, Robert Sim, Subhabrata Mukherjee, Victor R{\"u}hle, Laks V.~S. Lakshmanan, and Ahmed~Hassan Awadallah.
\newblock {Hybrid LLM: Cost-Efficient and Quality-Aware Query Routing}.
\newblock In {\em The Twelfth International Conference on Learning Representations}, 2024.
\newblock URL: \url{https://openreview.net/forum?id=02f3mUtqnM}.

\bibitem{model-pruning-for-mia-privacy}
S.~Ding, L.~Zhang, M.~Pan, and X.~Yuan.
\newblock {PATROL: Privacy-Oriented Pruning for Collaborative Inference Against Model Inversion Attacks}.
\newblock In {\em 2024 IEEE/CVF Winter Conference on Applications of Computer Vision (WACV)}, pages 4704--4713, Los Alamitos, CA, USA, jan 2024. IEEE Computer Society.
\newblock URL: \url{https://doi.ieeecomputersociety.org/10.1109/WACV57701.2024.00465}, \href {https://doi.org/10.1109/WACV57701.2024.00465} {\path{https://doi.org/10.1109/WACV57701.2024.00465}}.

\bibitem{on-device-llm-reuse}
Yucheng Ding, Chaoyue Niu, Fan Wu, Shaojie Tang, Chengfei Lyu, and Guihai Chen.
\newblock {Enhancing On-Device LLM Inference with Historical Cloud-Based LLM Interactions}.
\newblock In {\em Proceedings of the 30th ACM SIGKDD Conference on Knowledge Discovery and Data Mining}, KDD '24, page 597–608, New York, NY, USA, 2024. Association for Computing Machinery.
\newblock URL: \url{https://doi-org.ezproxy.bu.edu/10.1145/3637528.3671679}, \href {https://doi.org/10.1145/3637528.3671679} {\path{https://doi.org/10.1145/3637528.3671679}}.

\bibitem{model-inversion-attack-4}
Alexey Dosovitskiy and Thomas Brox.
\newblock {Inverting Visual Representations with Convolutional Networks}.
\newblock In {\em Proceedings of the IEEE conference on computer vision and pattern recognition}, pages 4829--4837, 2016.

\bibitem{survey-AI-EECS}
Sijing Duan, Dan Wang, Ju~Ren, Feng Lyu, Ye~Zhang, Huaqing Wu, and Xuemin Shen.
\newblock {Distributed Artificial Intelligence Empowered by End-Edge-Cloud Computing: A Survey}.
\newblock {\em IEEE Communications Surveys \& Tutorials}, 25(1):591--624, 2023.
\newblock \href {https://doi.org/10.1109/COMST.2022.3218527} {\path{https://doi.org/10.1109/COMST.2022.3218527}}.

\bibitem{dwork2014dp}
Cynthia Dwork, Aaron Roth, et~al.
\newblock The algorithmic foundations of differential privacy.
\newblock {\em Foundations and Trends{\textregistered} in Theoretical Computer Science}, 9(3--4):211--407, 2014.

\bibitem{combine-DNN-with-EE}
Maryam Ebrahimi, Alexandre da~Silva Veith, Moshe Gabel, and Eyal de~Lara.
\newblock {Combining DNN Partitioning and Early Exit}.
\newblock In {\em Proceedings of the 5th International Workshop on Edge Systems, Analytics and Networking}, EdgeSys '22, page 25–30, New York, NY, USA, 2022. Association for Computing Machinery.
\newblock \href {https://doi.org/10.1145/3517206.3526270} {\path{https://doi.org/10.1145/3517206.3526270}}.

\bibitem{HCI-responsetime-1}
Yasuhiro Endo, Zheng Wang, J.~Bradley Chen, and Margo Seltzer.
\newblock Using latency to evaluate interactive system performance.
\newblock In {\em Proceedings of the Second USENIX Symposium on Operating Systems Design and Implementation}, OSDI '96, page 185–199, New York, NY, USA, 1996. Association for Computing Machinery.
\newblock \href {https://doi.org/10.1145/238721.238775} {\path{https://doi.org/10.1145/238721.238775}}.

\bibitem{erdogan2022unsplit}
Ege Erdo\u{g}an, Alptekin K\"{u}p\c{c}\"{u}, and A.~Erc\"{u}ment \c{C}i\c{c}ek.
\newblock {UnSplit: Data‑Oblivious Model Inversion, Model Stealing, and Label Inference Attacks Against Split Learning}.
\newblock In {\em Proceedings of the 21st Workshop on Privacy in the Electronic Society (WPES '22)}, pages 115--124, Los Angeles, CA, USA, 2022. ACM.
\newblock \href {https://doi.org/10.1145/3559613.3563201} {\path{https://doi.org/10.1145/3559613.3563201}}.

\bibitem{AR-5G-edge-survey-1}
Melike Erol-Kantarci and Sukhmani Sukhmani.
\newblock {Caching and Computing at the Edge for Mobile Augmented Reality and Virtual Reality (AR/VR) in 5G}.
\newblock In Yifeng Zhou and Thomas Kunz, editors, {\em Ad Hoc Networks}, pages 169--177, Cham, 2018. Springer International Publishing.

\bibitem{BottleNet}
Amir~Erfan Eshratifar, Amirhossein Esmaili, and Massoud Pedram.
\newblock {BottleNet: A Deep Learning Architecture for Intelligent Mobile Cloud Computing Services}.
\newblock In {\em 2019 IEEE/ACM International Symposium on Low Power Electronics and Design (ISLPED)}, pages 1--6, 2019.
\newblock \href {https://doi.org/10.1109/ISLPED.2019.8824955} {\path{https://doi.org/10.1109/ISLPED.2019.8824955}}.

\bibitem{frantar2023optq}
Elias Frantar, Saleh Ashkboos, Torsten Hoefler, and Dan Alistarh.
\newblock {OPTQ: Accurate Quantization for Generative Pre-trained Transformers}.
\newblock In {\em The Eleventh International Conference on Learning Representations}, 2023.
\newblock URL: \url{https://openreview.net/forum?id=tcbBPnfwxS}.

\bibitem{model-inversion-attack-1}
Matt Fredrikson, Somesh Jha, and Thomas Ristenpart.
\newblock {Model Inversion Attacks That Exploit Confidence Information and Basic Countermeasures}.
\newblock In {\em Proceedings of the 22nd ACM SIGSAC Conference on Computer and Communications Security}, CCS '15, page 1322–1333, New York, NY, USA, 2015. Association for Computing Machinery.
\newblock \href {https://doi.org/10.1145/2810103.2813677} {\path{https://doi.org/10.1145/2810103.2813677}}.

\bibitem{model-inversion-attack-0}
Matt Fredrikson, Somesh Jha, and Thomas Ristenpart.
\newblock {Model Inversion Attacks That Exploit Confidence Information and Basic Countermeasures}.
\newblock In {\em Proceedings of the 22nd ACM SIGSAC Conference on Computer and Communications Security}, CCS '15, page 1322–1333, New York, NY, USA, 2015. Association for Computing Machinery.
\newblock \href {https://doi.org/10.1145/2810103.2813677} {\path{https://doi.org/10.1145/2810103.2813677}}.

\bibitem{CachedAttention}
Bin Gao, Zhuomin He, Puru Sharma, Qingxuan Kang, Djordje Jevdjic, Junbo Deng, Xingkun Yang, Zhou Yu, and Pengfei Zuo.
\newblock {Cost-Efficient} large language model serving for multi-turn conversations with {CachedAttention}.
\newblock In {\em 2024 USENIX Annual Technical Conference (USENIX ATC 24)}, pages 111--126, Santa Clara, CA, July 2024. USENIX Association.
\newblock URL: \url{https://www.usenix.org/conference/atc24/presentation/gao-bin-cost}.

\bibitem{gao2023pcat}
Xinben Gao and Lan Zhang.
\newblock {PCAT}: Functionality and data stealing from split learning by pseudo-client attack.
\newblock In {\em Proceedings of the 32nd USENIX Security Symposium (USENIX Security '23)}, pages 5271--5288, Anaheim, CA, August 2023. USENIX Association.
\newblock URL: \url{https://www.usenix.org/conference/usenixsecurity23/presentation/gao}.

\bibitem{geirhos2020shortcut}
Robert Geirhos, J{\"o}rn-Henrik Jacobsen, Claudio Michaelis, Richard Zemel, Wieland Brendel, Matthias Bethge, and Felix~A Wichmann.
\newblock {Shortcut Learning in Deep Neural Networks}.
\newblock {\em Nature Machine Intelligence}, 2(11):665--673, 2020.

\bibitem{memory-wall}
Amir Gholami, Zhewei Yao, Sehoon Kim, Coleman Hooper, Michael~W Mahoney, and Kurt Keutzer.
\newblock {AI and memory wall}.
\newblock {\em IEEE Micro}, 2024.

\bibitem{gcf_instance_lifespan}
{Google Cloud}.
\newblock {Cloud Functions Execution Environment: Instance Lifespan}.
\newblock \url{https://cloud.google.com/functions/docs/concepts/execution-environment#instance-lifespan}, 2025.
\newblock Accessed: 2025-05-01.

\bibitem{google-gen-privacy}
{Google Workspace Admin Help}.
\newblock Generative ai in google workspace privacy hub.
\newblock \url{https://support.google.com/a/answer/15706919?hl=en}, 2024.
\newblock Accessed: 2025‑04‑20.

\bibitem{gou2021knowledge}
Jianping Gou, Baosheng Yu, Stephen~J Maybank, and Dacheng Tao.
\newblock {Knowledge Distillation: A Survey}.
\newblock {\em International Journal of Computer Vision}, 129:1789--1819, 2021.

\bibitem{gross2017hard}
Sam Gross, Marc'Aurelio Ranzato, and Arthur Szlam.
\newblock {Hard Mixtures of Experts for Large Scale Weakly Supervised Vision}.
\newblock In {\em Proceedings of the IEEE Conference on Computer Vision and Pattern Recognition}, pages 6865--6873, 2017.

\bibitem{Spock}
Jashwant~Raj Gunasekaran, Prashanth Thinakaran, Mahmut~Taylan Kandemir, Bhuvan Urgaonkar, George Kesidis, and Chita Das.
\newblock {Spock: Exploiting Serverless Functions for SLO and Cost Aware Resource Procurement in Public Cloud}.
\newblock In {\em 2019 IEEE 12th International Conference on Cloud Computing (CLOUD)}, pages 199--208, 2019.
\newblock \href {https://doi.org/10.1109/CLOUD.2019.00043} {\path{https://doi.org/10.1109/CLOUD.2019.00043}}.

\bibitem{guo2017calibration}
Chuan Guo, Geoff Pleiss, Yu~Sun, and Kilian~Q Weinberger.
\newblock {On Calibration of Modern Neural Networks}.
\newblock In {\em International conference on machine learning}, pages 1321--1330. PMLR, 2017.

\bibitem{deepseek-r1}
Daya Guo, Dejian Yang, Haowei Zhang, Junxiao Song, Ruoyu Zhang, Runxin Xu, Qihao Zhu, Shirong Ma, Peiyi Wang, Xiao Bi, et~al.
\newblock {DeepSeek-R1: Incentivizing Reasoning Capability in LLMs via Reinforcement Learning}.
\newblock {\em arXiv preprint arXiv:2501.12948}, 2025.

\bibitem{flops-calculation}
Alexey Guzey.
\newblock {How to Measure FLOP/s for Neural Networks Empirically?}
\newblock \url{https://www.lesswrong.com/posts/jJApGWG95495pYM7C/how-to-measure-flop-s-for-neural-networks-empirically}, September 2021.
\newblock Accessed: 2025-09-09.

\bibitem{eie}
Song Han, Xingyu Liu, Huizi Mao, Jing Pu, Ardavan Pedram, Mark~A. Horowitz, and William~J. Dally.
\newblock {EIE: Efficient Inference Engine on Compressed Deep Neural Network}.
\newblock In {\em Proceedings of the 43rd International Symposium on Computer Architecture}, ISCA '16, page 243–254. IEEE Press, 2016.
\newblock \href {https://doi.org/10.1109/ISCA.2016.30} {\path{https://doi.org/10.1109/ISCA.2016.30}}.

\bibitem{Dynamic-Neural-Networks}
Yizeng Han, Gao Huang, Shiji Song, Le~Yang, Honghui Wang, and Yulin Wang.
\newblock {Dynamic Neural Networks: A Survey}.
\newblock {\em IEEE Transactions on Pattern Analysis and Machine Intelligence}, 44(11):7436--7456, 2022.
\newblock \href {https://doi.org/10.1109/TPAMI.2021.3117837} {\path{https://doi.org/10.1109/TPAMI.2021.3117837}}.

\bibitem{depth-dropout-privacy}
Zecheng He, Tianwei Zhang, and Ruby~B. Lee.
\newblock {Attacking and Protecting Data Privacy in Edge–Cloud Collaborative Inference Systems}.
\newblock {\em IEEE Internet of Things Journal}, 8(12):9706--9716, 2021.
\newblock \href {https://doi.org/10.1109/JIOT.2020.3022358} {\path{https://doi.org/10.1109/JIOT.2020.3022358}}.

\bibitem{hinton2015distilling}
Geoffrey Hinton, Oriol Vinyals, and Jeff Dean.
\newblock {Distilling the Knowledge in a Neural Network}.
\newblock {\em arXiv preprint arXiv:1503.02531}, 2015.

\bibitem{mobilenetv3}
Andrew Howard, Mark Sandler, Grace Chu, Liang-Chieh Chen, Bo~Chen, Mingxing Tan, Weijun Wang, Yukun Zhu, Ruoming Pang, Vijay Vasudevan, et~al.
\newblock {Searching for MobileNetV3}.
\newblock In {\em Proceedings of the IEEE/CVF international conference on computer vision}, pages 1314--1324, 2019.

\bibitem{efficient-moe}
Haiyang Huang, Newsha Ardalani, Anna Sun, Liu Ke, Shruti Bhosale, Hsien-Hsin~S. Lee, Carole-Jean Wu, and Benjamin Lee.
\newblock {Toward Efficient Inference for Mixture of Experts}.
\newblock In {\em The Thirty-eighth Annual Conference on Neural Information Processing Systems}, 2024.
\newblock URL: \url{https://openreview.net/forum?id=stXtBqyTWX}.

\bibitem{aws-edge-use-case-1}
Jim Huang and Philipp Landgraf.
\newblock {Remote Rendering for Real-time AR Applications at AWS Edge}, 2024.
\newblock URL: \url{https://aws.amazon.com/blogs/industries/remote-rendering-for-real-time-ar-applications-at-aws-edge/}.

\bibitem{CLIO}
Jin Huang, Colin Samplawski, Deepak Ganesan, Benjamin Marlin, and Heesung Kwon.
\newblock {CLIO: Enabling Automatic Compilation of Deep Learning Pipelines across IoT and Cloud}.
\newblock In {\em Proceedings of the 26th Annual International Conference on Mobile Computing and Networking}, MobiCom '20, New York, NY, USA, 2020. Association for Computing Machinery.
\newblock \href {https://doi.org/10.1145/3372224.3419215} {\path{https://doi.org/10.1145/3372224.3419215}}.

\bibitem{XAI-intermediate-data-compression}
Kai Huang and Wei Gao.
\newblock {Real-Time Neural Network Inference on Extremely Weak Devices: Agile Offloading with Explainable AI}.
\newblock In {\em Proceedings of the 28th Annual International Conference on Mobile Computing And Networking}, MobiCom '22, page 200–213, New York, NY, USA, 2022. Association for Computing Machinery.
\newblock \href {https://doi.org/10.1145/3495243.3560551} {\path{https://doi.org/10.1145/3495243.3560551}}.

\bibitem{LP-transmission-Edge-Cloud}
Yutao Huang, Yifei Zhu, Xiaoyi Fan, Xiaoqiang Ma, Fangxin Wang, Jiangchuan Liu, Ziyi Wang, and Yong Cui.
\newblock {Task Scheduling with Optimized Transmission Time in Collaborative Cloud-Edge Learning}.
\newblock In {\em 2018 27th International Conference on Computer Communication and Networks (ICCCN)}, pages 1--9, 2018.
\newblock \href {https://doi.org/10.1109/ICCCN.2018.8487352} {\path{https://doi.org/10.1109/ICCCN.2018.8487352}}.

\bibitem{huang2022making}
Zhenhua Huang, Shunzhi Yang, MengChu Zhou, Zheng Gong, Abdullah Abusorrah, Chen Lin, and Zheng Huang.
\newblock {Making Accurate Object Detection at the Edge: Review and New Approach}.
\newblock {\em Artificial Intelligence Review}, 55(3):2245--2274, 2022.

\bibitem{FaaS-dnn-serve}
Vatche Ishakian, Vinod Muthusamy, and Aleksander Slominski.
\newblock { Serving Deep Learning Models in a Serverless Platform }.
\newblock In {\em 2018 IEEE International Conference on Cloud Engineering (IC2E)}, pages 257--262, Los Alamitos, CA, USA, April 2018. IEEE Computer Society.
\newblock URL: \url{https://doi.ieeecomputersociety.org/10.1109/IC2E.2018.00052}, \href {https://doi.org/10.1109/IC2E.2018.00052} {\path{https://doi.org/10.1109/IC2E.2018.00052}}.

\bibitem{jacob2018quantization}
Benoit Jacob, Skirmantas Kligys, Bo~Chen, Menglong Zhu, Matthew Tang, Andrew Howard, Hartwig Adam, and Dmitry Kalenichenko.
\newblock {Quantization and Training of Neural Networks for Efficient Integer-Arithmetic-Only Inference}.
\newblock In {\em Proceedings of the IEEE conference on computer vision and pattern recognition}, pages 2704--2713, 2018.

\bibitem{Trustee}
Arthur~S. Jacobs, Roman Beltiukov, Walter Willinger, Ronaldo~A. Ferreira, Arpit Gupta, and Lisandro~Z. Granville.
\newblock {AI/ML for Network Security: The Emperor Has No Clothes}.
\newblock In {\em Proceedings of the 2022 ACM SIGSAC Conference on Computer and Communications Security}, CCS '22, page 1537–1551, New York, NY, USA, 2022. Association for Computing Machinery.
\newblock \href {https://doi.org/10.1145/3548606.3560609} {\path{https://doi.org/10.1145/3548606.3560609}}.

\bibitem{continuum}
Matthijs Jansen, Auday Al-Dulaimy, Alessandro~V Papadopoulos, Animesh Trivedi, and Alexandru Iosup.
\newblock {The SPEC-RG Reference Architecture for the Compute Continuum}.
\newblock In {\em 2023 IEEE/ACM 23rd International Symposium on Cluster, Cloud and Internet Computing (CCGrid)}, pages 469--484. IEEE, 2023.

\bibitem{DNN-partition-SLS}
Jananie Jarachanthan, Li~Chen, Fei Xu, and Bo~Li.
\newblock {AMPS-Inf: Automatic Model Partitioning for Serverless Inference with Cost Efficiency}.
\newblock In {\em 50th International Conference on Parallel Processing}, ICPP 2021, New York, NY, USA, 2021. Association for Computing Machinery.
\newblock \href {https://doi.org/10.1145/3472456.3472501} {\path{https://doi.org/10.1145/3472456.3472501}}.

\bibitem{PSL}
Joohyung Jeon and Joongheon Kim.
\newblock {Privacy-Sensitive Parallel Split Learning}.
\newblock In {\em 2020 International Conference on Information Networking (ICOIN)}, pages 7--9, 2020.
\newblock \href {https://doi.org/10.1109/ICOIN48656.2020.9016486} {\path{https://doi.org/10.1109/ICOIN48656.2020.9016486}}.

\bibitem{energy-edge-survey}
Congfeng Jiang, Tiantian Fan, Honghao Gao, Weisong Shi, Liangkai Liu, Christophe Cérin, and Jian Wan.
\newblock {Energy Aware Edge Computing: A Survey}.
\newblock {\em Computer Communications}, 151:556--580, 2020.
\newblock URL: \url{https://www.sciencedirect.com/science/article/pii/S014036641930831X}, \href {https://doi.org/https://doi.org/10.1016/j.comcom.2020.01.004} {\path{https://doi.org/https://doi.org/10.1016/j.comcom.2020.01.004}}.

\bibitem{jiang2024minference}
Huiqiang Jiang, Yucheng Li, Chengruidong Zhang, Qianhui Wu, Xufang Luo, Surin Ahn, Zhenhua Han, Amir~H. Abdi, Dongsheng Li, Chin-Yew Lin, Yuqing Yang, and Lili Qiu.
\newblock {MInference 1.0: Accelerating Pre-filling for Long-Context LLMs via Dynamic Sparse Attention}.
\newblock In {\em The Thirty-eighth Annual Conference on Neural Information Processing Systems}, 2024.
\newblock URL: \url{https://openreview.net/forum?id=fPBACAbqSN}.

\bibitem{jiang-etal-2024-longllmlingua}
Huiqiang Jiang, Qianhui Wu, , Xufang Luo, Dongsheng Li, Chin-Yew Lin, Yuqing Yang, and Lili Qiu.
\newblock {{LongLLMLingua: Accelerating and Enhancing LLMs in Long Context Scenarios via Prompt Compression}}.
\newblock In Lun-Wei Ku, Andre Martins, and Vivek Srikumar, editors, {\em Proceedings of the 62nd Annual Meeting of the Association for Computational Linguistics (Volume 1: Long Papers)}, pages 1658--1677, Bangkok, Thailand, August 2024. Association for Computational Linguistics.
\newblock URL: \url{https://aclanthology.org/2024.acl-long.91}.

\bibitem{jiang-etal-2023-llmlingua}
Huiqiang Jiang, Qianhui Wu, Chin-Yew Lin, Yuqing Yang, and Lili Qiu.
\newblock {{LLMLingua: Compressing Prompts for Accelerated Inference of Large Language Models}}.
\newblock In Houda Bouamor, Juan Pino, and Kalika Bali, editors, {\em Proceedings of the 2023 Conference on Empirical Methods in Natural Language Processing}, pages 13358--13376, Singapore, December 2023. Association for Computational Linguistics.
\newblock URL: \url{https://aclanthology.org/2023.emnlp-main.825}, \href {https://doi.org/10.18653/v1/2023.emnlp-main.825} {\path{https://doi.org/10.18653/v1/2023.emnlp-main.825}}.

\bibitem{faas-vs-iaas-ml-training}
Jiawei Jiang, Shaoduo Gan, Bo~Du, Gustavo Alonso, Ana Klimovic, Ankit Singla, Wentao Wu, Sheng Wang, and Ce~Zhang.
\newblock A systematic evaluation of machine learning on serverless infrastructure.
\newblock {\em The VLDB Journal}, 33(2):425–449, September 2023.
\newblock \href {https://doi.org/10.1007/s00778-023-00813-0} {\path{https://doi.org/10.1007/s00778-023-00813-0}}.

\bibitem{jiang2024megascale}
Ziheng Jiang, Haibin Lin, Yinmin Zhong, Qi~Huang, Yangrui Chen, Zhi Zhang, Yanghua Peng, Xiang Li, Cong Xie, Shibiao Nong, Yulu Jia, Sun He, Hongmin Chen, Zhihao Bai, Qi~Hou, Shipeng Yan, Ding Zhou, Yiyao Sheng, Zhuo Jiang, Haohan Xu, Haoran Wei, Zhang Zhang, Pengfei Nie, Leqi Zou, Sida Zhao, Liang Xiang, Zherui Liu, Zhe Li, Xiaoying Jia, Jianxi Ye, Xin Jin, and Xin Liu.
\newblock {MegaScale: Scaling Large Language Model Training to More Than 10,000 GPUs}, 2024.
\newblock \href {http://arxiv.org/abs/2402.15627} {\path{arXiv:2402.15627}}.

\bibitem{he-1}
Chiraag Juvekar, Vinod Vaikuntanathan, and Anantha Chandrakasan.
\newblock {GAZELLE: A Low Latency Framework for Secure Neural Network Inference}.
\newblock In {\em 27th USENIX Security Symposium (USENIX Security 18)}, pages 1651--1669, Baltimore, MD, August 2018. USENIX Association.
\newblock URL: \url{https://www.usenix.org/conference/usenixsecurity18/presentation/juvekar}.

\bibitem{kafetzis2025largelanguagemodelpartitioning}
Dimitrios Kafetzis, Ramin Khalili, and Iordanis Koutsopoulos.
\newblock Large language model partitioning for low-latency inference at the edge, 2025.
\newblock URL: \url{https://arxiv.org/abs/2505.02533}, \href {http://arxiv.org/abs/2505.02533} {\path{arXiv:2505.02533}}.

\bibitem{Neurosurgeon}
Yiping Kang, Johann Hauswald, Cao Gao, Austin Rovinski, Trevor Mudge, Jason Mars, and Lingjia Tang.
\newblock {Neurosurgeon: Collaborative Intelligence Between the Cloud and Mobile Edge}.
\newblock In {\em Proceedings of the Twenty-Second International Conference on Architectural Support for Programming Languages and Operating Systems}, ASPLOS '17, page 615–629, New York, NY, USA, 2017. Association for Computing Machinery.
\newblock \href {https://doi.org/10.1145/3037697.3037698} {\path{https://doi.org/10.1145/3037697.3037698}}.

\bibitem{ShallowDeepNet}
Yigitcan Kaya, Sanghyun Hong, and Tudor Dumitras.
\newblock {Shallow-Deep Networks: Understanding and Mitigating Network Overthinking}.
\newblock In {\em ICML}, 2019.

\bibitem{SplitNet}
Juyong Kim, Yookoon Park, Gunhee Kim, and Sung~Ju Hwang.
\newblock {SplitNet: Learning to Semantically Split Deep Networks for Parameter Reduction and Model Parallelization}.
\newblock In Doina Precup and Yee~Whye Teh, editors, {\em Proceedings of the 34th International Conference on Machine Learning}, volume~70 of {\em Proceedings of Machine Learning Research}, pages 1866--1874. PMLR, 06--11 Aug 2017.
\newblock URL: \url{https://proceedings.mlr.press/v70/kim17b.html}.

\bibitem{cifar10}
Alex Krizhevsky and Geoffrey Hinton.
\newblock {Learning Multiple Layers of Features from Tiny Images}.
\newblock Technical Report~0, University of Toronto, 2009.
\newblock URL: \url{https://www.cs.toronto.edu/~kriz/learning-features-2009-TR.pdf}.

\bibitem{ILSVRC12}
Alex Krizhevsky, Ilya Sutskever, and Geoffrey~E Hinton.
\newblock {ImageNet Classification with Deep Convolutional Neural Networks}.
\newblock {\em Advances in neural information processing systems}, 25, 2012.

\bibitem{edge-battery}
Karthik Kumar and Yung-Hsiang Lu.
\newblock {Cloud Computing for Mobile Users: Can Offloading Computation Save Energy?}
\newblock {\em Computer}, 43(4):51--56, 2010.
\newblock \href {https://doi.org/10.1109/MC.2010.98} {\path{https://doi.org/10.1109/MC.2010.98}}.

\bibitem{SPINN}
Stefanos Laskaridis, Stylianos~I. Venieris, Mario Almeida, Ilias Leontiadis, and Nicholas~D. Lane.
\newblock {SPINN: Synergistic Progressive Inference of Neural Networks over Device and Cloud}.
\newblock In {\em Proceedings of the 26th Annual International Conference on Mobile Computing and Networking}, MobiCom '20, New York, NY, USA, 2020. Association for Computing Machinery.
\newblock \href {https://doi.org/10.1145/3372224.3419194} {\path{https://doi.org/10.1145/3372224.3419194}}.

\bibitem{le2015tiny}
Ya~Le and Xuan Yang.
\newblock {Tiny ImageNet Visual Recognition Challenge}.
\newblock {\em CS 231N}, 7(7):3, 2015.

\bibitem{mnist}
Y.~Lecun, L.~Bottou, Y.~Bengio, and P.~Haffner.
\newblock Gradient-based learning applied to document recognition.
\newblock {\em Proceedings of the IEEE}, 86(11):2278--2324, 1998.
\newblock \href {https://doi.org/10.1109/5.726791} {\path{https://doi.org/10.1109/5.726791}}.

\bibitem{Bottle-Reconstruct-Coop}
Joo~Chan Lee, Yongwoo Kim, SungTae Moon, and Jong~Hwan Ko.
\newblock {A Splittable DNN-Based Object Detector for Edge-Cloud Collaborative Real-Time Video Inference}.
\newblock In {\em 2021 17th IEEE International Conference on Advanced Video and Signal Based Surveillance (AVSS)}, pages 1--8, 2021.
\newblock \href {https://doi.org/10.1109/AVSS52988.2021.9663806} {\path{https://doi.org/10.1109/AVSS52988.2021.9663806}}.

\bibitem{speculative-decoding-1}
Yaniv Leviathan, Matan Kalman, and Yossi Matias.
\newblock {Fast Inference from Transformers via Speculative Decoding}.
\newblock In Andreas Krause, Emma Brunskill, Kyunghyun Cho, Barbara Engelhardt, Sivan Sabato, and Jonathan Scarlett, editors, {\em Proceedings of the 40th International Conference on Machine Learning}, volume 202 of {\em Proceedings of Machine Learning Research}, pages 19274--19286. PMLR, 23--29 Jul 2023.
\newblock URL: \url{https://proceedings.mlr.press/v202/leviathan23a.html}.

\bibitem{ee-part-opt-form}
Chao Li, Hongli Xu, Yang Xu, Zhiyuan Wang, and Liusheng Huang.
\newblock {DNN Inference Acceleration with Partitioning and Early Exiting in Edge Computing}.
\newblock In {\em Wireless Algorithms, Systems, and Applications: 16th International Conference, WASA 2021, Nanjing, China, June 25–27, 2021, Proceedings, Part I}, page 465–478, Berlin, Heidelberg, 2021. Springer-Verlag.
\newblock \href {https://doi.org/10.1007/978-3-030-85928-2_37} {\path{https://doi.org/10.1007/978-3-030-85928-2_37}}.

\bibitem{li2019edge}
En~Li, Liekang Zeng, Zhi Zhou, and Xu~Chen.
\newblock {Edge AI: On-Demand Accelerating Deep Neural Network Inference via Edge Computing}.
\newblock {\em IEEE Transactions on Wireless Communications}, 19(1):447--457, 2019.

\bibitem{Edgent}
En~Li, Zhi Zhou, and Xu~Chen.
\newblock {Edge Intelligence: On-Demand Deep Learning Model Co-Inference with Device-Edge Synergy}.
\newblock In {\em Proceedings of the 2018 Workshop on Mobile Edge Communications}, MECOMM'18, page 31–36, New York, NY, USA, 2018. Association for Computing Machinery.
\newblock \href {https://doi.org/10.1145/3229556.3229562} {\path{https://doi.org/10.1145/3229556.3229562}}.

\bibitem{Auto-tuning-Quantization-NN}
Guangli Li, Lei Liu, Xueying Wang, Xiao Dong, Peng Zhao, and Xiaobing Feng.
\newblock {Auto-tuning Neural Network Quantization Framework for Collaborative Inference Between the Cloud and Edge}.
\newblock In V{\v{e}}ra K{\r{u}}rkov{\'a}, Yannis Manolopoulos, Barbara Hammer, Lazaros Iliadis, and Ilias Maglogiannis, editors, {\em Artificial Neural Networks and Machine Learning -- ICANN 2018}, pages 402--411, Cham, 2018. Springer International Publishing.

\bibitem{ResSFL}
Jingtao Li, Adnan~Siraj Rakin, Xing Chen, Zhezhi He, Deliang Fan, and Chaitali Chakrabarti.
\newblock {ResSFL: A Resistance Transfer Framework for Defending Model Inversion Attack in Split Federated Learning}.
\newblock In {\em Proceedings of the IEEE/CVF Conference on Computer Vision and Pattern Recognition (CVPR)}, pages 10194--10202, June 2022.

\bibitem{RAF-DB}
Shan Li, Weihong Deng, and JunPing Du.
\newblock {Reliable Crowdsourcing and Deep Locality-Preserving Learning for Expression Recognition in the Wild}.
\newblock In {\em Proceedings of the IEEE conference on computer vision and pattern recognition}, pages 2852--2861, 2017.

\bibitem{li2025scbench}
Yucheng Li, Huiqiang Jiang, Qianhui Wu, Xufang Luo, Surin Ahn, Chengruidong Zhang, Amir~H. Abdi, Dongsheng Li, Jianfeng Gao, Yuqing Yang, and Lili Qiu.
\newblock {SCBench: A {KV} Cache-Centric Analysis of Long-Context Methods}.
\newblock In {\em The Thirteenth International Conference on Learning Representations}, 2025.
\newblock URL: \url{https://openreview.net/forum?id=gkUyYcY1W9}.

\bibitem{li2025mminference}
Yucheng Li, Huiqiang Jiang, Chengruidong Zhang, Qianhui Wu, Xufang Luo, Surin Ahn, Amir~H Abdi, Dongsheng Li, Jianfeng Gao, Yuqing Yang, and Lili Qiu.
\newblock {MMIference: Accelerating Pre-filling for Long-Context VLMs via Modality-Aware Permutation Sparse Attention}.
\newblock In {\em Forty-second International Conference on Machine Learning}, 2025.
\newblock URL: \url{https://openreview.net/forum?id=me6PfbATWM}.

\bibitem{Pruning-and-quantization-survey}
Tailin Liang, John Glossner, Lei Wang, Shaobo Shi, and Xiaotong Zhang.
\newblock {Pruning and Quantization for Deep Neural Network Acceleration: A Survey}.
\newblock {\em Neurocomputing}, 461:370--403, 2021.
\newblock URL: \url{https://www.sciencedirect.com/science/article/pii/S0925231221010894}, \href {https://doi.org/https://doi.org/10.1016/j.neucom.2021.07.045} {\path{https://doi.org/https://doi.org/10.1016/j.neucom.2021.07.045}}.

\bibitem{lin2023awq}
Ji~Lin, Jiaming Tang, Haotian Tang, Shang Yang, Wei-Ming Chen, Wei-Chen Wang, Guangxuan Xiao, Xingyu Dang, Chuang Gan, and Song Han.
\newblock {AWQ: Activation-aware Weight Quantization for LLM Compression and Acceleration}.
\newblock In {\em MLSys}, 2024.

\bibitem{Edge-Systems-Survey-1}
Li~Lin, Xiaofei Liao, Hai Jin, and Peng Li.
\newblock {Computation Offloading Toward Edge Computing}.
\newblock {\em Proceedings of the IEEE}, 107(8):1584--1607, 2019.
\newblock \href {https://doi.org/10.1109/JPROC.2019.2922285} {\path{https://doi.org/10.1109/JPROC.2019.2922285}}.

\bibitem{COCO}
Tsung-Yi Lin, Michael Maire, Serge Belongie, Lubomir Bourdev, Ross Girshick, James Hays, Pietro Perona, Deva Ramanan, C.~Lawrence Zitnick, and Piotr Dollár.
\newblock {Microsoft COCO: Common Objects in Context}, 2015.
\newblock \href {http://arxiv.org/abs/1405.0312} {\path{arXiv:1405.0312}}.

\bibitem{context-aware-sentence-pruning}
Barys Liskavets, Maxim Ushakov, Shuvendu Roy, Mark Klibanov, Ali Etemad, and Shane~K. Luke.
\newblock {Prompt Compression with Context-aware Sentence Encoding for Fast and Improved LLM Inference}.
\newblock In {\em Proceedings of the Thirty-Ninth AAAI Conference on Artificial Intelligence and Thirty-Seventh Conference on Innovative Applications of Artificial Intelligence and Fifteenth Symposium on Educational Advances in Artificial Intelligence}, AAAI'25/IAAI'25/EAAI'25. AAAI Press, 2025.
\newblock \href {https://doi.org/10.1609/aaai.v39i23.34639} {\path{https://doi.org/10.1609/aaai.v39i23.34639}}.

\bibitem{Edge-Systems-Survey}
Fang Liu, Guoming Tang, Youhuizi Li, Zhiping Cai, Xingzhou Zhang, and Tongqing Zhou.
\newblock {A Survey on Edge Computing Systems and Tools}.
\newblock {\em Proceedings of the IEEE}, 107(8):1537--1562, 2019.
\newblock \href {https://doi.org/10.1109/JPROC.2019.2920341} {\path{https://doi.org/10.1109/JPROC.2019.2920341}}.

\bibitem{darts}
Hanxiao Liu, Karen Simonyan, and Yiming Yang.
\newblock {DARTS: Differentiable Architecture Search}.
\newblock In {\em International Conference on Learning Representations}, 2019.
\newblock URL: \url{https://openreview.net/forum?id=S1eYHoC5FX}.

\bibitem{model-freshness-deploy}
Haolin Liu, Sirui Liu, Saiqin Long, Qingyong Deng, and Zhetao Li.
\newblock {Joint Optimization of Model Deployment for Freshness-Sensitive Task Assignment in Edge Intelligence}.
\newblock In {\em IEEE INFOCOM 2024 - IEEE Conference on Computer Communications}, pages 1751--1760, 2024.
\newblock \href {https://doi.org/10.1109/INFOCOM52122.2024.10621314} {\path{https://doi.org/10.1109/INFOCOM52122.2024.10621314}}.

\bibitem{CE-AFL}
Jianchun Liu, Hongli Xu, Yang Xu, Zhenguo Ma, Zhiyuan Wang, Chen Qian, and He~Huang.
\newblock {Communication-Efficient Asynchronous Federated Learning in Resource-Constrained Edge Computing}.
\newblock {\em Comput. Netw.}, 199(C), apr 2022.
\newblock \href {https://doi.org/10.1016/j.comnet.2021.108429} {\path{https://doi.org/10.1016/j.comnet.2021.108429}}.

\bibitem{10.1016/j.comnet.2021.108468}
Juncai Liu, Jessie~Hui Wang, Chenghao Rong, Yuedong Xu, Tao Yu, and Jilong Wang.
\newblock {FedPA: An Adaptively Partial Model Aggregation Strategy in Federated Learning}.
\newblock {\em Comput. Netw.}, 199(C), apr 2022.
\newblock \href {https://doi.org/10.1016/j.comnet.2021.108468} {\path{https://doi.org/10.1016/j.comnet.2021.108468}}.

\bibitem{DRE+PSI+MvOT}
Luyang Liu, Hongyu Li, and Marco Gruteser.
\newblock {Edge Assisted Real-Time Object Detection for Mobile Augmented Reality}.
\newblock In {\em The 25th Annual International Conference on Mobile Computing and Networking}, MobiCom '19, New York, NY, USA, 2019. Association for Computing Machinery.
\newblock \href {https://doi.org/10.1145/3300061.3300116} {\path{https://doi.org/10.1145/3300061.3300116}}.

\bibitem{bi-quan}
Zechun Liu, Baoyuan Wu, Wenhan Luo, Xin Yang, Wei Liu, and Kwang-Ting Cheng.
\newblock {Bi-Real Net: Enhancing the Performance of 1-bit CNNs With Improved Representational Capability and Advanced Training Algorithm}.
\newblock In {\em Proceedings of the European conference on computer vision (ECCV)}, pages 722--737, 2018.

\bibitem{DVFS-0}
Daniel Lo, Taejoon Song, and G.~Edward Suh.
\newblock Prediction-guided performance-energy trade-off for interactive applications.
\newblock In {\em 2015 48th Annual IEEE/ACM International Symposium on Microarchitecture (MICRO)}, pages 508--520, 2015.
\newblock \href {https://doi.org/10.1145/2830772.2830776} {\path{https://doi.org/10.1145/2830772.2830776}}.

\bibitem{prompt-inversion-attack-1}
Xinjian Luo, Ting Yu, and Xiaokui Xiao.
\newblock Prompt inference attack on distributed large language model inference frameworks, 2025.
\newblock URL: \url{https://arxiv.org/abs/2503.09291}, \href {http://arxiv.org/abs/2503.09291} {\path{arXiv:2503.09291}}.

\bibitem{MMSL-LLM}
Mulei Ma, Chenyu Gong, Liekang Zeng, and Yang Yang.
\newblock Multi-tier multi-node scheduling of llm for collaborative ai computing.
\newblock In {\em IEEE INFOCOM 2025 - IEEE Conference on Computer Communications}, pages 1--10, 2025.
\newblock \href {https://doi.org/10.1109/INFOCOM55648.2025.11044698} {\path{https://doi.org/10.1109/INFOCOM55648.2025.11044698}}.

\bibitem{ma2025bitnetb1_58_2b4t}
Shuming Ma, Hongyu Wang, Shaohan Huang, Xingxing Zhang, Ying Hu, Ting Song, Yan Xia, and Furu Wei.
\newblock {BitNet b1.58 2B4T Technical Report}.
\newblock Technical Report arXiv:2504.12285, arXiv, Apr 2025.
\newblock URL: \url{https://arxiv.org/abs/2504.12285}, \href {https://doi.org/10.48550/arXiv.2504.12285} {\path{https://doi.org/10.48550/arXiv.2504.12285}}.

\bibitem{client-dp}
Yunlong Mao, Shanhe Yi, Qun Li, Jinghao Feng, Fengyuan Xu, and Sheng Zhong.
\newblock {Learning from Differentially Private Neural Activations with Edge Computing}.
\newblock In {\em 2018 IEEE/ACM Symposium on Edge Computing (SEC)}, pages 90--102, 2018.
\newblock \href {https://doi.org/10.1109/SEC.2018.00014} {\path{https://doi.org/10.1109/SEC.2018.00014}}.

\bibitem{matsubara2021neural}
Yoshitomo Matsubara and Marco Levorato.
\newblock {Neural Compression and Filtering for Edge-assisted Real-time Object Detection in Challenged Networks}.
\newblock In {\em 2020 25th International Conference on Pattern Recognition (ICPR)}, pages 2272--2279. IEEE, 2021.

\bibitem{survey-EE-SC}
Yoshitomo Matsubara, Marco Levorato, and Francesco Restuccia.
\newblock {Split Computing and Early Exiting for Deep Learning Applications: Survey and Research Challenges}.
\newblock {\em ACM Comput. Surv.}, 55(5), dec 2022.
\newblock \href {https://doi.org/10.1145/3527155} {\path{https://doi.org/10.1145/3527155}}.

\bibitem{FL}
H.~Brendan McMahan, Eider Moore, Daniel Ramage, Seth Hampson, and Blaise~Agüera y~Arcas.
\newblock {Communication-Efficient Learning of Deep Networks from Decentralized Data}.
\newblock {\em arXiv preprint arXiv:1602.05629}, 2023.
\newblock \href {http://arxiv.org/abs/1602.05629} {\path{arXiv:1602.05629}}.

\bibitem{llama-3-1-model-card}
Meta.
\newblock {MODEL\_CARD}, 2024.
\newblock URL: \url{https://github.com/meta-llama/llama-models/blob/main/models/llama3\_1/MODEL\_CARD.md}.

\bibitem{speculative-llm}
Xupeng Miao, Gabriele Oliaro, Zhihao Zhang, Xinhao Cheng, Zeyu Wang, Zhengxin Zhang, Rae Ying~Yee Wong, Alan Zhu, Lijie Yang, Xiaoxiang Shi, Chunan Shi, Zhuoming Chen, Daiyaan Arfeen, Reyna Abhyankar, and Zhihao Jia.
\newblock {SpecInfer: Accelerating Large Language Model Serving with Tree-based Speculative Inference and Verification}.
\newblock In {\em Proceedings of the 29th ACM International Conference on Architectural Support for Programming Languages and Operating Systems, Volume 3}, ASPLOS '24, page 932–949, New York, NY, USA, 2024. Association for Computing Machinery.
\newblock URL: \url{https://doi-org.ezproxy.bu.edu/10.1145/3620666.3651335}, \href {https://doi.org/10.1145/3620666.3651335} {\path{https://doi.org/10.1145/3620666.3651335}}.

\bibitem{Guidance}
{Microsoft}.
\newblock {Guidance: A guidance language for controlling large language models}.
\newblock \url{https://github.com/guidance-ai/guidance}, 2022.
\newblock Accessed: 2025-09-09.

\bibitem{azure_functions_hosting}
{Microsoft Learn}.
\newblock {Azure Functions Overview: Hosting Options}.
\newblock \url{https://learn.microsoft.com/en-us/azure/azure-functions/functions-overview#hosting-options}, 2025.
\newblock Accessed: 2025-05-01.

\bibitem{HCI-responsetime}
Robert~B. Miller.
\newblock {Response Time in Man-Computer Conversational Transactions}.
\newblock In {\em Proceedings of the December 9-11, 1968, Fall Joint Computer Conference, Part I}, AFIPS '68 (Fall, part I), page 267–277, New York, NY, USA, 1968. Association for Computing Machinery.
\newblock \href {https://doi.org/10.1145/1476589.1476628} {\path{https://doi.org/10.1145/1476589.1476628}}.

\bibitem{feat-selection-in-offloading}
Fatemehsadat Mireshghallah, Mohammadkazem Taram, Ali Jalali, Ahmed Taha~Taha Elthakeb, Dean Tullsen, and Hadi Esmaeilzadeh.
\newblock Not all features are equal: Discovering essential features for preserving prediction privacy.
\newblock In {\em Proceedings of the Web Conference 2021}, WWW '21, page 669–680, New York, NY, USA, 2021. Association for Computing Machinery.
\newblock \href {https://doi.org/10.1145/3442381.3449965} {\path{https://doi.org/10.1145/3442381.3449965}}.

\bibitem{Shredder-noise-tensor}
Fatemehsadat Mireshghallah, Mohammadkazem Taram, Prakash Ramrakhyani, Ali Jalali, Dean Tullsen, and Hadi Esmaeilzadeh.
\newblock {Shredder: Learning Noise Distributions to Protect Inference Privacy}.
\newblock In {\em Proceedings of the Twenty-Fifth International Conference on Architectural Support for Programming Languages and Operating Systems}, ASPLOS '20, page 3–18, New York, NY, USA, 2020. Association for Computing Machinery.
\newblock \href {https://doi.org/10.1145/3373376.3378522} {\path{https://doi.org/10.1145/3373376.3378522}}.

\bibitem{smpc-1}
Pratyush Mishra, Ryan Lehmkuhl, Akshayaram Srinivasan, Wenting Zheng, and Raluca~Ada Popa.
\newblock {Delphi: A Cryptographic Inference Service for Neural Networks}.
\newblock In {\em 29th USENIX Security Symposium (USENIX Security 20)}, pages 2505--2522. USENIX Association, August 2020.
\newblock URL: \url{https://www.usenix.org/conference/usenixsecurity20/presentation/mishra}.

\bibitem{mudvari2024splitllmcollaborativeinferencellms}
Akrit Mudvari, Yuang Jiang, and Leandros Tassiulas.
\newblock Splitllm: Collaborative inference of llms for model placement and throughput optimization, 2024.
\newblock URL: \url{https://arxiv.org/abs/2410.10759}, \href {http://arxiv.org/abs/2410.10759} {\path{arXiv:2410.10759}}.

\bibitem{NN-quant-survey}
James~O' Neill.
\newblock {An Overview of Neural Network Compression}.
\newblock {\em arXiv preprint arXiv:2006.03669}, 2020.
\newblock URL: \url{https://arxiv.org/abs/2006.03669}, \href {https://doi.org/10.48550/ARXIV.2006.03669} {\path{https://doi.org/10.48550/ARXIV.2006.03669}}.

\bibitem{aws-edge-use-case-2}
Netflix.
\newblock {Netflix Empowers Remote Artistry with Low-Latency Workstations Using AWS Local Zones}, 2024.
\newblock URL: \url{https://aws.amazon.com/solutions/case-studies/netflix-aws-local-zones-case-study/}.

\bibitem{privacy-edge-survey}
Lucien K.~L. Ng and Sherman S.~M. Chow.
\newblock Sok: Cryptographic neural-network computation.
\newblock In {\em 2023 IEEE Symposium on Security and Privacy (SP)}, pages 497--514, 2023.
\newblock \href {https://doi.org/10.1109/SP46215.2023.10179483} {\path{https://doi.org/10.1109/SP46215.2023.10179483}}.

\bibitem{PieSlicer}
Samuel~S. Ogden, Xiangnan Kong, and Tian Guo.
\newblock {PieSlicer: Dynamically Improving Response Time for Cloud-Based CNN Inference}.
\newblock In {\em Proceedings of the ACM/SPEC International Conference on Performance Engineering}, ICPE '21, page 249–256, New York, NY, USA, 2021. Association for Computing Machinery.
\newblock \href {https://doi.org/10.1145/3427921.3450256} {\path{https://doi.org/10.1145/3427921.3450256}}.

\bibitem{OpenAIChatGPT}
{OpenAI}.
\newblock Chatgpt.
\newblock \url{https://chat.openai.com}, 2024.
\newblock Accessed: 2025-09-09.

\bibitem{openai2024privacypolicy}
{OpenAI OpCo, LLC}.
\newblock Privacy policy.
\newblock \url{https://openai.com/policies/row-privacy-policy/}, 2024.
\newblock Published November 2024; Accessed: 2025‑04‑20.

\bibitem{pan-etal-2024-llmlingua}
Zhuoshi Pan, Qianhui Wu, Huiqiang Jiang, Menglin Xia, Xufang Luo, Jue Zhang, Qingwei Lin, Victor Ruhle, Yuqing Yang, Chin-Yew Lin, H.~Vicky Zhao, Lili Qiu, and Dongmei Zhang.
\newblock {{LLMLingua-2: Data Distillation for Efficient and Faithful Task-Agnostic Prompt Compression}}.
\newblock In Lun-Wei Ku, Andre Martins, and Vivek Srikumar, editors, {\em Findings of the Association for Computational Linguistics ACL 2024}, pages 963--981, Bangkok, Thailand and virtual meeting, August 2024. Association for Computational Linguistics.
\newblock URL: \url{https://aclanthology.org/2024.findings-acl.57}.

\bibitem{ChatGPT-Cost}
Dylan Patel and Afzal Ahmad.
\newblock {The Inference Cost Of Search Disruption – Large Language Model Cost Analysis}.
\newblock \url{https://www.semianalysis.com/p/the-inference-cost-of-search-disruption}, February 2023.
\newblock Accessed: 2025-09-09.

\bibitem{service-chain-mec-cost}
Kai Peng, Jiangtian Nie, Neeraj Kumar, Chao Cai, Jiawen Kang, Zehui Xiong, and Yang Zhang.
\newblock {Joint Optimization of Service Chain Caching and Task Offloading in Mobile Edge Computing}.
\newblock {\em Appl. Soft Comput.}, 103(C), May 2021.
\newblock \href {https://doi.org/10.1016/j.asoc.2021.107142} {\path{https://doi.org/10.1016/j.asoc.2021.107142}}.

\bibitem{efficient-transformer-infer}
Reiner Pope, Sholto Douglas, Aakanksha Chowdhery, Jacob Devlin, James Bradbury, Jonathan Heek, Kefan Xiao, Shivani Agrawal, and Jeff Dean.
\newblock {Efficiently Scaling Transformer Inference}.
\newblock {\em Proceedings of Machine Learning and Systems}, 5:606--624, 2023.

\bibitem{prompt-inversion-attack-0}
Wenjie Qu, Yuguang Zhou, Yongji Wu, Tingsong Xiao, Binhang Yuan, Yiming Li, and Jiaheng Zhang.
\newblock Prompt inversion attack against collaborative inference of large language models.
\newblock In {\em 2025 IEEE Symposium on Security and Privacy (SP)}, pages 1695--1712, 2025.
\newblock \href {https://doi.org/10.1109/SP61157.2025.00160} {\path{https://doi.org/10.1109/SP61157.2025.00160}}.

\bibitem{qwen2-speed-doc}
Qwen.
\newblock {Speed Benchmark}, 2024.
\newblock URL: \url{https://qwen.readthedocs.io/en/latest/benchmark/speed_benchmark.html}.

\bibitem{zero}
Samyam Rajbhandari, Jeff Rasley, Olatunji Ruwase, and Yuxiong He.
\newblock {ZeRO: Memory Optimizations Toward Training Trillion Parameter Models}.
\newblock In {\em Proceedings of the International Conference for High Performance Computing, Networking, Storage and Analysis}, SC '20. IEEE Press, 2020.

\bibitem{raza2021sok}
Ali Raza, Abraham Matta, Nabeel Akhtar, Vasiliki Kalavri, and Vatche Isahagian.
\newblock {SoK: Function-as-a-Service: From An Application Developer's Perspective}.
\newblock In {\em Journal of Systems Research - Mar 2021}, 2021.
\newblock URL: \url{https://openreview.net/forum?id=VdWaMgaTKtX}.

\bibitem{LIBRA}
Ali Raza, Zongshun Zhang, Nabeel Akhtar, Vatche Isahagian, and Ibrahim Matta.
\newblock {LIBRA: An Economical Hybrid Approach for Cloud Applications with Strict SLAs}.
\newblock In {\em 2021 IEEE International Conference on Cloud Engineering (IC2E)}, pages 136--146, 2021.
\newblock \href {https://doi.org/10.1109/IC2E52221.2021.00028} {\path{https://doi.org/10.1109/IC2E52221.2021.00028}}.

\bibitem{Redhat-White-Paper}
{Red Hat}.
\newblock {Bring Insights and Data Closer to Customers with Edge Computing}.
\newblock White paper, Red Hat, February 2022.
\newblock Accessed: 2025-09-09.
\newblock URL: \url{https://www.redhat.com/rhdc/managed-files/cl-bring-insight-data-customer-edge-computing-whitepaper-f30856pr-202202-en.pdf}.

\bibitem{Openshift-AI-press}
{Red Hat}.
\newblock {Red Hat OpenShift AI Accelerates Generative AI Adoption Across the Hybrid Cloud}.
\newblock \url{https://www.redhat.com/en/about/press-releases/red-hat-openshift-ai-accelerates-generative-ai-adoption-across-hybrid-cloud}, October 2023.
\newblock Accessed: 2025-09-09.

\bibitem{tinyyolo}
Joseph Redmon and Ali Farhadi.
\newblock {YOLO9000: Better, Faster, Stronger}, 2016.
\newblock \href {http://arxiv.org/abs/1612.08242} {\path{arXiv:1612.08242}}.

\bibitem{yolov3}
Joseph Redmon and Ali Farhadi.
\newblock {YOLOv3: An Incremental Improvement}, 2018.
\newblock \href {http://arxiv.org/abs/1804.02767} {\path{arXiv:1804.02767}}.

\bibitem{faster-r-cnn}
Shaoqing Ren, Kaiming He, Ross Girshick, and Jian Sun.
\newblock {Faster R-CNN: Towards Real-Time Object Detection with Region Proposal Networks}, 2016.
\newblock \href {http://arxiv.org/abs/1506.01497} {\path{arXiv:1506.01497}}.

\bibitem{AI-Accelerator-survey}
Albert Reuther, Peter Michaleas, Michael Jones, Vijay Gadepally, Siddharth Samsi, and Jeremy Kepner.
\newblock {AI Accelerator Survey and Trends}.
\newblock In {\em 2021 IEEE High Performance Extreme Computing Conference (HPEC)}, pages 1--9, 2021.
\newblock \href {https://doi.org/10.1109/HPEC49654.2021.9622867} {\path{https://doi.org/10.1109/HPEC49654.2021.9622867}}.

\bibitem{NN-bayesian-prob}
Michael~D Richard and Richard~P Lippmann.
\newblock {Neural Network Classifiers Estimate Bayesian a posteriori Probabilities}.
\newblock {\em Neural computation}, 3(4):461--483, 1991.

\bibitem{Llama}
Francisco Romero, Mark Zhao, Neeraja~J. Yadwadkar, and Christos Kozyrakis.
\newblock {Llama: A Heterogeneous Serverless Framework for Auto-Tuning Video Analytics Pipelines}.
\newblock In {\em Proceedings of the ACM Symposium on Cloud Computing}, SoCC '21, page 1–17, New York, NY, USA, 2021. Association for Computing Machinery.
\newblock \href {https://doi.org/10.1145/3472883.3486972} {\path{https://doi.org/10.1145/3472883.3486972}}.

\bibitem{autoencoder}
David~E. Rumelhart, James~L. McClelland, and PDP~Research Group.
\newblock {\em {Parallel Distributed Processing: Explorations in the Microstructure of Cognition, Vol. 1: Foundations}}, chapter Learning Internal Representations by Error Propagation, pages 318--362.
\newblock MIT Press, Cambridge, MA, 1986.

\bibitem{ILSVRC15}
Olga Russakovsky, Jia Deng, Hao Su, Jonathan Krause, Sanjeev Satheesh, Sean Ma, Zhiheng Huang, Andrej Karpathy, Aditya Khosla, Michael Bernstein, Alexander~C. Berg, and Li~Fei-Fei.
\newblock {ImageNet Large Scale Visual Recognition Challenge}.
\newblock {\em International Journal of Computer Vision (IJCV)}, 115(3):211--252, 2015.
\newblock \href {https://doi.org/10.1007/s11263-015-0816-y} {\path{https://doi.org/10.1007/s11263-015-0816-y}}.

\bibitem{AIoT-housekeeping}
Samsung.
\newblock {Samsung To Unveil New Vacuum Lineup That Redefines Home Cleaning With Enhanced AI at CES 2024}, 2024.
\newblock URL: \url{https://news.samsung.com/us/samsung-unveil-new-vacuum-lineup-redefines-home-cleaning-with-enhanced-ai-ces-2024/}.

\bibitem{mobilenetv2}
Mark Sandler, Andrew Howard, Menglong Zhu, Andrey Zhmoginov, and Liang-Chieh Chen.
\newblock {MobileNetV2: Inverted Residuals and Linear Bottlenecks}, 2019.
\newblock \href {http://arxiv.org/abs/1801.04381} {\path{arXiv:1801.04381}}.

\bibitem{SmartSwitch-AI-Accelerator}
Davide Sanvito, Giuseppe Siracusano, and Roberto Bifulco.
\newblock {Can the Network Be the AI Accelerator?}
\newblock In {\em Proceedings of the 2018 Morning Workshop on In-Network Computing}, NetCompute '18, page 20–25, New York, NY, USA, 2018. Association for Computing Machinery.
\newblock \href {https://doi.org/10.1145/3229591.3229594} {\path{https://doi.org/10.1145/3229591.3229594}}.

\bibitem{UniLCD-RL}
Kathakoli Sengupta, Zhongkai Shagguan, Sandesh Bharadwaj, Sanjay Arora, Eshed Ohn-Bar, and Renato Mancuso.
\newblock {UniLCD: Unified Local-Cloud Decision-Making via Reinforcement Learning}, 2024.
\newblock URL: \url{https://arxiv.org/abs/2409.11403}, \href {http://arxiv.org/abs/2409.11403} {\path{arXiv:2409.11403}}.

\bibitem{serverless-in-the-wild}
Mohammad Shahrad, Rodrigo Fonseca, Inigo Goiri, Gohar Chaudhry, Paul Batum, Jason Cooke, Eduardo Laureano, Colby Tresness, Mark Russinovich, and Ricardo Bianchini.
\newblock {Serverless in the Wild: Characterizing and Optimizing the Serverless Workload at a Large Cloud Provider}.
\newblock In {\em 2020 USENIX Annual Technical Conference (USENIX ATC 20)}, pages 205--218. USENIX Association, July 2020.
\newblock URL: \url{https://www.usenix.org/conference/atc20/presentation/shahrad}.

\bibitem{Task-Oriented-bottleneck}
Jiawei Shao, Yuyi Mao, and Jun Zhang.
\newblock {Learning Task-Oriented Communication for Edge Inference: An Information Bottleneck Approach}.
\newblock {\em IEEE Journal on Selected Areas in Communications}, 40(1):197--211, 2021.

\bibitem{shazeer2017}
Noam Shazeer, *Azalia Mirhoseini, *Krzysztof Maziarz, Andy Davis, Quoc Le, Geoffrey Hinton, and Jeff Dean.
\newblock {Outrageously Large Neural Networks: The Sparsely-Gated Mixture-of-Experts Layer}.
\newblock In {\em International Conference on Learning Representations}, 2017.
\newblock URL: \url{https://openreview.net/forum?id=B1ckMDqlg}.

\bibitem{AIoT-smartcity}
Siemens.
\newblock {From City Theory to Smart Tech Reality}, 2024.
\newblock URL: \url{https://www.siemens-advanta.com/whitepapers/smart-tech-reality}.

\bibitem{vgg}
Karen Simonyan and Andrew Zisserman.
\newblock {Very Deep Convolutional Networks for Large-Scale Image Recognition}.
\newblock {\em arXiv preprint arXiv:1409.1556}, 2015.
\newblock \href {http://arxiv.org/abs/1409.1556} {\path{arXiv:1409.1556}}.

\bibitem{AR-5G-edge-survey}
Yushan Siriwardhana, Pawani Porambage, Madhusanka Liyanage, and Mika Ylianttila.
\newblock {A Survey on Mobile Augmented Reality With 5G Mobile Edge Computing: Architectures, Applications, and Technical Aspects}.
\newblock {\em IEEE Communications Surveys \& Tutorials}, 23(2):1160--1192, 2021.
\newblock \href {https://doi.org/10.1109/COMST.2021.3061981} {\path{https://doi.org/10.1109/COMST.2021.3061981}}.

\bibitem{aws-edge-use-case-3}
SKT.
\newblock {SKT and AWS Launch the First 5G Edge Cloud Service in Korea}, 2024.
\newblock URL: \url{https://www.sktelecom.com/en/press/press_detail.do?page.page=1&idx=1494}.

\bibitem{lora-sparse}
Lin Song, Yukang Chen, Shuai Yang, Xiaohan Ding, Yixiao Ge, Ying-Cong Chen, and Ying Shan.
\newblock {Low-Rank Approximation for Sparse Attention in Multi-Modal LLMs}.
\newblock In {\em 2024 IEEE/CVF Conference on Computer Vision and Pattern Recognition (CVPR)}, pages 13763--13773, 2024.
\newblock \href {https://doi.org/10.1109/CVPR52733.2024.01306} {\path{https://doi.org/10.1109/CVPR52733.2024.01306}}.

\bibitem{ptflops}
Vladislav Sovrasov.
\newblock {ptflops: a FLOPs Counting Tool for Neural Networks in PyTorch Framework}.
\newblock \url{https://github.com/sovrasov/flops-counter.pytorch}, 2023.
\newblock Access Date: 2023-12-16.

\bibitem{Llumnix}
Biao Sun, Ziming Huang, Hanyu Zhao, Wencong Xiao, Xinyi Zhang, Yong Li, and Wei Lin.
\newblock Llumnix: dynamic scheduling for large language model serving.
\newblock In {\em Proceedings of the 18th USENIX Conference on Operating Systems Design and Implementation}, OSDI'24, USA, 2024. USENIX Association.

\bibitem{Integrated-Gradients}
Mukund Sundararajan, Ankur Taly, and Qiqi Yan.
\newblock {Axiomatic Attribution for Deep Networks}.
\newblock In {\em International conference on machine learning}, pages 3319--3328. PMLR, 2017.

\bibitem{geminiteam2023gemini}
Gemini Team et~al.
\newblock Gemini: A family of highly capable multimodal models, 2023.
\newblock \href {http://arxiv.org/abs/2312.11805} {\path{arXiv:2312.11805}}.

\bibitem{BranchyNet}
Surat Teerapittayanon, Bradley McDanel, and H.T. Kung.
\newblock {BranchyNet: Fast Inference via Early Exiting from Deep Neural Networks}.
\newblock In {\em 2016 23rd International Conference on Pattern Recognition (ICPR)}, pages 2464--2469, 2016.
\newblock \href {https://doi.org/10.1109/ICPR.2016.7900006} {\path{https://doi.org/10.1109/ICPR.2016.7900006}}.

\bibitem{DDNNs}
Surat Teerapittayanon, Bradley McDanel, and H.T. Kung.
\newblock {Distributed Deep Neural Networks Over the Cloud, the Edge and End Devices}.
\newblock In {\em 2017 IEEE 37th International Conference on Distributed Computing Systems (ICDCS)}, pages 328--339, 2017.
\newblock \href {https://doi.org/10.1109/ICDCS.2017.226} {\path{https://doi.org/10.1109/ICDCS.2017.226}}.

\bibitem{DDNN}
Surat Teerapittayanon, Bradley McDanel, and H.T. Kung.
\newblock Distributed deep neural networks over the cloud, the edge and end devices.
\newblock In {\em 2017 IEEE 37th International Conference on Distributed Computing Systems (ICDCS)}, pages 328--339, 2017.
\newblock \href {https://doi.org/10.1109/ICDCS.2017.226} {\path{https://doi.org/10.1109/ICDCS.2017.226}}.

\bibitem{google-autoencoder}
{TensorFlow}.
\newblock {Intro to Autoencoders}.
\newblock \url{https://www.tensorflow.org/tutorials/generative/autoencoder}, 2024.
\newblock Accessed: 2025-09-09.

\bibitem{CLONE}
Chunlin Tian, Xinpeng Qin, Kahou Tam, Li~Li, Zijian Wang, Yuanzhe Zhao, Minglei Zhang, and Chengzhong Xu.
\newblock Clone: customizing llms for efficient latency-aware inference at the edge.
\newblock In {\em Proceedings of the 2025 USENIX Conference on Usenix Annual Technical Conference}, USENIX ATC '25, USA, 2025. USENIX Association.

\bibitem{NoPeek}
Praneeth Vepakomma, Abhishek Singh, Otkrist Gupta, and Ramesh Raskar.
\newblock {NoPeek: Information leakage reduction to share activations in distributed deep learning}.
\newblock {\em arXiv preprint arXiv:2008.09161}, 2020.
\newblock URL: \url{https://arxiv.org/abs/2008.09161}, \href {https://doi.org/10.48550/ARXIV.2008.09161} {\path{https://doi.org/10.48550/ARXIV.2008.09161}}.

\bibitem{falcon}
Sameer Wagh, Shruti Tople, Fabrice Benhamouda, Eyal Kushilevitz, Prateek Mittal, and Tal Rabin.
\newblock {FALCON: Honest-Majority Maliciously Secure Framework for Private Deep Learning}.
\newblock {\em Proceedings on Privacy Enhancing Technologies}, 2021(2):323--343, 2021.
\newblock \href {https://doi.org/10.2478/popets-2021-0034} {\path{https://doi.org/10.2478/popets-2021-0034}}.

\bibitem{jpeg}
G.K. Wallace.
\newblock {The JPEG Still Picture Compression Standard}.
\newblock {\em IEEE Transactions on Consumer Electronics}, 38(1):xviii--xxxiv, 1992.
\newblock \href {https://doi.org/10.1109/30.125072} {\path{https://doi.org/10.1109/30.125072}}.

\bibitem{wang2020survey}
Bo~Wang, Changhai Wang, Wanwei Huang, Ying Song, and Xiaoyun Qin.
\newblock {A Survey and Taxonomy on Task Offloading for Edge-Cloud Computing}.
\newblock {\em IEEE Access}, 8:186080--186101, 2020.

\bibitem{survey-offload-EC}
Bo~Wang, Changhai Wang, Wanwei Huang, Ying Song, and Xiaoyun Qin.
\newblock {A Survey and Taxonomy on Task Offloading for Edge-Cloud Computing}.
\newblock {\em IEEE Access}, 8:186080--186101, 2020.
\newblock \href {https://doi.org/10.1109/ACCESS.2020.3029649} {\path{https://doi.org/10.1109/ACCESS.2020.3029649}}.

\bibitem{model-inversion-attack-2}
Kuan-Chieh Wang, YAN FU, Ke~Li, Ashish Khisti, Richard Zemel, and Alireza Makhzani.
\newblock {Variational Model Inversion Attacks}.
\newblock In M.~Ranzato, A.~Beygelzimer, Y.~Dauphin, P.S. Liang, and J.~Wortman Vaughan, editors, {\em Advances in Neural Information Processing Systems}, volume~34, pages 9706--9719. Curran Associates, Inc., 2021.
\newblock URL: \url{https://proceedings.neurips.cc/paper/2021/file/50a074e6a8da4662ae0a29edde722179-Paper.pdf}.

\bibitem{SkipNet}
Xin Wang, Fisher Yu, Zi-Yi Dou, Trevor Darrell, and Joseph~E. Gonzalez.
\newblock {SkipNet: Learning Dynamic Routing in Convolutional Networks}.
\newblock In {\em Proceedings of the European Conference on Computer Vision (ECCV)}, September 2018.

\bibitem{prompt-inversion-attack-lora}
Yiming Wang, Yu~Lin, Xiaodong Zeng, and Guannan Zhang.
\newblock Privatelora for efficient privacy preserving llm, 2023.
\newblock URL: \url{https://arxiv.org/abs/2311.14030}, \href {http://arxiv.org/abs/2311.14030} {\path{arXiv:2311.14030}}.

\bibitem{survey-collaborate-EC}
Yingchao Wang, Chen Yang, Shulin Lan, Liehuang Zhu, and Yan Zhang.
\newblock {End-Edge-Cloud Collaborative Computing for Deep Learning: A Comprehensive Survey}.
\newblock {\em IEEE Communications Surveys \& Tutorials}, pages 1--1, 2024.
\newblock \href {https://doi.org/10.1109/COMST.2024.3393230} {\path{https://doi.org/10.1109/COMST.2024.3393230}}.

\bibitem{DDI}
Yue Wang, Jianghao Shen, Ting-Kuei Hu, Pengfei Xu, Tan Nguyen, Richard~G. Baraniuk, Zhangyang Wang, and Yingyan Lin.
\newblock {Dual Dynamic Inference: Enabling More Efficient, Adaptive and Controllable Deep Inference}.
\newblock {\em IEEE Journal of Selected Topics in Signal Processing}, 2020.
\newblock URL: \url{https://par.nsf.gov/biblio/10159763}, \href {https://doi.org/10.1109/JSTSP.2020.2979669} {\path{https://doi.org/10.1109/JSTSP.2020.2979669}}.

\bibitem{workday-hybridcloud-infra}
Campbell Webb.
\newblock {Unleashing the Power of Innovation with Public Cloud}, 2024.
\newblock URL: \url{https://blog.workday.com/en-us/unleashing-the-power-innovation-with-public-cloud.html}.

\bibitem{survey-trustworthy-dist-ai}
Wenqi Wei and Ling Liu.
\newblock {Trustworthy Distributed AI Systems: Robustness, Privacy, and Governance}.
\newblock {\em ACM Comput. Surv.}, February 2024.
\newblock Just Accepted.
\newblock \href {https://doi.org/10.1145/3645102} {\path{https://doi.org/10.1145/3645102}}.

\bibitem{serverless-metric-benchmark}
Jinfeng Wen, Zhenpeng Chen, Jianshu Zhao, Federica Sarro, Haodi Ping, Ying Zhang, Shangguang Wang, and Xuanzhe Liu.
\newblock Scope: Performance testing for serverless computing.
\newblock {\em ACM Trans. Softw. Eng. Methodol.}, February 2025.
\newblock Just Accepted.
\newblock \href {https://doi.org/10.1145/3717609} {\path{https://doi.org/10.1145/3717609}}.

\bibitem{Zero-Time-Waste}
Maciej Wo\l~czyk, Bartosz W\'{o}jcik, Klaudia Ba\l~azy, Igor~T Podolak, Jacek Tabor, Marek \'{S}mieja, and Tomasz Trzcinski.
\newblock {Zero Time Waste: Recycling Predictions in Early Exit Neural Networks}.
\newblock In M.~Ranzato, A.~Beygelzimer, Y.~Dauphin, P.S. Liang, and J.~Wortman Vaughan, editors, {\em Advances in Neural Information Processing Systems}, volume~34, pages 2516--2528. Curran Associates, Inc., 2021.
\newblock URL: \url{https://proceedings.neurips.cc/paper/2021/file/149ef6419512be56a93169cd5e6fa8fd-Paper.pdf}.

\bibitem{xiao2023smoothquant}
Guangxuan Xiao, Ji~Lin, Mickael Seznec, Hao Wu, Julien Demouth, and Song Han.
\newblock {SmoothQuant: Accurate and Efficient Post-Training Quantization for Large Language Models}.
\newblock In {\em International Conference on Machine Learning}, pages 38087--38099. PMLR, 2023.

\bibitem{MLaaS-federation}
Shuzhao Xie, Yuan Xue, Yifei Zhu, and Zhi Wang.
\newblock {Cost Effective MLaaS Federation: A Combinatorial Reinforcement Learning Approach}.
\newblock In {\em IEEE INFOCOM 2022 - IEEE Conference on Computer Communications}, page 2078–2087. IEEE Press, 2022.
\newblock \href {https://doi.org/10.1109/INFOCOM48880.2022.9796701} {\path{https://doi.org/10.1109/INFOCOM48880.2022.9796701}}.

\bibitem{gating-DRL}
Xiong Xiong, Kan Zheng, Lei Lei, and Lu~Hou.
\newblock {Resource Allocation Based on Deep Reinforcement Learning in IoT Edge Computing}.
\newblock {\em IEEE Journal on Selected Areas in Communications}, 38(6):1133--1146, 2020.
\newblock \href {https://doi.org/10.1109/JSAC.2020.2986615} {\path{https://doi.org/10.1109/JSAC.2020.2986615}}.

\bibitem{Xiph}
{Xiph.Org Foundation}.
\newblock {Xiph.org Video Test Media [derf's collection]}.
\newblock \url{https://media.xiph.org/video/derf/}, 2024.
\newblock Accessed: 2025-09-09.

\bibitem{xu2024stealthy}
Xiaoyang Xu, Mengda Yang, Wenzhe Yi, Ziang Li, Juan Wang, Hongxin Hu, Yong Zhuang, and Yaxin Liu.
\newblock {A Stealthy Wrongdoer: Feature-Oriented Reconstruction Attack against Split Learning}.
\newblock In {\em Proceedings of the IEEE/CVF Conference on Computer Vision and Pattern Recognition (CVPR)}, 2024.
\newblock Accessed: 2025-07-10.
\newblock URL: \url{https://arxiv.org/abs/2405.04115}.

\bibitem{EOP}
Yuanjia Xu, Heng Wu, Wenbo Zhang, and Yi~Hu.
\newblock {EOP: Efficient Operator Partition for Deep Learning Inference over Edge Servers}.
\newblock In {\em Proceedings of the 18th ACM SIGPLAN/SIGOPS International Conference on Virtual Execution Environments}, VEE 2022, page 45–57, New York, NY, USA, 2022. Association for Computing Machinery.
\newblock \href {https://doi.org/10.1145/3516807.3516820} {\path{https://doi.org/10.1145/3516807.3516820}}.

\bibitem{NN_Structure_Search}
Zhao Yang, Shengbing Zhang, Ruxu Li, Chuxi Li, Miao Wang, Danghui Wang, and Meng Zhang.
\newblock {Efficient Resource-Aware Convolutional Neural Architecture Search for Edge Computing with Pareto-Bayesian Optimization}.
\newblock {\em Sensors}, 21(2), 2021.
\newblock URL: \url{https://www.mdpi.com/1424-8220/21/2/444}, \href {https://doi.org/10.3390/s21020444} {\path{https://doi.org/10.3390/s21020444}}.

\bibitem{model-inversion-attack-3}
Ziqi Yang, Jiyi Zhang, Ee-Chien Chang, and Zhenkai Liang.
\newblock {Neural Network Inversion in Adversarial Setting via Background Knowledge Alignment}.
\newblock In {\em Proceedings of the 2019 ACM SIGSAC Conference on Computer and Communications Security}, CCS '19, page 225–240, New York, NY, USA, 2019. Association for Computing Machinery.
\newblock \href {https://doi.org/10.1145/3319535.3354261} {\path{https://doi.org/10.1145/3319535.3354261}}.

\bibitem{Deep-Compressive-Offloading}
Shuochao Yao, Jinyang Li, Dongxin Liu, Tianshi Wang, Shengzhong Liu, Huajie Shao, and Tarek Abdelzaher.
\newblock {Deep Compressive Offloading: Speeding up Neural Network Inference by Trading Edge Computation for Network Latency}.
\newblock In {\em Proceedings of the 18th Conference on Embedded Networked Sensor Systems}, SenSys '20, page 476–488, New York, NY, USA, 2020. Association for Computing Machinery.
\newblock \href {https://doi.org/10.1145/3384419.3430898} {\path{https://doi.org/10.1145/3384419.3430898}}.

\bibitem{MLaaS-complexity-performance}
Yuanshun Yao, Zhujun Xiao, Bolun Wang, Bimal Viswanath, Haitao Zheng, and Ben~Y. Zhao.
\newblock {Complexity vs. Performance: Empirical Analysis of Machine Learning as a Service}.
\newblock In {\em Proceedings of the 2017 Internet Measurement Conference}, IMC '17, page 384–397, New York, NY, USA, 2017. Association for Computing Machinery.
\newblock \href {https://doi.org/10.1145/3131365.3131372} {\path{https://doi.org/10.1145/3131365.3131372}}.

\bibitem{ye2025flashinfer}
Zihao Ye, Lequn Chen, Ruihang Lai, Wuwei Lin, Yineng Zhang, Stephanie Wang, Tianqi Chen, Baris Kasikci, Vinod Grover, Arvind Krishnamurthy, and Luis Ceze.
\newblock {FlashInfer: Efficient and Customizable Attention Engine for LLM Inference Serving}.
\newblock In {\em Proceedings of the 8th Conference on Machine Learning and Systems (MLSys)}, Santa Clara, CA, USA, 2025.
\newblock To appear.
\newblock URL: \url{https://arxiv.org/abs/2501.01005}, \href {https://doi.org/10.48550/arXiv.2501.01005} {\path{https://doi.org/10.48550/arXiv.2501.01005}}.

\bibitem{opacus}
Ashkan Yousefpour, Igor Shilov, Alexandre Sablayrolles, Davide Testuggine, Karthik Prasad, Mani Malek, John Nguyen, Sayan Ghosh, Akash Bharadwaj, Jessica Zhao, Graham Cormode, and Ilya Mironov.
\newblock {Opacus: User-Friendly Differential Privacy Library in PyTorch}, 2022.
\newblock \href {http://arxiv.org/abs/2109.12298} {\path{arXiv:2109.12298}}.

\bibitem{yuan-etal-2025-native}
Jingyang Yuan, Huazuo Gao, Damai Dai, Junyu Luo, Liang Zhao, Zhengyan Zhang, Zhenda Xie, Yuxing Wei, Lean Wang, Zhiping Xiao, Yuqing Wang, Chong Ruan, Ming Zhang, Wenfeng Liang, and Wangding Zeng.
\newblock Native sparse attention: Hardware-aligned and natively trainable sparse attention.
\newblock In Wanxiang Che, Joyce Nabende, Ekaterina Shutova, and Mohammad~Taher Pilehvar, editors, {\em Proceedings of the 63rd Annual Meeting of the Association for Computational Linguistics (Volume 1: Long Papers)}, pages 23078--23097, Vienna, Austria, July 2025. Association for Computational Linguistics.
\newblock URL: \url{https://aclanthology.org/2025.acl-long.1126/}, \href {https://doi.org/10.18653/v1/2025.acl-long.1126} {\path{https://doi.org/10.18653/v1/2025.acl-long.1126}}.

\bibitem{label-smoothing-regularization-KD}
Li~Yuan, Francis~EH Tay, Guilin Li, Tao Wang, and Jiashi Feng.
\newblock {Revisiting Knowledge Distillation via Label Smoothing Regularization}.
\newblock In {\em Proceedings of the IEEE/CVF Conference on Computer Vision and Pattern Recognition (CVPR)}, June 2020.

\bibitem{FedMEC-conv-cli-dense-edge-dp}
Jiale Zhang, Yanchao Zhao, Junyu Wang, and Bing Chen.
\newblock {FedMEC: Improving Efficiency of Differentially Private Federated Learning via Mobile Edge Computing}.
\newblock {\em Mobile Networks and Applications}, 25(6):2421--2433, Dec 2020.
\newblock \href {https://doi.org/10.1007/s11036-020-01586-4} {\path{https://doi.org/10.1007/s11036-020-01586-4}}.

\bibitem{zhang2024sageattention2}
Jintao Zhang, Haofeng Huang, Pengle Zhang, Jia Wei, Jun Zhu, and Jianfei Chen.
\newblock Sageattention2: Efficient attention with thorough outlier smoothing and per-thread int4 quantization.
\newblock In {\em International Conference on Machine Learning (ICML)}, 2025.

\bibitem{zhang2025sageattention}
Jintao Zhang, Jia Wei, Pengle Zhang, Jun Zhu, and Jianfei Chen.
\newblock Sageattention: Accurate 8-bit attention for plug-and-play inference acceleration.
\newblock In {\em International Conference on Learning Representations (ICLR)}, 2025.

\bibitem{zhang2025spargeattn}
Jintao Zhang, Chendong Xiang, Haofeng Huang, Jia Wei, Haocheng Xi, Jun Zhu, and Jianfei Chen.
\newblock Spargeattn: Accurate sparse attention accelerating any model inference.
\newblock In {\em International Conference on Machine Learning (ICML)}, 2025.

\bibitem{Self-Distillation}
Linfeng Zhang, Chenglong Bao, and Kaisheng Ma.
\newblock {Self-Distillation: Towards Efficient and Compact Neural Networks}.
\newblock {\em IEEE Transactions on Pattern Analysis and Machine Intelligence}, 44(8):4388--4403, 2022.
\newblock \href {https://doi.org/10.1109/TPAMI.2021.3067100} {\path{https://doi.org/10.1109/TPAMI.2021.3067100}}.

\bibitem{GPT4RoI}
Shilong Zhang, Peize Sun, Shoufa Chen, Min Xiao, Wenqi Shao, Wenwei Zhang, Yu~Liu, Kai Chen, and Ping Luo.
\newblock {GPT4RoI: Instruction Tuning Large Language Model on Region-of-Interest}.
\newblock In {\em Computer Vision – ECCV 2024 Workshops: Milan, Italy, September 29–October 4, 2024, Proceedings, Part VIII}, page 52–70, Berlin, Heidelberg, 2025. Springer-Verlag.
\newblock \href {https://doi.org/10.1007/978-3-031-91813-1_4} {\path{https://doi.org/10.1007/978-3-031-91813-1_4}}.

\bibitem{praxipaas}
Zongshun Zhang, Rohan Kumar, Jason Li, Lisa Korver, Anthony Byrne, Gianluca Stringhini, Ibrahim Matta, and Ayse Coskun.
\newblock {PraxiPaaS: A Decomposable Machine Learning System for Efficient Container Package Discovery}.
\newblock In {\em 12th IEEE International Conference on Cloud Engineering}, 2024.

\bibitem{FSL-journal}
Zongshun Zhang, Andrea Pinto, Valeria Turina, Flavio Esposito, and Ibrahim Matta.
\newblock {Privacy and Efficiency of Communications in Federated Split Learning}.
\newblock {\em IEEE Transactions on Big Data}, 9(5):1380--1391, 2023.
\newblock \href {https://doi.org/10.1109/TBDATA.2023.3280405} {\path{https://doi.org/10.1109/TBDATA.2023.3280405}}.

\bibitem{label-smoothing-bias-variance-KD}
Helong Zhou, Liangchen Song, Jiajie Chen, Ye~Zhou, Guoli Wang, Junsong Yuan, and Qian Zhang.
\newblock {Rethinking Soft Labels for Knowledge Distillation: A Bias-Variance Tradeoff Perspective}.
\newblock {\em arXiv preprint arXiv:2102.00650}, 2021.

\bibitem{BBNet}
Hongbo Zhou, Weiwei Zhang, Chengwei Wang, Xin Ma, and Haoran Yu.
\newblock {BBNet: A Novel Convolutional Neural Network Structure in Edge-Cloud Collaborative Inference}.
\newblock {\em Sensors}, 2021.

\bibitem{mono-service-mec-cost}
Huan Zhou, Zhenning Wang, Hantong Zheng, Shibo He, and Mianxiong Dong.
\newblock {Cost Minimization-Oriented Computation Offloading and Service Caching in Mobile Cloud-Edge Computing: An A3C-Based Approach}.
\newblock {\em IEEE Transactions on Network Science and Engineering}, 10(3):1326--1338, 2023.
\newblock \href {https://doi.org/10.1109/TNSE.2023.3255544} {\path{https://doi.org/10.1109/TNSE.2023.3255544}}.

\bibitem{BertLosePatience}
Wangchunshu Zhou, Canwen Xu, Tao Ge, Julian McAuley, Ke~Xu, and Furu Wei.
\newblock {BERT Loses Patience: Fast and Robust Inference with Early Exit}.
\newblock {\em Advances in Neural Information Processing Systems}, 33:18330--18341, 2020.

\end{thebibliography}

\end{document}